\documentclass[aps,prx,10pt,twocolumn,floatfix,superscriptaddress,longbibliography]{revtex4-2}

\usepackage[utf8]{inputenc}
\usepackage{natbib}
\usepackage[normalem]{ulem}
\usepackage{graphicx}
\usepackage{mathtools}
\usepackage{siunitx}
\usepackage{amsmath,amssymb}
\usepackage{braket}
\usepackage{nicefrac}
\usepackage{xcolor}
\usepackage[backref=none,
bookmarksnumbered=true,
bookmarks=true,
bookmarksopen=true,
colorlinks=true,
citecolor=blue,
linkcolor=blue,
anchorcolor=green,
urlcolor=blue,unicode=false]{hyperref}
\newcommand{\mote}{{MoTe\textsubscript{2}\,}}
\newcommand{\wse}{{WSe\textsubscript{2}}}

\newcommand{\nuc}{N_{\text{UC}}}

\newcommand{\gmoire}{\mathbf{g}}

\newcommand{\layerpotreal}[2]{V^{#1}(#2)}

\newcommand{\tunnelreal}[1]{T(#1)}
\newcommand{\tunnelrealdagger}[1]{T^\dagger(#1)}

\newcommand{\cth}{C_{3z}}
\newcommand{\ctz}{C_{2z}}
\newcommand{\csz}{C_{6z}}

\newcommand{\trs}{\mathcal{T}}

\newcommand{\cty}{C_{2y}}
\newcommand{\hc}{h.\,c.}

\def\uc{\mathrm{UC}}
\def\blochpart{u_{\mathbf k}(\mathbf r)}
\def\blochpartideal{u^\mathrm{Ideal}_{\mathbf k}(\mathbf r)}

\def\blochpartidealz{u^\mathrm{Ideal}_{\mathbf k_0}(\mathbf r)}
\def\blochpartdelta{\Delta u_{\mathbf k}(\mathbf r)}
\def\blochpartdeltapara{\Delta u^\parallel_{\mathbf k}(\mathbf r)}
\def\blochpartdeltaperp{\Delta u^\perp_{\mathbf k}(\mathbf r)}
\def\spinor{\chi_{\mathbf k}(\mathbf r)}

\def\scalarpart{\psi}
\def\gaugephase{\gamma(\mathbf r)}

\def\berrycur{B_{\mathbf k} (\mathbf r)}
\def\chernnumber{C_{\mathbf k}}

\def\llwf{\mathfrak{u}}

\def\nguy{\boldsymbol{\hat n}(\mathbf r)}
\def\mguy{\boldsymbol{\hat m}(\mathbf r)}
\def\mguynor{\boldsymbol{\hat m}}

\newcommand{\removeKK}[1]{}

\usepackage{ifthen}

\def\indexsolution{\alpha}

\newcommand{\deltazero}[1][]{
   \ifthenelse{ \equal{#1}{} }
    {\delta (E_{\mathbf k, \indexsolution}- eV)}
    {\delta (E_{\mathbf k, #1}- eV)}
}

\newcommand{\rmrk}[2]{}
\newcommand{\comment}[1]{}

\UseRawInputEncoding

\graphicspath{}
\newcommand{\br}{\mathbf r}
\newcommand{\bk}{\mathbf k}

\newcommand{\manuscript}{manuscript}

\begin{document}
\title{Robustness of real-space topology in moir\'e systems 
}

\begin{abstract}
The appearance of fractional Chern insulators in moir\'e systems can be rationalized by 
the presence of a fictitious 
magnetic field associated with 
the spatial texture of layer-resolved electronic wavefunctions. Here, we present a systematic study of real-space topology and the associated fictitious magnetic fields in moir\'e systems.
We first show that at the level of individual Bloch wavefunctions, the
real-space Chern number, akin to a Pontryagin index, is a fragile marker. 
It generically vanishes except for specific limits where the Bloch functions exhibit fine-tuned zeroes within the unit cell, such as the chiral limit of twisted bilayer graphene (TBG) or the adiabatic regime of twisted homobilayer transition metal dichalcogenides (TMD). 
We then show that these limitations do not apply to textures associated with ensembles of Bloch wavefunctions, such as entire bands or the ensemble of states at a given energy. The Chern number of these textures defines a robust topological index protected by a spectral gap. We find that symmetries constrain it to be nonzero for both twisted TMDs and TBG across all twist angles and levels of corrugation, which can be verified in scanning tunneling microscopy measurements. By projection to the band ensemble texture, a single-component Hamiltonian under fictitious magnetic field emerges in a broad regime, enabling a direct comparison
of multicomponent bands to Landau-level wavefunctions.
We also study real-space topology within the topological heavy fermion model of TBG, finding that the real-space topological features are supported only by the light c-electrons.
  \end{abstract}

\author{Kry\v{s}tof Kol\'a\v{r}}
\affiliation{\mbox{Dahlem Center for Complex Quantum Systems and Fachbereich Physik, Freie Universit\"at Berlin, 14195 Berlin, Germany}}
\affiliation{Department of Applied Physics, Aalto University School of Science, FI-00076 Aalto, Finland}
\author{Kang Yang}
\thanks{Current address: Department of Physics, Westlake University, Hangzhou 310030, Zhejiang, China}
\affiliation{\mbox{Dahlem Center for Complex Quantum Systems and Fachbereich Physik, Freie Universit\"at Berlin, 14195 Berlin, Germany}}
\author{Felix von Oppen}
\affiliation{\mbox{Dahlem Center for Complex Quantum Systems and Fachbereich Physik, Freie Universit\"at Berlin, 14195 Berlin, Germany}}
\author{Christophe Mora}
\affiliation{Universit\'e Paris Cit\'e, CNRS,  Laboratoire  Mat\'eriaux  et  Ph\'enom\`enes  Quantiques, 75013  Paris,  France}
        
\date{\today}

\maketitle
\section{Introduction}

Moir\'e materials are now well-established as a platform to study correlated electronic phenomena.  
Two prime examples are twisted bilayer graphene (TBG) \cite{macdonaldBistritzerMoireBandsTwisted2011} and twisted bilayer transition metal dichalcogenides (TMDs)\cite{macdonaldWuTopologicalInsulatorsTwisted2019}, hosting correlated insulating as well as superconducting phases \cite{jarillo-herreroCaoCorrelatedInsulatorBehaviour2018,deanYankowitzTuningSuperconductivityTwisted2019,kimHaoElectricFieldTunable2021,yazdaniOhEvidenceUnconventionalSuperconductivity2021,efetovLuSuperconductorsOrbitalMagnets2019,jarillo-herreroCaoNematicityCompetingOrders2021,liLiuTuningElectronCorrelation2021,nadj-pergeAroraSuperconductivityMetallicTwisted2020,efetovStepanovUntyingInsulatingSuperconducting2020,youngSaitoIndependentSuperconductorsCorrelated2020,ilaniZondinerCascadePhaseTransitions2020a,yazdaniWongCascadeElectronicTransitions2020,deanWangCorrelatedElectronicPhases2020,makXiaUnconventionalSuperconductivityTwisted2024,deanGuoSuperconductivityTwistedBilayer2024,pasupathyGhiottoQuantumCriticalityTwisted2021}. Moreover, three moir\'e systems -- rhombohedral multilayer graphene~\cite{juLuFractionalQuantumAnomalous2024,Zhengguang2025,xie2024fqah}, twisted bilayer \mote{} ~\cite{xuCaiSignaturesFractionalQuantum2023,xuParkObservationFractionallyQuantized2023,shanZengThermodynamicEvidenceFractional2023,liXuObservationIntegerFractional2023,feldmanFouttyMappingTwisttunedMultiband2024,makKangEvidenceFractionalQuantum2024}, and TBG above $5$T~\cite{yacobyXieFractionalChernInsulators2021} -- have been demonstrated to feature the fractional quantum anomalous Hall effect ~\cite{Neupertprl_2011, zhao_review, regnault2011, tang2011, Sheng2011, qi2011}.

The emergence of these fractional quantum Hall states at zero magnetic field~\cite{Neupertprl_2011, zhao_review, regnault2011, tang2011, Sheng2011, qi2011} suggests the presence of a mechanism that generates a fictitious magnetic field experienced by the electrons. Support for this intuitive expectation has been provided for twisted bilayer TMDs within the so-called adiabatic approximation \cite{yaoYuGiantMagneticField2020,yaoZhaiTheoryTunableFlux2020,macdonaldMorales-DuranMagicAnglesFractional2024}, where a fictitious magnetic field arises from the geometric phase associated with the spatial rotation of the layer spinor. Experimentally observed in TMDs~\cite{shihZhangExperimentalSignatureLayer2025,yankowitzThompsonMicroscopicSignaturesTopology2025}, such a layer-skyrmion texture has also been predicted in chiral TBG (and other multilayer graphene systems)~\cite{moraGuerciLayerSkyrmionsIdeal2024} and is expected to arise quite generally in ideal Chern bands of multicomponent systems. In the adiabatic limit of both , TMDs and chiral TBG, the two-component spinor multiplies a Landau level wavefunction~\cite{vishwanathTarnopolskyOriginMagicAngles2019,yangWangExactLandauLevel2021,canoWangChiralApproximationTwisted2021,vishwanathLedwithFractionalChernInsulator2020,liuWangOriginModelFractional2023,parkerDongCompositeFermiLiquid2023}, whose associated effective magnetic field precisely compensates for the skyrmion texture in a sense that we will make precise.

The texture-induced fictitious magnetic field provides a useful intuitive understanding of the properties of moir\'e systems. However, realistic experimental settings are typically beyond these 
idealized limits. 
In this paper, we explore the robustness of the layer skyrmions away from these   
limits, and establish a general method to quantify the real-space topology in multicomponent (layer or sublattice)  Bloch bands. We propose a real-space Chern number obtained from an ensemble of wavefunctions and show that it is both robust and nontrivial for a wide range of parameters in both TBG and TMD systems, implying a generic fictitious magnetic field
in these systems. 

\begin{figure}[t]
    \centering
    \includegraphics[width=.99\columnwidth]{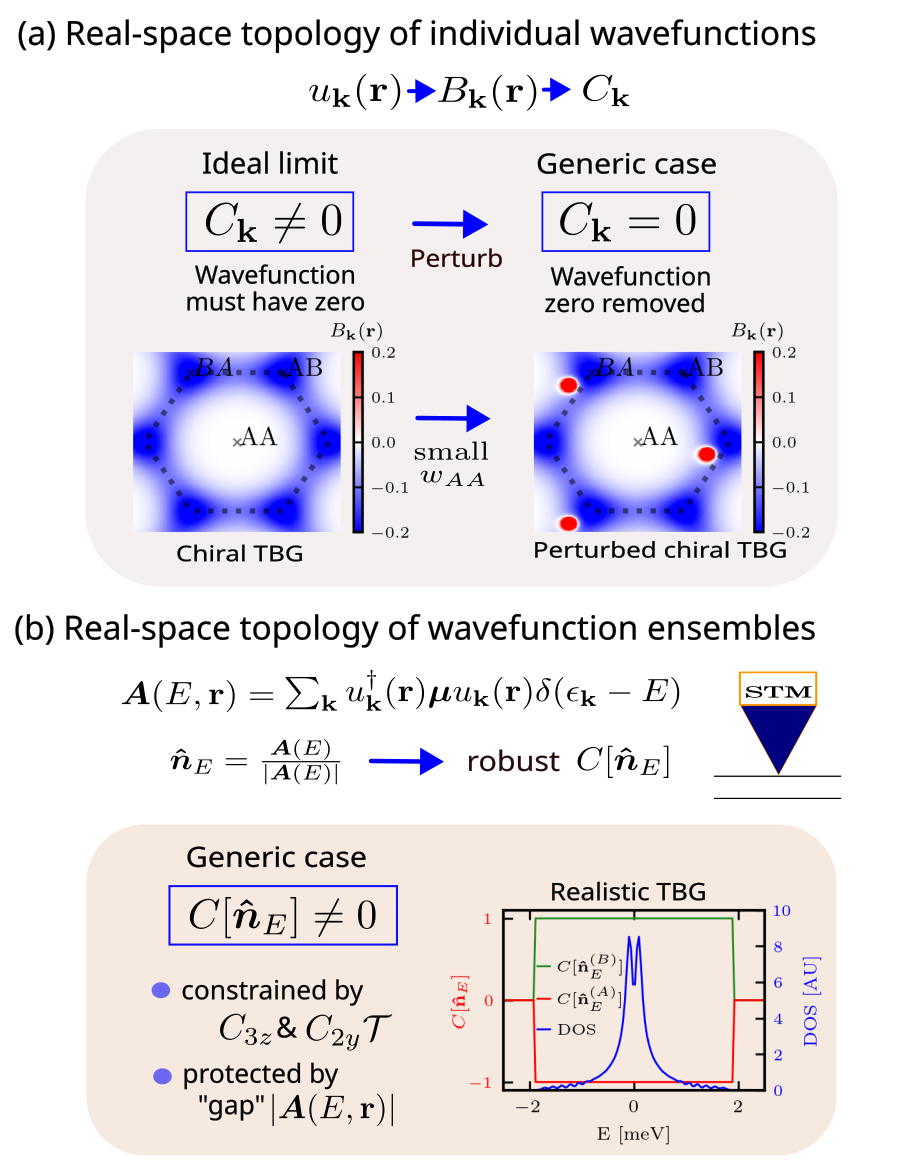}
    \caption{(a) Illustration of real-space topology of individual wavefunctions at a given momentum. 
     The real-space Chern number $C_{\mathbf k}$ for a spinor wavefunction $u_{\mathbf k}(\mathbf r)$ 
     is obtained by integrating the real-space Berry curvature $B_{\mathbf k}$ over the unit cell.
     At zero applied magnetic field, the real-space Chern number is nonzero
    only in fine-tuned cases, requiring the full spinor wavefunction to vanish identically at some point in the unit cell. 
    Infinitesimal perturbations generically remove this zero and cause the real-space 
    Chern number $C_{\mathbf k}$ to vanish.
    Left inset shows the real-space Berry curvature $B_{\mathbf k}$ for a wavefunction in the chiral limit, integrating to $C_{\mathbf k}=-1$.
    Right inset shows the real-space Berry curvature for a perturbed chiral TBG, leading to the vanishing of the Chern number.
(b) Real-space topology of wavefunction ensembles as probed in scanning tunneling microscopy. 
In contrast to (a), the real-space Chern number is robust in this case, 
and protected by symmetries for both TBG and twisted TMDs.
Inset shows the energy-dependent sublattice-projected Chern numbers 
$C[\boldsymbol{\hat n}^{(A)}_E]$ and $C[\boldsymbol{\hat n}^{(B)}_E]$
for a realistic model of TBG. }
\label{fig:figzero}
\end{figure}

We first quantify the skyrmion winding by constructing a real-space Chern number for individual Bloch wavefunctions. However, we find that this Chern number generically vanishes at zero applied magnetic field. Only in certain special situations such as high-symmetry momenta or the ideal Chern-band and adiabatic limits, this real-space Chern number differs from zero. In particular, a nonzero Chern number requires that the multicomponent Bloch wavefunction exhibits zeroes in the unit cell. Since the zeroes are generically lifted by perturbations, this only occurs in fined-tuned situations, as illustrated in Fig.~\ref{fig:figzero}a.

We find that the situation changes markedly if one considers textures formed by ensembles of wavefunctions, for example within a given band or at a given energy.
The texture is defined by a spatially varying three-dimensional vector (Fig.~\ref{fig:figzero}b), with the topology remaining robust as long as the vector does not vanish. At zero applied flux, this texture can exhibit a nonzero real-space Chern number without fine tuning, with spatial symmetries imposing general constraints on it. In particular, we demonstrate that both twisted TMDs and sublattice-projected TBG always possess a nontrivial real-space topology as a consequence of their $\cth$ and $\cty \trs$ symmetries. 
Our proposed ensemble real-space texture for bands allows one to directly 
associate a well-defined fictitious magnetic field to it, forming a complementary characterization beyond conventional band geometry.
Crucially, this association does not place any conditions on the underlying Hamiltonian
and allows a direct mapping to Landau levels to be established,
making it a versatile method to rationalize the emergence of the fractional quantum anomalous Hall effect in moir\'e materials beyond ideal limits.

We demonstrate this approach by applying it to TBG, finding that the chiral limit topological texture
is present even at realistic corrugation.
For a realistic model of twisted bilayer \mote{} at $\theta=2.1^\circ$ \cite{choAhnFirstLandauLevel2024,xiaoWangHigherLandauLevelAnalogues2024}, we find a robust well-defined texture for the relevant band, even though the adiabatic approximation \cite{yaoYuGiantMagneticField2020,yaoZhaiTheoryTunableFlux2020,macdonaldMorales-DuranMagicAnglesFractional2024} breaks down in that model.
We show that robust textures allow the multi-component problem to be naturally mapped to a single-component problem
with a magnetic field defined by the texture, allowing a natural interpretation for the experimentally observed fractional Chern insulators in terms of Landau levels.
We find large Landau level overlaps for the two topmost bands in twisted bilayer \mote{} at $\theta=2.1^\circ$. 

Nontrivial fixed-energy textures are observable in scanning tunneling experiments,
and our results for TMDs explain recent experimental observations~\cite{shihZhangExperimentalSignatureLayer2025,yankowitzThompsonMicroscopicSignaturesTopology2025}.
Remarkably, we find that the same phenomenology is expected for TBG when considering sublattice-projected textures. 
Furthermore, we find that layer-projected textures in TBG, which are better accessible in STM experiments, are typically also nontrivial provided the $\ctz \trs$ symmetry is broken, for instance by a sublattice term.

We also study real-space textures in the presence of an integer number of flux quanta $\Phi$ threading the unit cell.
We find that in this case, generically, the real-space Chern number equals the applied flux, $C_{\mathbf k} = \Phi$.
We show that close to ideal limits, this result arises from localized skyrmions 
that are generated when overall wavefunction zeroes are removed.

Finally, we examine the real-space topology of TBG in the context of the topological heavy fermion model~\cite{bernevigSongMagicAngleTwistedBilayer2022,calugaru2023,Shi2022}. We find that the  Wannier states localized around the AA sites, accounting for most of the central bands, do not exhibit any real-space topology. Instead, the nontrivial topology originates from the itinerant electrons, which are concentrated in the AB and BA regions where the ensemble texture exhibits significant winding.

This \manuscript{} is structured as follows. In Sec.~\ref{sec:realspacetextures}, we present a general theory of real-space topology of individual multicomponent wavefunctions. In Sec.~\ref{sec:windinglllikewf}, we introduce Landau-level-like wavefunctions, which feature nonzero real-space Chern numbers. In Sec.~\ref{sec:windingstabilityzeroes}, we show that the real-space Chern number generically vanishes for wavefunctions at zero applied magnetic field, and discuss twisted bilayer TMDs and TBG for illustration. In Sec.~\ref{sec:topoofensembles}, we introduce a theory of 
real-space topology of ensembles of wavefunctions, and in 
Sec.~\ref{sec:topoofensembles-appl}, we demonstrate its robustness on twisted bilayer TMDs and TBG.
In Sec.~\ref{sec:bandlandaumapping}, we show how band ensemble textures 
allow multicomponent bands to be understood in terms of single-component wavefunctions
experiencing a fictitious magnetic field.
In Sec.~\ref{sec:heavyfermion}, we discuss the textures of TBG in the context of the heavy fermion model.
We conclude with a discussion in Sec.~\ref{sec:windingdiscussion}.

\section{Real-space textures of wavefunctions}
\label{sec:realspacetextures}

Consider a general $N \times N$ continuum Hamiltonian in two spatial dimensions, 
where the $N$ components typically correspond to layer or sublattice degrees of freedom.
We consider a situation where the unit cell is threaded by 
an integer number of $\Phi$ flux quanta, so that the system 
retains its original lattice periodicity with lattice vectors $\mathbf a_1, \mathbf a_2$, 
and the eigenstates are (magnetic) translation eigenstates labeled 
by a band index $n$, a flavor (spin and/or valley) index $\lambda$,  and a momentum $\mathbf k$ in the Brillouin zone.
While below, we apply this framework to single-particle models, 
it applies equally well to many-body models where interactions are treated within the mean-field approximation.

\subsection{Real-space topology in terms of wavefunctions}\label{sec:realspace-topology}

The eigenstates in a given band are expressed as $e^{i\mathbf k\cdot \mathbf r} \blochpart$ with the $N$-component Bloch spinors
\begin{equation}
u_{\mathbf k}(\mathbf r) =  [u^{1}_{\mathbf k}(\mathbf r),u^{2}_{\mathbf k}(\mathbf r), \ldots, u^{N}_{\mathbf k}(\mathbf r)]^T,
\end{equation}
where the band $n$ and flavor index $\lambda$ are omitted for brevity. We can use standard considerations based on the magnetic translation algebra to define Bloch states in the presence of a uniform magnetic field corresponding to an integer number $\Phi$ of flux quanta per unit cell, see App.\ \ref{sec:BlochMagnField} for details. We use a generalized Landau gauge, which is fully specified by the conditions that 
(i) $\mathbf{A}$ is linear in the coordinates,  (ii) the  translation operator along the $\mathbf{a}_1$-direction does not involve an Aharonov-Bohm phase, and (iii) the translation operator in the $\mathbf{a}_2$ direction is a product of commuting exponentials, which depend on position and momentum, respectively. In
this gauge, the Bloch spinors
are (quasi-)periodic functions over the unit cell with 
\begin{equation}
\label{eq:ucbc}
\begin{split}
u_{\mathbf k}(\mathbf r+ \mathbf a_1 ) & =
u_{\mathbf k}(\mathbf r) \\
u_{\mathbf k}(\mathbf r+ \mathbf a_2 ) & = e^{i\gaugephase} u_{\mathbf k}(\mathbf r). 
\end{split}
    \end{equation}
Here, the uniform applied magnetic field leads to the Aharonov-Bohm phase 
\begin{equation} 
\label{eq:windinggaugephasedef}
\gaugephase = -\Phi \mathbf G_1 \cdot \mathbf r ,
\end{equation}
where $\mathbf{G}_1=2\pi \frac{\mathbf{a}_2\times\mathbf{\hat z}}{\mathbf{\hat z}\cdot(\mathbf{a}_1 \times\mathbf{a}_2)}$ is a reciprocal lattice vector.
Importantly, the Aharonov-Bohm phase $\gaugephase$ introduces $\Phi$ additional phase windings of the Bloch functions $u_\mathbf{k}(\mathbf{r})$ along the boundary of the unit cell. This follows since 
\begin{equation}
    \int_{\mathbf R+\mathbf a_1}^{\mathbf R}d\mathbf r \cdot \,\nabla_{\br} \gaugephase = 2\pi\Phi,
\end{equation}
encoding the integer magnetic flux enclosed by the unit cell.

We now consider implications for scalar ($N=1$) and spinor ($N>1$) Bloch functions:\\

(i) Scalar wavefunctions can be expressed in terms of modulus and phase factor, 
\begin{equation}
    \blochpart = e^{i \theta_{\bk} (\br)} | \blochpart |.
\end{equation}
Since the Bloch wavefunction is smooth as a function of $\mathbf r$, its phase $\theta_{\mathbf k} (\br)$  is also smooth and
well-defined, except at vortex cores where $\blochpart$ vanishes and around which $\theta_{\bk} (\br)$ winds by multiples of $2 \pi$. The Aharanov-Bohm phase $\gamma (\br)$ imposes that 
$\theta_{\bk} (\br)$ winds by $2\pi \Phi$ along the unit cell boundary. For nonzero $\Phi$, this implies that $\blochpart$ must have vortices (and zeroes), with a total vorticity of $\Phi$~\footnote{More precisely, the wavefunction $\blochpart$ defines a nonperiodic map from the unit cell to complex numbers $\mathbb C$. Outside vortex cores where $\blochpart$ vanishes, it defines a map from part of the unit cell to nonzero complex numbers $\mathbb C^\times=\mathbb
C-\{0\}$. The fundamental group of $\mathbb C^\times$, given $\pi_1(\mathbb
C^\times)=\mathbb Z$, captures the phase winding along a closed loop. Since we can deform the loop along the unit cell border to a set of loops encircling
each vortex core, it follows that the total vorticity $\Phi$ is given by the
sum of the vorticities associated with each zero of $\blochpart$.}. We note that Landau levels provide a natural basis for describing such wavefunctions at finite flux~\cite{wangLiuTheoryGeneralizedLandau2024}. \\

(ii) For spinor wavefunctions, we first consider the case where $\blochpart$ is nonzero everywhere. Then, we can define the normalized spinor 
\begin{equation}
\label{eq:defspinorfromwf}
\spinor =  \frac{\blochpart}{\left|\blochpart \right|}
\end{equation}
and a real-space Berry connection 
\begin{equation}
\label{eq:berryconnection}
A_{\mathbf k}(\mathbf r) = -i\chi^\dagger_{\mathbf k}(\mathbf r) \nabla_{\br}   \chi_{\mathbf k}(\mathbf r),
\end{equation}
treating the position $\mathbf r$ as a parameter. This Berry connection describes the additional geometric phase obtained by $\chi^{\phantom{\dagger}}_{\mathbf k}(\mathbf r)$ during adiabatic variations of the 
real-space position $\mathbf{r}$. The associated Berry curvature 
\begin{equation}
\label{eq:realspaceberry}
B_{\mathbf k} (\mathbf r)= \nabla_{\mathbf r} \times A_{\mathbf k}(\mathbf r)
\end{equation}
is invariant under a change of phase, $\blochpart \to  e^{i \alpha_{\mathbf k}(\mathbf r)}\blochpart$.
Upon integrating over a unit cell ($\uc$), this defines the real-space Chern number
\begin{equation}
\label{eq:realspacechern}
C_{\mathbf k} = \frac{1}{2\pi} \int_{\uc} d \mathbf r B_{\mathbf k}(\mathbf r) \in \mathbb{Z}
\end{equation}
quantifying the skyrmion winding of the spinor wavefunction.
Importantly, the (quasi-)periodicity expressed in Eq.~\eqref{eq:ucbc} fixes the integrated Berry connection along the boundary of the unit cell as
\begin{equation}
\label{eq:aharonovbohm}
 \oint_{\partial \uc} d \mathbf r \cdot A_{\mathbf k}(\mathbf r) = 2\pi \Phi,
\end{equation}
corresponding to the Aharonov-Bohm phase when going around the unit cell. Since the spinor $\spinor$ is a smooth function of $\mathbf r$ throughout the unit cell, Stokes' theorem relates the two integrals in Eqs.~\eqref{eq:realspacechern} and~\eqref{eq:aharonovbohm} and proves that $C_{\mathbf k} = \Phi$. The real-space Chern number of wavefunctions, which are nonzero everywhere, is thus strictly tied to the number of flux quanta per unit cell. In particular, it has to vanish at zero magnetic field.
\\

(iii) We now consider spinor wavefunctions that have zeroes in the unit cell. Then,
the normalized spinor introduced in Eq.~\eqref{eq:defspinorfromwf} is ill-defined at the zeroes of $\blochpart$, which prevents a direct use of the Berry connection. Near each zero, the phase of the spinor in Eq.~\eqref{eq:defspinorfromwf} can wind, forming a vortex or anti-vortex in real space. 
We explicitly extract the geometric content of the wavefunction by factoring out these zeroes and phase vortices using the decomposition
\begin{equation}
\label{eq:generalwfdecomp}
u_{\mathbf k}(\mathbf r) = \chi_{\mathbf k}^{\Phi_1}(\mathbf r)  \scalarpart^{\Phi_2}_{\mathbf k} (\mathbf r).
\end{equation}
Here,  $\chi_{\mathbf k}^{\Phi_1}$ is a normalized spinor wavefunction, and
$\scalarpart^{\Phi_2}_{\mathbf k}$ is a scalar (single-component) wavefunction
with the same norm as $\blochpart$, or $|\scalarpart^{\Phi_2}_{\mathbf k}|^2 =
u^{\dagger}_{\mathbf k} u_{\mathbf k}$, in particular the same zeroes and vortices.
The new spinor $\chi_{\mathbf k}^{\Phi_1}$ is now a
smooth function of $\mathbf{r}$, even at the zeroes of $\blochpart$, and defines a real-space connection
as well as a Berry curvature following Eqs.~\eqref{eq:berryconnection}
and~\eqref{eq:realspaceberry}. It can thus be characterized by a real-space Chern number $C_{\mathbf k} =
\Phi_1$~\footnote{There is some arbitrariness
(gauge-freedom) in the decomposition of Eq.~\eqref{eq:generalwfdecomp} as a
phase term $e^{i \alpha_{\mathbf k} (\mathbf r)}$ can always be transferred
between the spinor and the scalar parts. Although this shift of phase  can
change the Berry connection, it does not affect the gauge-invariant Berry
curvature, yielding a uniquely defined real-space Chern number $C_{\bk}$, fixed by the winding $\Phi_1$ of $\boldsymbol{\hat n_{\mathbf k}}$.}. 
From Stokes' theorem, one obtains that the line integral of the Berry connection along the unit cell boundary yields $2 \pi \Phi_1$. 
This implies that the spinor part can be chosen to satisfy the boundary conditions in  Eq.~\eqref{eq:ucbc} with flux $\Phi_1$. Consequently, the scalar part $\scalarpart^{\Phi_2}_{\mathbf k}$ satisfies the same boundary conditions with the integer flux 
\begin{equation}
\Phi_2 = \Phi - \Phi_1,
\end{equation}
in order for the Bloch wavefunction $\blochpart$ to satisfy Eq.~\eqref{eq:ucbc} with the applied magnetic flux $\Phi$. For the scalar function $\scalarpart^{\Phi_2}_{\mathbf k}$, we return to case (i), according to which the integer $\Phi_2$ determines its total vorticity, corresponding to the sum of the vorticities at all isolated zeroes of the function. 

Physically, the factorization employed in Eq.~\eqref{eq:generalwfdecomp} thus corresponds to a splitting of the magnetic flux into a skyrmion winding $\Phi_1$ carried by the spinor part and a set of real-space vortices with total vorticity $\Phi_2$ of the scalar part. At zero external flux, we must have $\Phi_1 = -\Phi_2$, implying that the skyrmion texture exactly compensates for the magnetic phase of the scalar part. As we will elaborate below, this structure is precisely the one found in the adiabatic limit of TMDs~\cite{macdonaldMorales-DuranMagicAnglesFractional2024} and in  chiral TBG~\cite{vishwanathTarnopolskyOriginMagicAngles2019,moraGuerciLayerSkyrmionsIdeal2024}. For nonzero external flux $\Phi$, we have that $C_{\mathbf k} = \Phi_1 = \Phi-\Phi_2$, so that 
$C_{\mathbf k} \neq \Phi$ is possible if the wavefunction contains vortices.

\subsection{Formulation in terms of Bloch sphere}\label{sec:blochsphere}

For two-component wavefunctions ($N=2$), 
the spinor winding in real space can be equivalently described using the three-dimensional  unit vector~\footnote{In contrast to $\frac{\blochpart}{\left|\blochpart \right|}$ and its phase vortices, the ratio $\frac{u_{\mathbf k}^\dagger (\mathbf r) \boldsymbol{\mu} \blochpart}{|u_{\mathbf k}^\dagger (\mathbf r)   \blochpart |}$ can be extended continuously at the zeroes of $\blochpart$. We also note that the scalar part $\scalarpart^{\Phi_2}_{\mathbf k}$ drops out from this ratio.}
\begin{equation}
\label{eq:defvectorcoefficient}
\boldsymbol{\hat n_{\mathbf k}}(\mathbf r) = \frac{u_{\mathbf k}^\dagger (\mathbf r) \boldsymbol{\mu} \blochpart}{u_{\mathbf k}^\dagger (\mathbf r)   \blochpart }
= \chi^\dagger_{\mathbf k}(\mathbf r) \boldsymbol{\mu} \chi^{\phantom{\dagger}}_{\mathbf k}(\mathbf r),
\end{equation}
where $\boldsymbol{\mu}=(\mu_x,\mu_y,\mu_z)$ is the vector of Pauli matrices 
and the superscript $\Phi_1$ is omitted for simplicity.
Vector fields $\boldsymbol{\hat n}(\mathbf r)$ on the Bloch sphere can be characterized by 
their Pontryagin density
\begin{equation}\label{eq:pontryagin-density}
B\left[\boldsymbol{\hat n}\right](\mathbf r)= \frac{1}{2}{\boldsymbol{\hat n}}(\mathbf{r})\cdot \partial_x \boldsymbol{\hat n}(\mathbf{r})\times \partial_y \boldsymbol{\hat n}(\mathbf{r})
\end{equation}
and the associated index
\begin{equation}
\label{eq:realspacechernpontryagin}
C\left[\boldsymbol{\hat n}\right] =\, \frac{1}{4\pi} \int_{\uc} d \mathbf r \, {\boldsymbol{\hat n}}(\mathbf{r})\cdot \partial_x \boldsymbol{\hat n}(\mathbf{r})\times \partial_y \boldsymbol{\hat n}(\mathbf{r}).
\end{equation}
If we compute the Pontryagin density for $\boldsymbol{\hat n}= \boldsymbol{\hat n_{\mathbf k}}$ as defined in Eq.~\eqref{eq:defvectorcoefficient}, we recover the Berry curvature in Eq.~\eqref{eq:realspaceberry},
\begin{equation}
B\left[\boldsymbol{\hat n}_{\mathbf k}\right](\mathbf r)=B_{\mathbf k}(\mathbf r),
\end{equation}
which also implies 
\begin{equation}
C\left[\boldsymbol{\hat n_{\mathbf k}}\right]  = C_{\mathbf k} = \Phi_1
\end{equation}
for the Pontryagin index. The advantage of this formulation is that it allows the real-space Chern number to be defined also for vector fields that do not arise out of individual wavefunctions. We will return to this possibility in Sec.~\ref{sec:topoofensembles}, where we consider textures of wavefunction ensembles.

The vector-field formulation naturally extends to multi-component wavefunctions with $N>2$ using the properties of the $\mathfrak{su}(N)$ Lie algebra \cite{piechonGrafBerryCurvatureQuantum2021,unalKempNestedsphereDescription$N$level2022}. One defines a $N(N+1)/2$-dimensional vector 
$\boldsymbol{\hat n_{\mathbf k}}(\mathbf r)|_a =\chi^\dagger_{\mathbf k}(\mathbf r) \lambda_a \chi^{\phantom{\dagger}}_{\mathbf k}(\mathbf r)$, where $\lambda_a$
are the generators of $\mathfrak{su}(N)$. The real-space Chern number is then defined using Eq.~\eqref{eq:realspacechernpontryagin}
with a suitable generalization of the triple product to the $\mathfrak{su}(N)$ case \cite{piechonGrafBerryCurvatureQuantum2021,unalKempNestedsphereDescription$N$level2022}.

Physical insight into $N>2$ component wavefunctions can be gleaned by projecting onto specific components,
obtaining  effective two-component wavefunctions 
and permitting an interpretation in terms of the Bloch sphere as defined above. For instance, $N=4$ systems comprising layer and sublattice can be projected on a single sublattice to obtain a two-component spinor. We will later perform this projection for TBG.

\section{Landau-level-like wavefunctions}
\label{sec:windinglllikewf}

We illustrate the decomposition of the Bloch wavefunction into spinor and scalar components by two examples. The first example considers twisted bilayer transition metal dichalcogenides within the adiabatic approximation. The second corresponds to the abstract class of ideal wavefunctions, exemplified by twisted bilayer graphene in the chiral limit. In both cases, the scalar part experiences an
effective magnetic field and can be expressed in terms of Landau levels.

\subsection{Twisted bilayer TMDs}
\label{sec:windinglllikewftmd}

Hole-doped twisted bilayer TMDs can be understood in terms of parabolically dispersing holes moving in a layer-Zeeman field denoted $\boldsymbol{\Delta}(\mathbf r)$
and a scalar potential $\Delta_0(\mathbf r)$.
In the $K$-valley, the Hamiltonian reads
\begin{equation}
\label{eq:sphammanewgauge}
H_{\text{tTMD}}^{K}= -\frac{\hbar^2 }{2m^*}\begin{pmatrix}\mathbf k^2 &0\\0& (\mathbf k -\mathbf q_1)^2\end{pmatrix} + \boldsymbol{\Delta}(\mathbf r)\cdot \boldsymbol{\mu} +\Delta_0(\mathbf r)\mu_0,
\end{equation}
where $m^*$ is the effective mass of the holes, $\mu_{0,x,y,z}$ denotes the Pauli matrices in layer space and ${\mathbf k} = - i \nabla_{\mathbf r}$ the spatial derivative. Moreover, 
$\mathbf q_1=(0,4\pi \theta/(3a_0))$ is the momentum offset between the valence band maxima of the two layers, with $a_0$ the lattice constant and $\theta$ the twist angle.

\begin{figure}[t]
    \centering
    \includegraphics[width=\columnwidth]{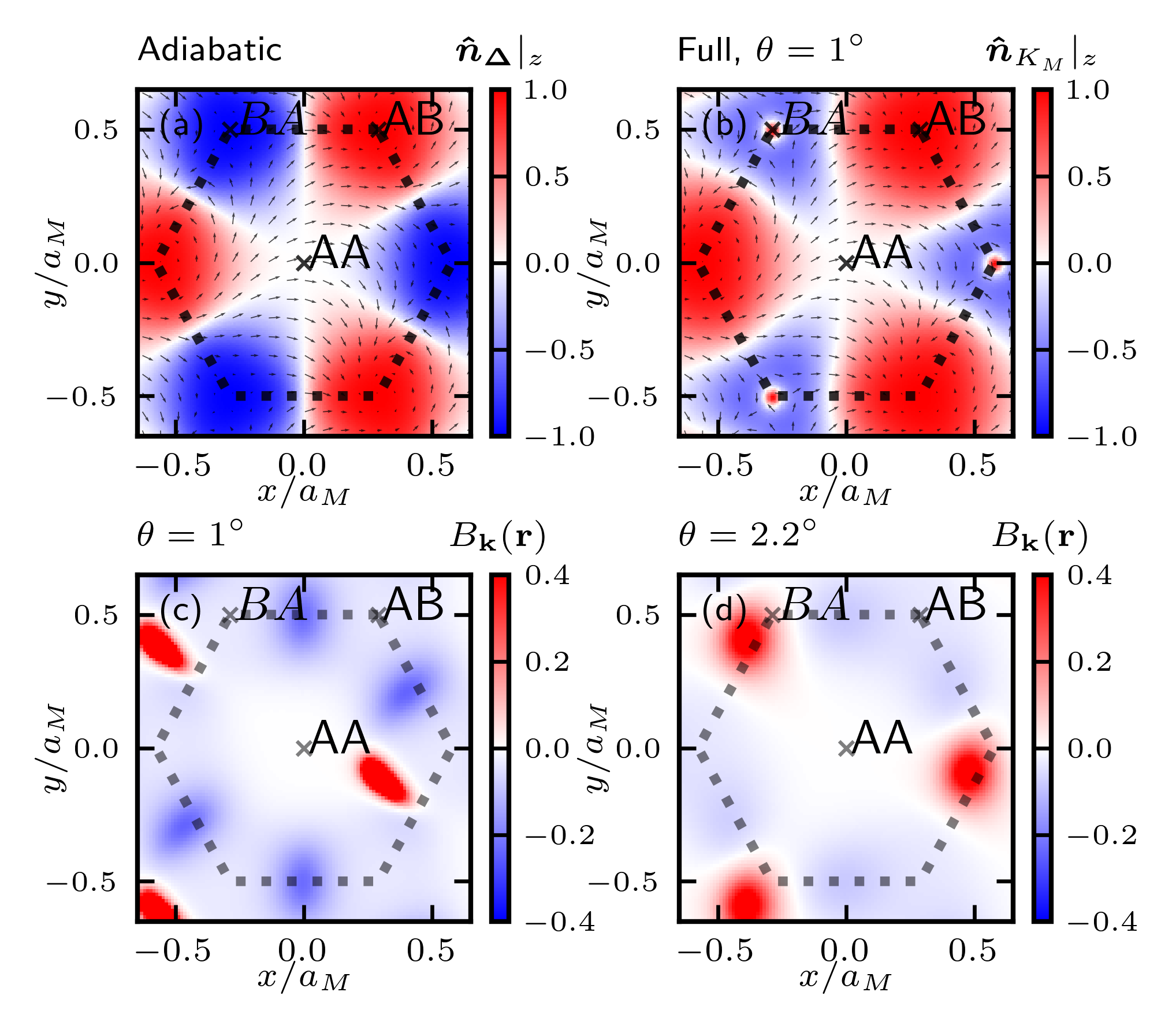}
    \caption{(a) Unit cell map  of the texture $\boldsymbol{\hat n_{\boldsymbol{\Delta}}}(\mathbf r) = \boldsymbol{\Delta}(\mathbf r)/|\boldsymbol{\Delta}(\mathbf r)|$ [Eq.~\eqref{eq:defndelta}]. 
    Colormap shows the $z$ component, while the arrows denote the in-plane direction of $\boldsymbol{\hat n}$.
    (b) Same as (a) but for a texture for the wavefunction of the topmost band wavefunction of \wse{} at $\theta=1^\circ$, defined in in Eq.~\eqref{eq:defvectorcoefficient}.
    (c) Real space Berry curvature at at $\theta=1^\circ$ at a generic momentum $\mathbf k = [-0.2,0.2] \, q_1$, with $q_1 = |\mathbf q_1|$.
    (d) Same as (c), but at $\theta=2.2^\circ$.}
    \label{fig:figone}
\end{figure}

The layer Zeeman field can have a nonzero real-space Pontryagin index
$C[\boldsymbol{\hat n}_{\boldsymbol{\Delta}}]$ 
of the unit vector 
\begin{equation}
\label{eq:defndelta}
\boldsymbol{\hat n_{\boldsymbol{\Delta}}}(\mathbf r) = \boldsymbol{\Delta}(\mathbf r)/|\boldsymbol{\Delta}(\mathbf r)|.\end{equation}
For illustration, Fig.\ \ref{fig:figone}a exhibits the texture corresponding to the $K$-valley Hamiltonian  of \wse{}, using model parameters from Ref.~\cite{fuDevakulMagicTwistedTransition2021} [see App.~\ref{app:tmdintro} for explicit forms of $\Delta_0(\mathbf r)$ and $\boldsymbol{\Delta}(\mathbf r)$]. 
This texture has $C[\boldsymbol{\hat n}_{\boldsymbol{\Delta}}]=-1$, so that the vector $\boldsymbol{\hat n}_{\boldsymbol{\Delta}}(\mathbf r)$ covers the entire unit sphere as the unit cell is traversed.

In the adiabatic approach \cite{yaoYuGiantMagneticField2020,yaoZhaiTheoryTunableFlux2020,macdonaldMorales-DuranMagicAnglesFractional2024}, one assumes that the layer pseudospin aligns with the local layer-Zeeman field at every $\mathbf r$. Focusing on the low-energy subspace, we can then write the wavefunction as
\begin{eqnarray}
\label{eq:wfinadiabaticapprox}
 u^{Ad}_{\mathbf k}(\mathbf r)= \chi^+(\mathbf r) \, \psi_{\mathbf k} (\mathbf r),
\end{eqnarray}
where we use the eigenspinors of the local layer-Zeeman field,
\begin{equation}
\boldsymbol{\Delta}(\mathbf r)\cdot \boldsymbol{\mu}\, \chi^\pm(\mathbf r) =\pm |\boldsymbol{\Delta}(\mathbf r)| \chi^\pm(\mathbf r)
\end{equation}
Thus, the spinor inherits a skyrmion winding directly from the winding of the layer Zeeman field as described by $\boldsymbol{\hat n_{\boldsymbol{\Delta}}}(\mathbf r)$. In practice, this approximation holds for sufficiently small twist angles, where the kinetic energy remains perturbative \cite{wuLiVariationalMappingChern2024,macdonaldShiAdiabaticApproximationAharonovCasher2024}.

We remark that the same approximation has been used to understand the properties of electrons in a skyrmion lattice
\cite{taillefumierBrunoTopologicalHallEffect2004,kohnoNakazawaTopologicalHallEffect2019,aritaMatsuiSkyrmionsizeDependenceTopological2021}, which can be described using a Hamiltonian such as Eq.~\eqref{eq:sphammanewgauge}.

The scalar wavefunction $\psi_{\mathbf k}$ in Eq.\ \eqref{eq:wfinadiabaticapprox} is an eigenfunction of the effective Hamiltonian
\begin{equation}
\label{eq:sphammacdoapp}
H^{K}_{\text{Adiabatic}}= -\frac{(\hbar \mathbf k-e\tilde{\mathbf{A}}(\mathbf r))^2}{2m^*} + \tilde V(\mathbf r).
\end{equation}
Importantly, this Hamiltonian includes a fictitious vector potential. The associated fictitious magnetic field is also determined by the texture of $\boldsymbol{\hat n_{\boldsymbol{\Delta}}}$, 
\begin{equation}
    \nabla\times \tilde{\mathbf{A}}(\mathbf r) = -\frac{\hbar}{e} B\left[\boldsymbol{\hat n_{\boldsymbol{\Delta}}}\right](\mathbf r).
\end{equation}
The fictitious field is nonuniform but due to the winding of $\boldsymbol{\hat n_{\boldsymbol{\Delta}}}$, its average corresponds to one flux quantum threading the moir\'e unit cell. The additional scalar potential is given by $\tilde V(\mathbf r) = - D(\mathbf r) + |\boldsymbol \Delta (\mathbf r)| + \Delta_0(\mathbf r)$ including the  correction 
\begin{equation} \label{eq:expressionD}
D(\mathbf{r}) = \frac{\hbar^2}{8m^*}\sum_{i=x,y}[\partial_{i} \mathbf{\hat n_{\boldsymbol{\Delta}}}(\mathbf{r})]^2
\end{equation}
associated with the quantum geometry.

Thus, the wavefunction \eqref{eq:wfinadiabaticapprox} in the adiabatic approximation has precisely the structure of Eq.\ \eqref{eq:generalwfdecomp}. The spinor winding of $\chi_+(\mathbf r)$ gives $\Phi_1 = -1$, while the scalar wavefunction has $\Phi_2 = 1$ and can be decomposed into Landau-level wavefunctions, $\llwf^{\Phi=1}_{n,\mathbf k}$ with Landau-level index $n$ and wave vector $\mathbf{k}$,
\begin{equation}  
\label{eq:LLmixture}
\scalarpart_{\mathbf k}^{\Phi_2=1}(\mathbf r) =  \sum_{n=0}^{+\infty} c^{\mathbf k}_{n} \, \llwf^{\Phi=1}_{n,\mathbf k}(\mathbf r)
\end{equation}
which are eigenstates of the magnetic translations across the moir\'e unit cell~\cite{moraKolarHofstadterSpectrumChern2024}. Consequently, the fictitious magnetic field exactly compensates for the spinor winding, consistent with the absence of an applied magnetic field, $\Phi=0$. We note that in this adiabatic approximation, $\boldsymbol{\hat n}_{\boldsymbol{\Delta}}(\mathbf r)$ of Eq.~\eqref{eq:defndelta} and 
$\boldsymbol{\hat n}_{\mathbf k}(\mathbf r)$ of Eq.~\eqref{eq:defvectorcoefficient} coincide, since 
the spinor part of the wavefunction precisely follows the texture of $\boldsymbol{\Delta}(\mathbf r)$.

\subsection{Ideal bands}
\label{sec:windinglllikewfideal}
Ideal Chern bands are believed to be well suited for hosting fractional Chern insulator phases \cite{royRoyBandGeometryFractional2014,yangWangExactLandauLevel2021,vishwanathLedwithFractionalChernInsulator2020,bergholtzAbouelkomsanQuantumMetricInduced2023,ledwithFujimotoHigherVortexabilityZero2024,wangLiuTheoryGeneralizedLandau2024,liuWangOriginModelFractional2023,crepelEstienneIdealChernBands2023,PhysRevB.111.L201105}
allowing for the construction of Laughlin-like trial states \cite{vishwanathLedwithFractionalChernInsulator2020}.
At zero applied magnetic field $\Phi=0$, their wavefunctions possess an exact Landau-level representation \cite{yangWangExactLandauLevel2021}
\begin{equation}
\blochpartideal = N_{\mathbf k} \mathcal{B}(\mathbf r) \llwf^{\Phi=1}_{0,\mathbf k}(\mathbf r),
\label{eq:idealwf}
\end{equation}
and carry a momentum-space Chern number of $1$. Here, $\llwf^{\Phi=1}_{0,\mathbf k}(\mathbf r)$ is the zeroth Landau-level wavefunction at momentum $\mathbf k$ for a magnetic field of one flux quantum per unit cell,
$N_{\mathbf k}$ is a normalization constant, and $\mathcal{B}(\mathbf r)$ is a scalar or vector function.
Note that the lowest-Landau-level wavefunction $\llwf^{\Phi=1}_{0,\mathbf k}(\mathbf r)$ is quasiperiodic, satisfying Eq.~\eqref{eq:ucbc} with $\Phi=1$.
The Landau-level quasiperiodicity of $\llwf^{\Phi=1}_{0,\mathbf k}(\mathbf r)$ implies that $\mathcal{B}(\mathbf r)$ is necessarily also quasiperiodic, with an effective flux $\Phi=-1$. For certain models with double Dirac cones~\cite{sunWanTopologicalExactFlat2023,Eugenio_2023}, $\mathcal{B}(\mathbf r)$ may have a single component and behave as a Landau level with negative flux, compensating the flux of $\llwf^{\Phi=1}_{0,\mathbf k}$. However, for single-Dirac models, $\mathcal{B}(\mathbf r)$ is nonvanishing everywhere and necessarily possesses $N>1$ components. We prove this result in App.~\ref{app:zero}. Then, the negative flux $\Phi=-1$ arises as the geometric phase induced by a real-space vector winding. In this case, the ideal-band wavefunction of Eq.~\eqref{eq:idealwf} reproduces the form of Eq.~\eqref{eq:generalwfdecomp} with the identification
\begin{eqnarray}
\scalarpart^{\Phi_2=1}_{\mathbf k} (\mathbf r) &=& N_{\mathbf k}|\mathcal{B}(\mathbf r)| \llwf^{\Phi=1}_{0,\mathbf k}(\mathbf r)\\
\chi^{\Phi_1=-1} (\mathbf r) &=& \mathcal{B}(\mathbf r)/|\mathcal{B}(\mathbf r)|
\end{eqnarray}
and therefore possesses the real-space Chern number $C_{\mathbf k}=-1$ for all $\mathbf k$. For TBG, the decomposition also applies~\cite{sternShefferChiralMagicangleTwisted2021} 
to a nonzero magnetic field, $\Phi \ne 0$: the spinor part $\chi^{\Phi_1=-1}$  is unchanged, while the scalar part evolves as $N_{\mathbf k}|\mathcal{B}(\mathbf r)| \llwf^{1+\Phi}_{0,\mathbf k}(\mathbf r)$, where the lowest-Landau-level wavefunction carries the flux $\Phi_2 = 1+ \Phi$. Ideal Chern bands with a momentum Chern number of $-1$ exhibit the same  decomposition with the opposite values $\Phi_1=1$ and $\Phi_2=-1$.

An important example of ideal bands is twisted bilayer graphene in the chiral limit.
The $K$-valley
Hamiltonian of twisted bilayer graphene is 
\begin{equation}
\label{eq:tbgham}
H_{TBG}^{K}=\begin{pmatrix}
v_F \mathbf k \cdot \boldsymbol{\sigma} & \tunnelreal{\mathbf r}\\
\tunnelrealdagger{\mathbf r}&v_F (\mathbf k- \mathbf q_1) \cdot \boldsymbol{\sigma}
\end{pmatrix},
\end{equation}
where $v_F$ is the  Dirac velocity of monolayer graphene and the tunneling matrix is 
\begin{multline}
\label{eq:tbgtunnel}
\tunnelreal{\mathbf r} = w_{AA}\sigma_0  (1 + e^{i\gmoire_5\cdot \mathbf r} + e^{i\gmoire_6\cdot \mathbf r} )+ \\
+ w_{AB}\sigma_x (1 +e^{2\pi i /3 \sigma_z} e^{i\gmoire_5\cdot \mathbf r} +e^{4\pi i /3 \sigma_z} e^{i\gmoire_6\cdot \mathbf r} ).
\end{multline}
Here, $w_{AA}$ ($w_{AB}$) is the intra- (inter-)sublattice tunneling 
and the moir\'e reciprocal vectors $\gmoire_j$ are defined as the $(j-1)$th counterclockwise $\csz$ rotations of $\gmoire_1 =(4\pi \theta /(\sqrt{3}a_0),0)$.

In the chiral limit, the intra-sublattice tunneling $w_{AA}$ is set to zero, $w_{AA}=0$ \cite{vishwanathTarnopolskyOriginMagicAngles2019}, while the inter-sublattice is $w_{AB} = \SI{110}{meV}$. At the magic angle $\theta= 1.09^\circ$, one obtains two sublattice polarized ideal bands at zero energy.
In what follows, we will use 
an explicit sublattice mass $m_s>0$ to tune the Hamiltonian of Eq.~\eqref{eq:tbgham},
\begin{equation}
\label{eq:tbghamsubl}
H_{TBG;m_s}^{K}= H_{TBG}^{K} + m_s \sigma_z.
\end{equation}
Such a term can arise either due to sublattice-polarizing interactions or an aligned hBN layer.
In this model, the two zero-energy sublattice-polarized flat bands are split and have energies $\pm m_s$, allowing them to be studied independently. The two flat bands are ideal bands. The A-polarized band at energy $m_s$ has a momentum-space Chern number of \( 1 \) and follows the form of Eq.~\eqref{eq:idealwf}, while the B-polarized band at energy $-m_s$ has opposite Chern numbers and effective fluxes. The real-space Berry curvature of the $A$-band is illustrated in Fig.~\ref{fig:figzero}a and integrates to $C_{\mathbf k} = -1$ over the unit cell.
We note that other twisted multilayer configurations also exhibit ideal bands in the chiral limit~\cite{vishwanathDongExactManyBodyGround2022,khalafLedwithFamilyIdealChern2022,moraGuerciChernMosaicIdeal2024,guerci2023nature}.

\section{Stability of real-space Chern numbers for individual wavefunctions}
\label{sec:windingstabilityzeroes}

While these two examples exhibit a nontrivial decomposition, with $\Phi_1$ and $\Phi_2$ both being nonzero at vanishing magnetic field ($\Phi=0$), we will now show that these cases are fine-tuned. Broadly speaking, in the classification of Sec.~\ref{sec:realspacetextures}, case (iii) is fine-tuned, whereas case (ii) represents the generic situation. Generically, the real-space Chern number is zero, $C_{\mathbf k}=0$, in the absence of a magnetic field.
As a direct consequence, the wavefunctions of twisted bilayer TMDs beyond the adiabatic approximation and twisted bilayer graphene away from the chiral limit both exhibit a vanishing real-space Chern number.

\subsection{Stability to perturbations of the zeroes of the wavefunction in Eq.~\eqref{eq:generalwfdecomp}}
\label{subsec:stabilitywf}
A crucial feature of the wavefunctions in case (iii) of Sec.~\ref{sec:realspacetextures} is that they vanish at certain points in the unit cell. We therefore consider the stability of the zeroes of multicomponent wavefunctions. We write a general perturbed wavefunction
\begin{equation} 
\label{eq:windingperturbedwf}
u_{\mathbf k}(\mathbf r) = \chi_{\mathbf k}^{\Phi_1}(\mathbf r)  \scalarpart^{\Phi_2}_{\mathbf k} (\mathbf r)+ \epsilon 
\blochpartdelta,
\end{equation}
where $\blochpartdelta$ perturbs weakly ($\epsilon\ll 1$) about a wavefunction with a zero in the unit cell. Importantly, we expect $\blochpartdelta$ to be a generic wavefunction, 
which is not necessarily aligned with $\chi_{\mathbf k}^{\Phi_1}(\mathbf r)$.
We decompose 
\begin{equation} 
\blochpartdelta = \blochpartdeltapara + \blochpartdeltaperp.
\end{equation}
into two orthogonal components, where
$\blochpartdeltapara$ is aligned with the unperturbed spinor $\chi_{\mathbf k}^{\Phi_1}(\mathbf r)$ and $\blochpartdeltaperp$ is orthogonal to it at every $\mathbf{r}$.
The admixture of $\blochpartdeltapara$ preserves the structure of Eq.~\eqref{eq:generalwfdecomp},
and therefore also the real-space Chern number, so that we can absorb it into 
the unperturbed wavefunction.
In contrast, the admixture of $\blochpartdeltaperp$ will generically remove any zeroes.
To see that, consider \begin{equation}
|u_{\mathbf k}(\mathbf r)|^2 =|\scalarpart^{\Phi_2}_{\mathbf k} (\mathbf r)|^2+ \epsilon^2|\blochpartdeltaperp|^2,
\end{equation}
where we used the orthogonality property of $\blochpartdeltaperp$. 
The full wavefunction can only have a zero if both 
$\scalarpart^{\Phi_2}_{\mathbf k} (\mathbf r)$ and $\blochpartdeltaperp$  vanish simultaneously. We do not expect the zeroes of a perturbing wavefunction to coincide with those of
the original wavefunction so that generically, the zeroes of $\scalarpart^{\Phi_2}_{\mathbf k} (\mathbf r)$ will be removed.

Thus, the zeroes of the wavefunction in
Eq.~\eqref{eq:generalwfdecomp} are generically lifted under perturbations.
Their removal corresponds to a discontinuous transition from case (iii) to case
(ii), where the real-space Chern number becomes identically zero for zero applied flux (or $C_{\mathbf k}=\Phi$ for general fluxes). Exceptions to
this behavior occur at high-symmetry momenta in the Brillouin zone, where
real-space zeroes can be protected by symmetries and thus remain stable under
symmetry-preserving perturbations. However, such symmetry-protected zeroes arise
only at isolated points (sets of measure zero) in the Brillouin zone.

\subsection{Localized skyrmions generated by perturbed zeroes}
\label{subsec:skyrmdiscussion}

We now turn to an explicit illustration of the connection between the removal
of vortex zeroes and a change in real-space Chern number. Effectively, removing a zero
of the wavefunction introduces an additional skyrmion-winding of the spinor
wavefunction, which changes the real-space Chern number. To see this intuitively, we
consider a two-component spinor. The zero of the wavefunction
is removed by displacing the zeroes of the two components relative to each
other. At the zero of the bottom layer, the spinor has all its weight in the
upper layer, i.e., at the north pole of the Bloch sphere. Conversely, at the
zero of the top layer, the spinor has all its weight in the lower layer, i.e.,
at the south pole. Intuitively,
thus, we expect the spinor to cover the entire Bloch sphere in the vicinity of these two points. 

We illustrate this further with the (unnormalized) two-component spinor 
\begin{equation}
u_{\epsilon,\mathbf k'}(\mathbf r) =  [\llwf^{\Phi=1}_{0,\mathbf k=0},\epsilon \llwf^{\Phi=1}_{0,\mathbf k'}]^T
\end{equation}
at applied flux $\Phi=1$. Here, $\llwf^{\Phi=1}_{0,\mathbf k}$ denotes the lowest-Landau-level wavefunction, and we assume that most of the weight is in the upper component, $\epsilon \ll1$. Since the spinor components are equal to Landau-level wavefunctions at different wavevectors, the positions of their zeroes are different. The position of these zeroes is known analytically~\cite{vishwanathTarnopolskyOriginMagicAngles2019},
\begin{eqnarray}
\label{eq:landaulevelzeroposition}
    {\bf r}_0 = {\bf r}_{BA} + \frac{|{\bf a}_1|}{|{\bf G}_1|} \, \hat{\mathbf{z}}\times {\bf k}, 
\end{eqnarray}
where the basis vectors ${\bf a}_1$ and ${\bf G}_1$ are defined below Eq.~\eqref{eq:ucbc} and Eq.~\eqref{eq:windinggaugephasedef}, respectively. Thus, for $\mathbf k=0$, the Landau-level wavefunction has a zero at $\mathbf r_0=\bf r_{BA}$, while for nonzero $\mathbf k$ it shifts away from $\bf r_{BA}$. At $\epsilon=0$, the spinor $u_{0,\mathbf k'}(\mathbf r)$ points upwards everywhere in the unit cell,
implying a real-space Chern number $C_{0,\mathbf k'}=0$.  
For finite $\epsilon$, but $\mathbf k'=0$, the spinor direction $u_{\epsilon,0}(\mathbf r) \propto [1,\epsilon]^T$ is still constant everywhere, so that  
$C_{\epsilon,0}=0$. In both of these cases, the entire wavefunction vanishes at $\mathbf r=\bf r_{BA}$, which  corresponds to case (iii) of our wavefunction classification in Sec,\ \ref{sec:realspace-topology}. The overall zero is removed when $\epsilon \neq 0$ and $\mathbf k' \neq 0$, corresponding to case (ii). Since $\Phi=1$, we thus expect $C_{\epsilon \neq 0 ,\mathbf k' \neq 0 }=1$, suggesting that the two Landau-level-like spinor components produce a skyrmion. For $\epsilon \ll 1$, the wavefunction is pointing up everywhere except close to $\mathbf r = \bf r_{BA}$, where it can be  approximated as $u_{\epsilon,\mathbf k'}(\mathbf r \sim {\bf r_{BA}}) \approx  [B_1z ,\epsilon \llwf^{\Phi=1}_{0,\mathbf k'}(\bf r_{BA})]^T$, where $B_1$ is a constant and $z=(x +iy)-(x_{BA}+iy_{BA})$ is the vortex of the zeroth Landau level. For any finite $\epsilon$, the spinor points down at $\mathbf r=\bf r_{BA}$. Away from $\mathbf r =\bf r_{BA}$, the spinor tilts upward, winding in-plane due to the factor of $z$ in the up component. For large $|\mathbf r-\bf r_{BA}|$, the spinor points up, giving rise to a Chern number $C_{\mathbf k}=1$ as expected.

\subsection{Fragility in twisted bilayer TMDs} 
\label{sec:fragilityTMD}

We proceed to illustrate  explicitly how the removal of zeroes leads to a sudden change in the real-space Chern number for twisted bilayer TMDs. The adiabatic approximation assumes a wavefunction of the form of Eq.~\eqref{eq:wfinadiabaticapprox},
effectively projecting on the Hilbert space of wavefunctions with spinor components given by $\chi^+(\mathbf r)$.
Going beyond the adiabatic approximation by allowing the admixture of wavefunctions with spinors 
proportional to $\chi^-(\mathbf r)$ leads to a Bloch wavefunction of the form
\begin{equation}
    u_{\mathbf k}(\mathbf r)= \chi^+(\mathbf r) \, \psi_{\mathbf k} (\mathbf r) + \beta_{\mathbf k} \chi^-(\mathbf r) \, \varphi_{\mathbf k} (\mathbf r).
\end{equation}
This introduces a second scalar function $\varphi_{\mathbf k} (\mathbf r)$.
The second term, with $|\beta_{\mathbf k}| \ll 1$, is small close to the adiabatic limit and scales with the twist angle. The scalar wavefunction $\psi_{\mathbf k}(\mathbf r) $ is a mixture of Landau-level wavefunctions for nonzero flux, see Eq.~\eqref{eq:LLmixture}, and thus exhibits a zero at some position $r_{\mathbf k} $ in the unit cell, which depends on the momentum ${\mathbf k}$. In general, $\varphi_{\mathbf k} (\mathbf r_{\mathbf k}) \ne 0$ meaning that while the Bloch wavefunction is nearly aligned with the spinor $\chi^+(\mathbf r)$ across most of the unit cell, it becomes dominated by the oppositely oriented spinor $\chi^-(\mathbf r)$ in the vicinity of $\mathbf r_{\mathbf k}$ and thus never vanishes completely. The spinor $\boldsymbol{\hat n}_{\mathbf k}(\mathbf r)$, obtained from Eqs.~\eqref{eq:defvectorcoefficient} and \eqref{eq:defspinorfromwf},  closely follows the texture of $\boldsymbol{\Delta}(\mathbf r)$, such that $\boldsymbol{\hat n}_{\mathbf k}(\mathbf r) \simeq \boldsymbol{\hat n}_{\boldsymbol{\Delta}}(\mathbf r)$, except near $\mathbf r_{\mathbf k}$ where it flips sign, $\boldsymbol{\hat n}_{\mathbf k}(\mathbf r_{\mathbf k}) = - \boldsymbol{\hat n}_{\boldsymbol{\Delta}}(\mathbf r_{\mathbf k})$. 

To make this concrete, in Fig.~\ref{fig:figone}b we show a real-space texture 
$\boldsymbol{\hat n}_{K_{M}}(\mathbf r)$ of the full wavefunction
at the $K_{M}$ point for the topmost band of \wse{} at $\theta=1^\circ$. 
Note that nominally, the adiabatic approximation is expected to be very accurate at these small twist angles \cite{yaoZhaiTheoryTunableFlux2020,macdonaldMorales-DuranMagicAnglesFractional2024}. 
And indeed, the texture $\boldsymbol{\hat n}_{K_{M}}(\mathbf r)$ largely follows the texture 
$\boldsymbol{\hat n}_{\boldsymbol{\Delta}}(\mathbf r)$ defined by the layer Zeeman field, which is shown in Fig.~\ref{fig:figone}a.
However, exactly at the AB stacking point, where $\psi_{K_{M}} (\mathbf r_{AB}) =0$, rather than being fully polarized to the top layer like the underlying texture $\boldsymbol{\hat n}_{\boldsymbol{\Delta}}(\mathbf r)$,
it has the opposite layer polarization, yielding a vanishing real-space Chern number $C_{K_M}=0$, while $C[\boldsymbol{\hat n}_{\boldsymbol{\Delta}}] = -1$.

The same effect occurs also for other momenta $\mathbf k$ in the Brillouin zone.
We show the real-space Berry curvature $\berrycur$ at $\mathbf k = [-0.2,0.2] \, q_1$ in Fig.~\ref{fig:figone}c, where the admixture of wavefunctions beyond the adiabatic limit leads to a localized region of positive Berry curvature, which  cancels the negative contribution of the rest of the unit cell, leading to a zero real-space Chern number.
This positive Berry curvature, which cancels the total Chern number, can be understood as arising from a strongly localized skyrmion which emerges when the position of the zeroes in the two layers is split as discussed above in Sec.~\ref{subsec:skyrmdiscussion}. This interpretation is confirmed by plotting the layer resolved densities in Figs.~\ref{fig:finitefluxtmd}a,b, showing additional zeroes near the center of the unit cell. While these zeroes are close to each other, they are nevertheless split by a small amount, generating a localized skyrmion (see Sec.~\ref{subsec:skyrmdiscussion}), and ensuring that the total real-space Chern number satisfies the relation $C_{\mathbf k} =\Phi$.

For larger twist angles ($\theta=2.2^\circ$), the adiabatic approximation becomes less accurate,
leading to larger admixtures of the $\chi^-(\mathbf r)$ component. In this regime, the sharp feature is smeared out, as seen in Fig.~\ref{fig:figone}d, but the real-space Chern number still vanishes. 

We have numerically verified that this vanishing of the real-space Chern number is generic for states across the Brillouin zone except for the high-symmetry $\Gamma$ point. Here, we obtain $C_\Gamma=-1$ with its associated  zero even away from the adiabatic approximation. We show in Sec.~\ref{sec:topoofensembles} that the stability of the real-space Chern number at the $\Gamma$ point
is in fact a result of its symmetry properties under the $\cty \trs$ and $\cth$ transformations. Unlike for generic points in the Brillouin zone, these leave the $\Gamma$ point invariant, so that they imply symmetry constraints on the $\Gamma$-point wavefunction. We will see below that the same symmetries allow one to define a robust real-space Chern number for ensembles of wavefunctions.

\subsection{Finite magnetic flux in TMDs}
\begin{figure}[t]
    \centering
    \includegraphics[width=\columnwidth]{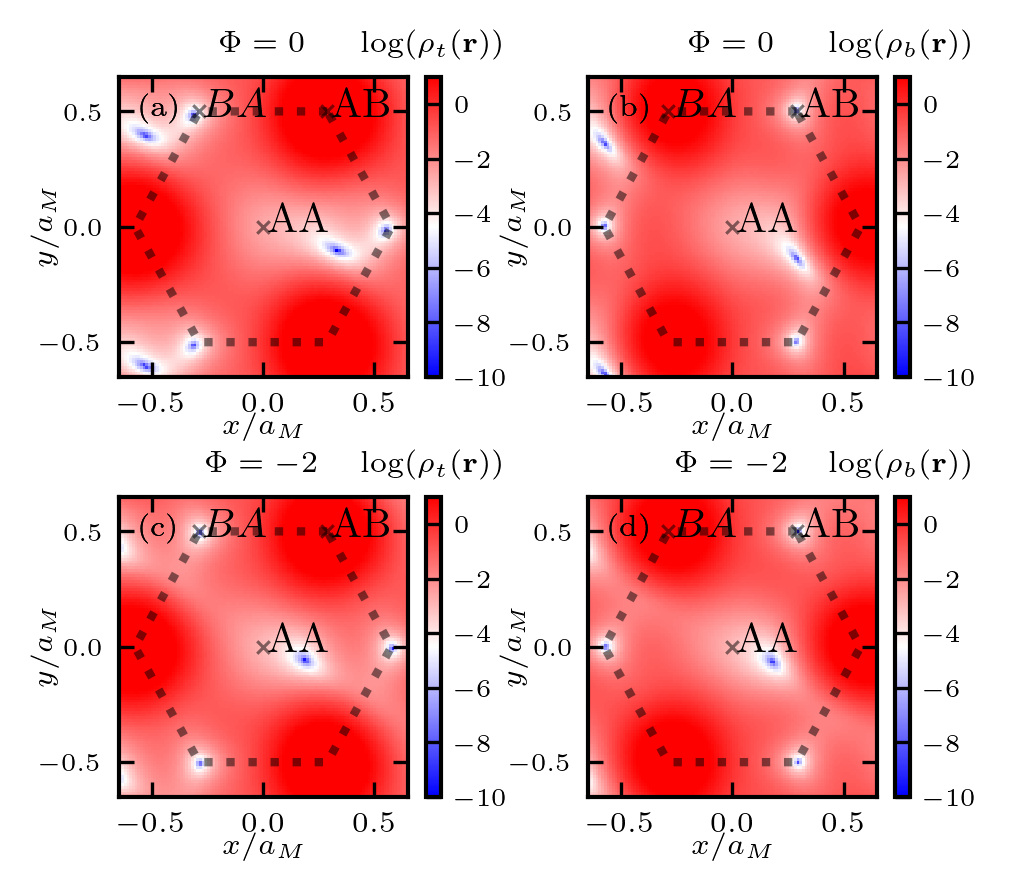}
    \caption{Wavefunction layer densities for \wse{} at $\theta=1^\circ$.
    (a) Unit cell map  of the log of the top layer density $\rho_t(\mathbf r)$ at $\Phi=0$ at a generic momentum $\mathbf k = [-0.2,0.2] \, q_1$.
    (b) Same as (a) but for the bottom layer density $\rho_t(\mathbf r)$
    (c-d) Same as (a-b), but at flux $\Phi=-2$. The layer zeroes are located at the centers of the dark blue regions. Those near the BA and AB sites are inherited from the skyrmion texture of TMDs, which is layer-polarized at these positions. At $\Phi = -2$, a second zero per unit cell appears in each layer, as required by the external magnetic flux. Its position changes with the momentum $\mathbf{k}$ and it originates from the zero of the adiabatic scalar function Eq.~\eqref{eq:LLmixture}. The (weak) deviation from adiabaticity slightly separates the second zeroes in the two layers.
     A second zero is also present at $\Phi = 0$, but in this case, it forms an anti-vortex, ensuring that the total vorticity across the unit cell vanishes. }
    \label{fig:finitefluxtmd}
\end{figure}

The adiabatic approximation is also applicable to twisted TMDs in the presence of a finite number of applied flux quanta $\Phi$~\cite{moraKolarHofstadterSpectrumChern2024},
and is expected to be accurate at small twist angles. The spinor-scalar decomposition in Eq.~\eqref{eq:wfinadiabaticapprox} still holds, while the applied magnetic field adds to the fictitious field in the Hamiltonian in Eq.~\eqref{eq:sphammacdoapp}, which  governs the scalar part. For the $K$-valley of twisted \wse{} for instance, the texture of the spinor part is quantized to a negative winding  $C[\boldsymbol{\hat n}_{\boldsymbol{\Delta}}]=-1$. The scalar wavefunction then experiences an effective flux of $1+\Phi$ quanta per unit cell. 
When perturbing away from the adiabatic limit, the zeroes in the full Bloch wavefunction are generically lifted, and $C_{\mathbf k}= \Phi$ is expected.

Our numerical simulations confirm this expectation, and show that the same mechanism of zero removal is at play at finite applied flux. As an illustration, in Figs.~\ref{fig:finitefluxtmd}c,d, we plot the layer-projected densities at flux $\Phi=2$. In the adiabatic approximation, the full wavefunction has a zero in the middle of the unit cell. In the full model, however, the zeroes of the two layers are split, and generate a localized skyrmion following the mechanism of Sec.~\ref{subsec:skyrmdiscussion}. This effect reduces the adiabatic $C[\boldsymbol{\hat n}_{\boldsymbol{\Delta}}]=-1$ to $C_{\mathbf k}= -2$ expected generically at $\Phi=-2$.

\subsection{Fragility in realistic TBG}
\label{sec:chiraltbgbreakdown}

As discussed in Sec.~\ref{sec:windinglllikewfideal}, TBG in the chiral
limit exhibits two ideal bands, each localized on a distinct sublattice. They
have opposite Chern numbers, both in momentum and real space. The real-space
Berry curvature for the A-polarized band is shown in the left inset of Fig.~\ref{fig:figzero}a, and integrated to $C_{\mathbf k}=-1$.

Away from the chiral limit, based on our arguments above, the real-space Chern number $\chernnumber$ is expected to vanish due to wavefunction admixture. The right inset of Fig.~\ref{fig:figzero}a illustrates this, showing the real-space Berry curvature of the predominantly A-polarized four-component wavefunction of TBG. There we introduce weak intra-sublattice tunneling $w_{AA}=\SI{5}{meV}$ and a sublattice mass $m_s=\SI{5}{meV}$ to split the two bands.  The texture remains nearly identical to that in the chiral limit, except for a small region in real space near the zero of the chiral-limit Landau-level wavefunction  (whose position is given in Eq.~\eqref{eq:landaulevelzeroposition}), where the Berry curvature is large and positive. Although highly localized, this sharp peak nonetheless cancels the total real-space Chern number $\chernnumber=0$.  Increasing $w_{AA}$ and deviating further from the chiral limit smoothens out the sharp singularity. 

We note that, alternatively, the $N = 4$ component Bloch wavefunctions can be
reduced to $N = 2$ by projecting onto a given sublattice. This makes the
profile of the chiral texture much more robust towards increasing $w_{AA}$,
even though the real-space Chern number collapses to $C_{\mathbf k} =0$ for
infinitesimal $w_{AA}$. Despite the vanishing total real-space Chern number of
non-ideal bands, the real-space texture inherited from the chiral limit remains
clearly visible as illustrated in Fig.~\ref{fig:figzero}a, suggesting that the
chiral-limit texture is still relevant even away from the ideal limit. In Sec.\
\ref{sec:topoofensembles-appl}, we will clarify this observation and 
show that the chiral-limit texture is in fact recovered when one considers wavefunction ensembles, rather than individual
wavefunctions.

\section{Real-space textures of ensembles of wavefunctions}
\label{sec:topoofensembles}

A generalization to wavefunction ensembles allows one to describe more physically relevant textures. First, it permits the description of the real-space topology of a given band. This ensemble encodes the number of flux quanta per unit cell associated with a fictitious magnetic field affecting the band. Second, we consider the real-space topology of the electronic states at a given energy. This ensemble is relevant for describing tunneling experiments.

In this section, we construct a topological index for such wavefunction ensembles.  In contrast to the case of individual wavefunctions, its nontriviality does not require fine-tuned zeroes in the unit cell.

\subsection{Topology of ensembles}\label{sec:topology-ensemble}

We generalize the unit vector introduced in Eq.\ \eqref{eq:defvectorcoefficient} for $N=2$ to describe ensembles of states,
\begin{equation}
\label{eq:energyresolved}
\boldsymbol A(\mathbf r)= 
\sum_{\lambda,n,\mathbf k \in BZ}u^\dagger_{\mathbf k,n,\lambda}(\mathbf r) \boldsymbol{\mu}
u_{\mathbf k,n,\lambda}(\mathbf r)  p_{\mathbf k,n,\lambda}.
\end{equation}
The sum is over the Bloch  wavefunctions $u^\dagger_{\mathbf k,n,\lambda}(\mathbf r)$ in the first Brillouin zone, indexed by the band index $n$ and the spin/valley flavor index $\lambda$. The ensemble is specified by the nonnegative coefficients $p_{\mathbf k,n,\lambda}$  describing the weight with which different wavefunctions contribute to the ensemble. We define the associated normalized vector 
\begin{equation}\label{eq:vectornE}
\boldsymbol{\hat n}(\mathbf r)= 
\frac{\boldsymbol A(\mathbf r)}{|\boldsymbol A(\mathbf r)|},
\end{equation}
which inherits the spatial periodicity of the triangular moir\'e lattice. Other lattice symmetries will be discussed in Sec.~\ref{sec:SymmetriesInRealSpace} below. This normalized vector generalizes Eq.\ \eqref{eq:defvectorcoefficient} and 
can thus be used, via Eq.~\eqref{eq:realspacechernpontryagin}, to compute a corresponding real-space Chern number $C[\boldsymbol{\hat n}]$. 

The robustness of the topological invariant is conveniently described by introducing the $2\times 2$-Hamiltonian 
\begin{equation}
\label{eq:energyresolvedh}
H({\mathbf r})=  {\mathbf A} ({\mathbf r}) \cdot \boldsymbol{\mu},
\end{equation}
which has two bands as a function of the position ${\mathbf r}$. The Chern number of the upper band of this Hamiltonian is equal to $C[\boldsymbol{\hat n}]$, which is topologically protected by the spectral gap $2|{\mathbf A} ({\mathbf r})|$ between the two bands. 

If the sum in Eq.~\eqref{eq:energyresolved} were to reduce to a single momentum, this definition would coincide with the one used for individual wavefunctions, as given in Eq.~\eqref{eq:defvectorcoefficient}. A nonzero real-space Chern number at zero flux would only be possible if the wavefunction vanishes somewhere, i.e., if the real-space gap $2|{\mathbf A} ({\mathbf r})|$ vanishes in the unit cell, implying a lack of robustness. In contrast, when $\boldsymbol{\hat n}(\mathbf r)$ is constructed from an ensemble of wavefunctions, the arguments presented in Sec.~\ref{sec:realspacetextures} do not apply, and a nonzero real-space Chern number $C[\boldsymbol{\hat n}]$ for zero applied flux no longer requires the presence of fine-tuned zeroes of the wavefunctions.

A nonzero value of the texture $\boldsymbol A(\mathbf r)$
implies a tendency of the wavefunctions in the ensemble to follow a common preferred spinor direction. 
A large value of $\boldsymbol A(\mathbf r)$ implies that, while the underlying system has multiple components, at that point in space, the ensemble wavefunctions are behaving like those of a single-component system aligned with the direction defined by $\boldsymbol A(\mathbf r)$. 
A topological phase transition is only possible when $\boldsymbol A(\mathbf r)$ vanishes somewhere.
At such a point, the wavefunctions in the ensemble are, on average, not aligned with any particular direction.

This definition can be generalized to $N>2$ in two ways.  First,  $u_{\mathbf k}$ can be projected on two components, reducing the $N>2$ problem to an effective two-component problem. In this case, a generalization of Eq.~\eqref{eq:energyresolved} reads  
\begin{equation}
\label{eq:energyresolved2}
\boldsymbol A^{(\alpha)}(\mathbf r)= 
\sum_{\lambda,n,\mathbf k \in BZ}u^\dagger_{\mathbf k,n,\lambda}(\mathbf r) \boldsymbol \mu^{(\alpha)}
u_{\mathbf k,n,\lambda}(\mathbf r)   p_{\mathbf k,n,\lambda},
\end{equation}
where $\boldsymbol \mu^{\alpha}$ denote the Pauli matrices in the projected subspace. Then, the analysis follows the discussion for two components. 
Second, one can directly consider the properties of an $N\times N$ real-space effective Hamiltonian $H({\mathbf r})$
defined from the texture. 
Depending on the number of spectral gaps, there are at most $N$ real-space Chern numbers whose sum is constrained to be zero. We leave the detailed discussion to App.~\ref{app:Ncomponent}. In what follows, we focus on the first approach, which leads to a more straightforward classification.

\subsubsection{Textures at a fixed energy}

We first consider textures of wavefunction ensembles consisting of states at a fixed energy. These are obtained by taking
\begin{equation}
    p_{\mathbf k,n,\lambda}= \delta (\epsilon_{\mathbf k,n,\lambda}-E)
\end{equation}
and provide energy-resolved real-space textures, which we denote ${\mathbf A} (E,{\mathbf r})$. This corresponds to the density of states typically probed in tunneling experiments~\cite{nadj-pergeChoiElectronicCorrelationsTwisted2019,pasupathyKerelskyMaximizedElectronInteractions2019,bernevigCalugaruSpectroscopyTwistedBilayer2021,zaletelHongDetectingSymmetryBreaking2021,yazdaniNuckollsQuantumTexturesManybody2023,nadj-pergeKimImagingIntervalleyCoherent2023},

The associated real-space topology is obtained from the normalized vector
\begin{equation}
\boldsymbol{\hat n}_E(\mathbf r)= 
\frac{\boldsymbol A(E,\mathbf r)}{|\boldsymbol A(E,\mathbf r)|},
\end{equation}
which leads to a real-space Chern number $C[\boldsymbol{\hat n}_E]$.
The components of $\boldsymbol{\hat n}_E(\mathbf r)$ are accessible in tunneling experiments~\cite{shihZhangExperimentalSignatureLayer2025,yankowitzThompsonMicroscopicSignaturesTopology2025}. 

When the Pauli matrices $\boldsymbol \mu^{(\alpha)}$ are in layer space (as for twisted bilayer TMDs and sublattice-projected TBG), only the $z$-component of $\boldsymbol{\hat{n}}$ is 
directly accessible, and corresponds to local density of states imbalance between the two layers. 
The $x$-component would be recoverable from the combination of the $z$-component and the value of the density of states at the midpoint between the two layers. However, we note that the latter is not immediately accessible in STM.
On the other hand, when the $\boldsymbol \mu^{\alpha}$ refer to sublattice space (as we will discuss below for layer-projected twisted bilayer graphene), both the $z$ and $x$ components are accessible in STM experiments \cite{yazdaniLiuVisualizingBrokenSymmetry2021}. The $z$-component corresponds to sublattice polarization in the density of states, while the $x$-component is recoverable from a combination of the $z$-component and the local the density of states at the midpoint between the two sublattices. We remark that the $y$-component does not have a direct signature in STM, leading instead to atomic-scale currents.

\subsubsection{Textures of a band}
\label{subsec:texturesofbands}

Next, we consider textures associated with bands. By choosing
\begin{equation}
    p_{\mathbf k,n,\lambda} = \frac{1}{\nuc}\delta_{n,b}\delta_{\lambda, \lambda_b},
\end{equation}
where $\nuc$ is the number of unit cells (number of $\mathbf k$-points),
we sum over all momenta within the Brillouin zone for a  given band $b$ and  flavor $\lambda_b$, and  obtain the intrinsic texture associated with that band.
The corresponding real-space Chern number encodes the number of flux quanta of the fictitious magnetic field experienced by the electrons in that band. 
We postpone a detailed discussion of the physical implications of this field to Sec.~\ref{sec:bandlandaumapping}.

\subsection{Symmetries in real space}
\label{sec:SymmetriesInRealSpace}

Spatial symmetries are generally not very informative for individual wavefunctions, as such symmetries act not only on the spatial position $\mathbf{r}$ but also on the momentum $\mathbf{k}$. Thus, a symmetry operation maps $u_{\mathbf{k}}$ to a different Bloch function, unless $\mathbf{k}$ is a high-symmetry momentum such as $\Gamma$. 
In contrast, the ensemble average of Eq.~\eqref{eq:energyresolved} sums over different momenta weighted 
with $p_{\mathbf k,n,\lambda}$.
Thus, provided that $p_{\mathbf k,n,\lambda}$ is invariant under a symmetry of the system, the real-space Hamiltonian introduced in Eq.~\eqref{eq:energyresolvedh} inherits that symmetry, transferring it to the space-dependent vector $\boldsymbol{\hat n}(\mathbf r)$. 

Specifically, a (not necessarily spatial) symmetry operation $\mathcal S$ acts on general multicomponent wavefunctions 
$\Psi_\lambda(\mathbf r)$ 
as
\begin{equation}
\mathcal S \Psi_\lambda(\mathbf r) = S(\mathbf r) \Psi_{\mathcal S(\lambda)}({\mathcal R}_{\mathcal S} \mathbf r),
\end{equation}
transforming both the real-space position, induced by ${\mathcal R}_{\mathcal S}$, the flavor index $\lambda$, and the wavefunction with some possibly space-dependent (anti)unitary matrix $S(\mathbf r)$. 
Provided $p_{\mathbf k,n,\lambda}$ respects the symmetry $\mathcal S$,
the symmetry properties of the wavefunctions translate to the real-space Hamiltonian.
Explicitly
\begin{equation}
\mathcal S H(\mathbf r) \mathcal S^{-1}=  H(\mathbf r),
\end{equation}
which can be rewritten as 
\begin{equation}
 S(\mathbf r) H({\mathcal R}_{\mathcal S}  \mathbf r) S^{-1}(\mathbf r)=  H(\mathbf r).
\end{equation}
Inserting this expression into the definition Eq.~\eqref{eq:energyresolvedh} of $H(\mathbf r)$, we obtain
\begin{equation}
    \boldsymbol{\hat n}({\mathcal R}_{\mathcal S} \mathbf r) = {\mathcal R}_{\mu,\mathcal S} \, \boldsymbol{\hat n}( \mathbf r),
\end{equation}
where ${\mathcal R}_{\mu,\mathcal S} {\boldsymbol \mu} = S(\mathbf r) {\boldsymbol \mu}  S^{-1}(\mathbf r)$ denotes the rotation of the Pauli matrices ${\boldsymbol \mu}$ induced by the matrix $S(\mathbf r)$.

We now discuss the role of time-reversal symmetry, which enforces $H(\mathbf r)$ to be real. This  in turn implies that the vector $\boldsymbol{\hat{n}}(\mathbf{r})$ has a vanishing $y$-component and lies entirely in the $xz$-plane. As a result, the real-space Chern number vanishes identically in time-reversal symmetric systems, $C[\boldsymbol{\hat{n}}] = 0$. We conclude that nontrivial real-space topology, $C[\boldsymbol{\hat n}] \neq 0$, requires broken time-reversal symmetry.
In what follows, we will assume that time-reversal symmetry is broken by spontaneous valley polarization. We note, however, that in the presence of time-reversal symmetry, one may still define a valley-resolved Chern number.

\section{Robustness of real-space topology of wavefunction ensembles}

\label{sec:topoofensembles-appl}

We now apply the concept of the topology of ensembles to twisted TMDs and TBG, demonstrating the robustness of their real-space topology. We will exploit the $\cth$ and $\cty \trs$ symmetries and show that they
impose the real-space topology to be nontrivial.

\subsection{Robustness in twisted bilayer TMDs}\label{sec:twistedTMDs}
For twisted bilayer TMDs, 
we now use symmetry arguments to show that the real-space Chern number at a given energy is typically nonzero, $C[\boldsymbol{\hat n}_E] \neq 0$,
provided the system valley polarizes due to interactions. The spontaneous valley polarization breaks time-reversal symmetry, but leaves the intravalley $\cth$ and $\cty \trs$ symmetries intact. A possible flavor-polarized state is the $\nu=1$ valley-polarized state, observed in many experiments to date. The $\cth$ and $\cty \trs$ symmetries transform wavefunctions in the $K$-valley as  
\begin{eqnarray}
\cth \Psi_K(\mathbf r) = \begin{pmatrix}1&0\\ 0 & e^{i \mathbf g_5 \cdot \mathbf r}\end{pmatrix} \Psi_K( \cth^{-1} \mathbf r), \\
\cty \trs  \Psi_K(\mathbf r)= e^{i \mathbf q_1 \cdot \mathbf r} \mu_x \Psi_K^*(\cty \mathbf r).
\end{eqnarray}
Assuming that the ensemble of electronic states is invariant under these symmetries,
they impose constraints on the effective Hamiltonian of Eq.~\eqref{eq:energyresolvedh},
\begin{eqnarray}
H(E,\cth \mathbf r) &=& \begin{pmatrix}1&0\\ 0 & e^{i \mathbf g_5 \cdot \mathbf r}\end{pmatrix}H(E,\mathbf r)\begin{pmatrix}1&0\\ 0 & e^{-i \mathbf g_5 \cdot \mathbf r}\end{pmatrix}\\
H(E,\cty \mathbf r) &= &\mu_x H^*(E,\mathbf r)\mu_x,
\end{eqnarray}
where the second equation uses that  $e^{i \mathbf q_1 \cdot \mathbf r}$ commutes with the local Hamiltonian. This in turn imposes conditions on the normalized vector $\boldsymbol{\hat n}_E(\mathbf r)$ on the Bloch sphere,
\begin{eqnarray}
\boldsymbol{\hat n}_E(\cth \mathbf r)= 
R_z(\mathbf g_5 \cdot \mathbf r) \boldsymbol{\hat n}_E(\mathbf r), \\
\boldsymbol{\hat n}_E(\cty \mathbf r)= 
M_z \boldsymbol{\hat n}_E(\mathbf r),
\end{eqnarray}
where $M_z$ denotes the mirror symmetry about the $xy$ plane, and $R_z(\theta)$ a rotation by the angle $\theta$ about the $z$-axis. At high-symmetry stacking points, these relations, together with the moir\'e periodicity, imply that 
\begin{eqnarray}
\boldsymbol{\hat n}_E(AA)|_z=0, \label{eq:naa}\\
\boldsymbol{\hat n}_E(AB)= (0,0,\pm 1),\label{eq:nab}\\
\boldsymbol{\hat n}_E(BA)= (0,0,\mp 1)\label{eq:nba},
\end{eqnarray}
so that the vector is purely in-plane at AA sites and purely out-of-plane at AB and BA sites. Moreover, the layer polarizations are opposite at the AB and BA sites.
Remarkably, these properties imply that
\begin{equation}\label{eq:real-space-chern-TMD}
C[\boldsymbol{\hat n}_E] = \pm 1 \mod 3
\end{equation}
is necessarily nonvanishing. We obtain this result in App.~\ref{app:symmetryindicators} 
using symmetry indicators of band topology \cite{bernevigFangBulkTopologicalInvariants2012}.

\begin{figure}[t]
    \centering
    \includegraphics[width=\columnwidth]{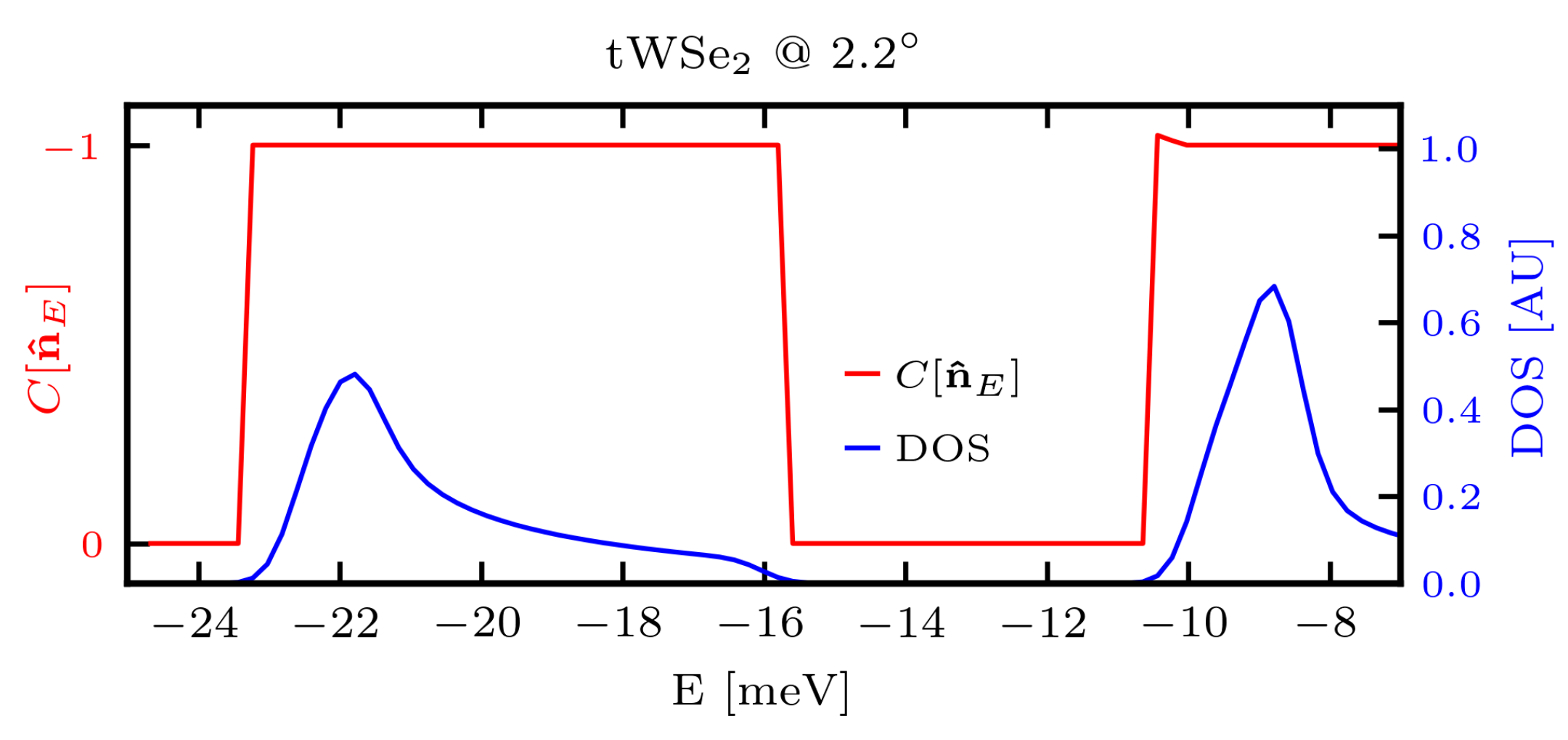}
    \caption{
     $C[\boldsymbol{\hat n}_E]$ (left axis) and DOS (right axis) for twisted \wse{} at $\theta =2.2^\circ$, far away from the adiabatic limit.
The real-space Chern number is everywhere nonvanishing in this case. Note that we consider a fully valley-polarized state.}
    \label{fig:figthree}
\end{figure}

We now focus separately on energy-resolved textures and textures of entire bands. \subsubsection{Energy-resolved textures}
As a specific example, we consider homobilayer \wse{} twisted by an angle $\theta = 2.2^\circ$ far from the adiabatic limit. A numerical evaluation of the real-space Chern number, shown in Fig.~\ref{fig:figthree}a, confirms our prediction. We find $C[\boldsymbol{\hat n}_E] = -1$ for all energies within the two top bands. Notably, these bands each possess a momentum-space Chern number of $C = 1$~\cite{moraKolarHofstadterSpectrumChern2024}.

We note that these arguments based on $\cty \trs$ and $\cth$ can also be applied to the topology of the wavefunction at $\Gamma$, the only point in the BZ that remains invariant under these two symmetries. They explain why, at $\Gamma$, the finite real-space Chern number for an individual wavefunction persists even away from the adiabatic limit.

Our arguments also apply in the valley unpolarized state, where time-reversal precludes a nonzero Chern number. 
In this case, $\cth$ and $\cty$ symmetries still impose Eqs.~\eqref{eq:nab} and~\eqref{eq:nba}, 
implying opposite layer polarizations at the BA and AB points, consistent with recent experiments~\cite{shihZhangExperimentalSignatureLayer2025,yankowitzThompsonMicroscopicSignaturesTopology2025}.

\subsubsection{Textures of full bands}
Turning to real-space textures of individual bands as introduced in Sec.~\ref{subsec:texturesofbands}, the same symmetry arguments imply that, generically, the real-space Chern number of every band is nonzero, $C[\boldsymbol{\hat n}]\neq 0$. This implies the generic presence of a net fictitious magnetic field in TMDs. 
Specifically, we find for the top band of twisted \wse{} that $C[\boldsymbol{\hat n}]= -1$.
The same analysis can be applied to other models of twisted TMDs, such as models of \mote{}, suggested to host states analogous to Landau levels \cite{choAhnFirstLandauLevel2024,xiaoWangHigherLandauLevelAnalogues2024,fuReddyNonAbelianFractionalizationTopological2024,zhangXuMultipleChernBands2024,xiaoZhangPolarizationdrivenBandTopology2024,chen2025robust,wuLiVariationalMappingChern2024}. Focusing on a model of twisted \mote{} at $2.1^\circ$, taking model parameters from Refs.~\cite{choAhnFirstLandauLevel2024,xiaoWangHigherLandauLevelAnalogues2024}, we find that the textures of the two bands are consistent with a fictitious magnetic field picture, possessing a real-space Chern number $C[\boldsymbol{\hat n}]= -1$.
Furthermore, the band follows the texture extremely well, and allows a mapping to Landau levels to be made
as we detail in Sec.~\ref{sec:bandlandaumapping} below.
\\
\subsection{Robustness in TBG}

We now consider TBG, where $\boldsymbol{\mu}$ and $\sigma$ denote the layer and sublattice Pauli matrices, respectively. The relevant symmetries are $\cth$, $\cty \trs$ and $\ctz \trs$, which act on wavefunctions within the $K$-valley as 
\begin{align}
\cth \Psi_K(\mathbf r) = 
\begin{pmatrix}1&0\\ 0 & e^{i \mathbf g_5 \cdot \mathbf r}\end{pmatrix} \otimes e^{i \frac{2\pi}{3} \sigma_z} 
\Psi_K(\cth^{-1} \mathbf r), \label{eq:defcthtbg}\\
\cty \trs  \Psi_K(\mathbf r) =  \mu_x \Psi_K^*(\cty \mathbf r), \\
\ctz \trs  \Psi_K(\mathbf r)=  \sigma_x \Psi_K^*(- \mathbf r).
\end{align}
There are two natural ways to turn $N=4$ components into a two-component spinor. First, one may project on a given sublattice. Then, the remaining degree of freedom is layer.
Alternatively, one projects onto a given layer, and the remaining degree of freedom is sublattice. These two cases determine the available symmetries, which we now analyze in detail.

\begin{figure}[t]
    \centering
    \includegraphics[width=\columnwidth]{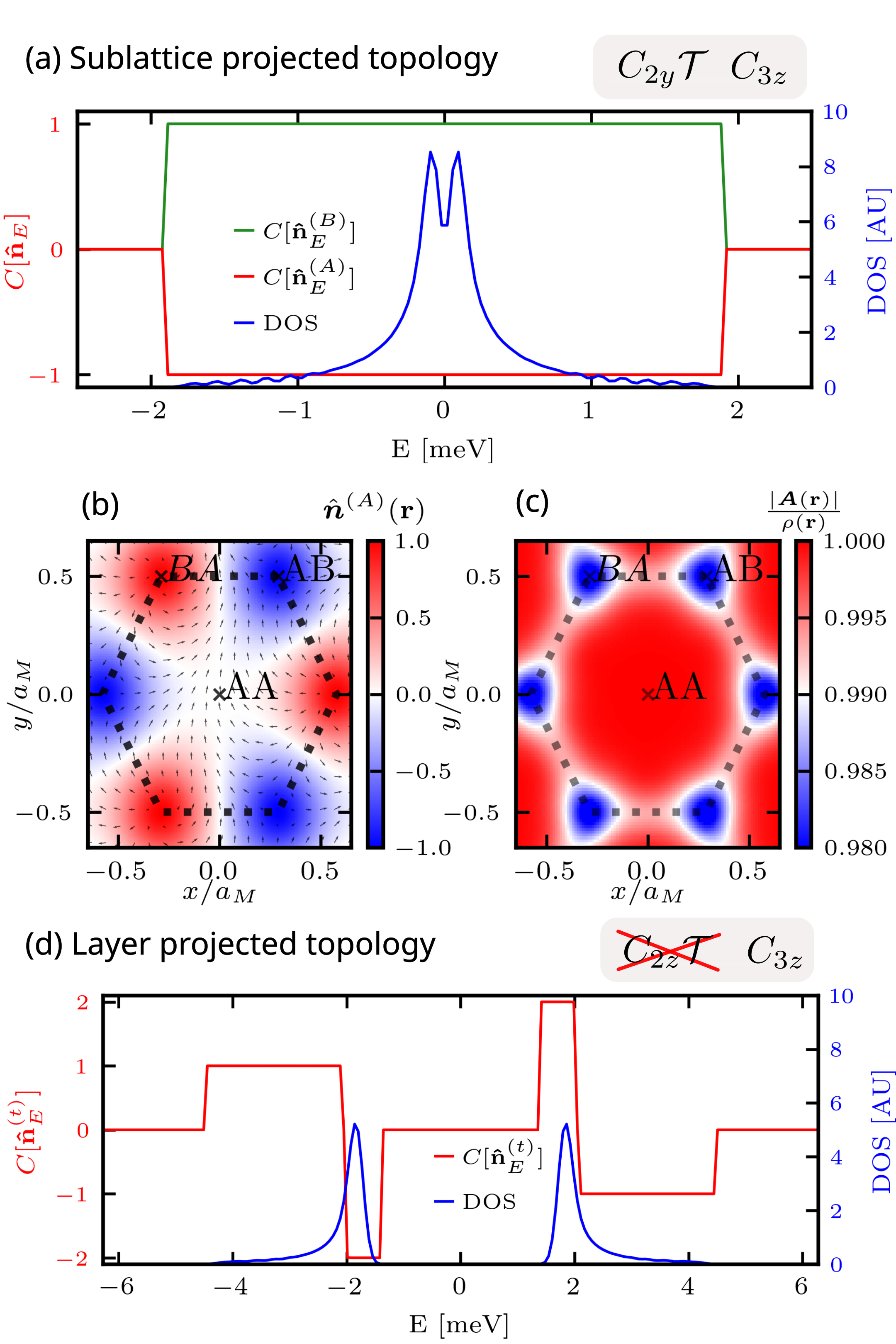}
    \caption{TBG: (a) Sublattice projected topology.
    In a realistic TBG model ($m_s=0, w_{AA}=\SI{80}{meV}$), the sublattice projected Chern number $C[\boldsymbol{\hat n}^{(A/B/t)}_E]$ is necessarily nonzero
    as a result of $\cth$ and $\cty \trs$ symmetries.
     (b) A-sublattice projected texture of the TBG flat bands.
    (c) Value of $|\boldsymbol{A}(\mathbf r)|/\rho(\mathbf r)$ in the unit cell
    for the sublattice projected texture of the TBG flat bands.
    (d) Layer-projected topology. 
    A finite Chern number arises when $\ctz \trs$ is broken by
a sublattice mass $m_s = \SI{5}{meV}$. 
We keep a realistic corrugation $w_{AA}=\SI{80}{meV}$.}
    \label{fig:figfour}
\end{figure}

 \subsubsection{Sublattice-resolved textures}

We define the sublattice-projected Pauli matrices for the two sublattices $A$ and $B$ as
\begin{eqnarray}
{\boldsymbol{\mu}}^{(A)} = \boldsymbol{\mu} \otimes (1+\sigma_z)/2, \\
{\boldsymbol{\mu}}^{(B)} = \boldsymbol\mu \otimes (1-\sigma_z)/2.
\end{eqnarray}
Using these, we introduce (for energy-resolved ensembles) 
$\boldsymbol A^{(A)}(E,\mathbf r)$ and
$\boldsymbol A^{(B)}(E,\mathbf r)$, 
along the corresponding renormalized vectors $\boldsymbol{\hat n}^{(A)}_E$ and $\boldsymbol{\hat n}^{(B)}_E$.
They define two real-space Chern numbers $C[\boldsymbol{\hat n}^{(A)}_E]$ and $C[\boldsymbol{\hat n}^{(B)}_E]$ which are opposite due to $\ctz \trs$ symmetry.
After the sublattice projection, the symmetry analysis previously carried out for twisted TMDs in Sec.~\ref{sec:twistedTMDs} remains applicable, as the two key symmetries, $\cty\trs$ and $\cth$, are diagonal in sublattice space. This allows us to recover the same symmetry-based arguments and formulas derived for twisted TMDs, with in-plane layer polarization at AA sites and out-of-plane polarization at AB and BA sites. The only difference with twisted TMDs lies in the sublattice-dependent phase in Eq.~\eqref{eq:defcthtbg}, which disappears when analyzing how $H$ and $\boldsymbol{\hat n}_E$ transform under $\cth$.  By employing symmetry indicators, we find — just as in twisted TMDs — that the real-space Chern number is always nonzero and follows Eq.~\eqref{eq:real-space-chern-TMD}. 

This remarkable result is illustrated for TBG in Fig.~\ref{fig:figfour}a, where we compute the real-space Chern numbers of the two sublattices at the magic angle, $\theta = 1.09^\circ$, using a realistic $w_{AA} = \SI{80}{meV}$. We find that the real-space Chern numbers for the two sublattices are consistently nonzero and opposite in sign. This demonstrates that the real-space topology of the chiral limit extends to the original Bistritzer-Macdonald model \cite{macdonaldBistritzerMoireBandsTwisted2011} with a realistic corrugation.
We expect this topology to manifest in scanning tunneling experiments when focusing on a single sublattice.

Considering the sublattice-projected textures of the entire TBG flat bands, we find a robust fictitious magnetic field, opposite on each sublattice.
This allows the mapping to an effective two-band Landau level problem, as detailed in Sec.~\ref{sec:bandlandaumapping} below.

 \subsubsection{Layer-projected textures}
Another natural option is to project onto a single layer, say the top layer, by defining
\begin{eqnarray}
\boldsymbol \mu^{(t)} =[(1+\mu_z)/2]\otimes \boldsymbol\sigma .
\end{eqnarray}
Using this, we can define
$\boldsymbol A^{(t)}(E,\mathbf r)$. However, now the symmetries available within a single valley reduce to $\ctz \trs$ and $\cth$.
These constrain the vector $\boldsymbol A^{(t)}(E,\mathbf r)$ through
\begin{equation}\label{eq:sym-constrains}
\begin{split}
\boldsymbol A^{(t)}(E, \mathbf r) &= 
R_z(2\pi/3) \boldsymbol A^{(t)}(E,\mathbf r)\\[1mm]
\boldsymbol A^{(t)}(E,- \mathbf r) &= 
 M_z  \boldsymbol A^{(t)}(E,\mathbf r),    
\end{split}
\end{equation}
respectively, and similarly for the normalized vector $\boldsymbol{\hat n}^{(t)}_E(\mathbf r)$. Crucially, $\cth$ requires that, at the high-symmetry $AA$, $AB$, and $BA$ sites, the normalized vector is fully sublattice polarized, $\boldsymbol{\hat n}^{(t)}_E(\mathbf r) = (0,0,\pm 1)$. Furthermore, $\ctz \trs$ constrains  
\begin{equation}\label{eq:opposite-pol}
\boldsymbol{\hat n}^{(t)}_E(AB)= -\boldsymbol{\hat n}^{(t)}_E(BA),    
\end{equation}
such that the sublattice polarizations are opposite at the AB and BA sites. At AA sites, the two symmetries constraints of Eq.~\eqref{eq:sym-constrains} are only compatible with a vanishing vector, $\boldsymbol A^{(t)}(E,AA)=0$, corresponding to a gap closing for the effective Hamiltonian Eq.~\eqref{eq:energyresolvedh} and an ill-defined real-space Chern number.

A spectral gap of $H(E,{\mathbf r})$ is opened  by applying a sublattice mass which breaks $\ctz \trs$ but preserves the threefold rotation symmetry $\cth$ and the complete sublattice  polarization at AB and BA points. Therefore, the constraint of Eq.~\eqref{eq:opposite-pol} still applies by continuity despite the breaking of $\ctz \trs$. 
Using symmetry indicators as in the previous section, we therefore obtain the real-space Chern number
\begin{equation}
\label{eq:symmind}
    C[\boldsymbol{\hat n}^{(t)}_E]=\pm 1 \mod 3.
\end{equation}
This prediction is confirmed by a numerical evaluation of the real-space Chern number presented in Fig.~\ref{fig:figfour}d for the two central bands of TBG with a realistic value for the corrugation. A sublattice potential $m_s = \SI{5}{meV}$ has been added to break $\ctz \trs$. Note that the jump by three of the Chern number is 
consistent with Eq.~\eqref{eq:symmind}, which only constrains it $\mod 3$. It corresponds to a gap closing of the fictitious Hamiltonian $H(E,{\mathbf r})$ near $E=\pm\SI{2}{meV}$ at the three $\cth$-related $AB-BA$ midpoints in the unit cell.
We expect this topology to become manifest in scanning tunneling experiments. Intriguingly, all three components of
$\boldsymbol{\hat n}^{(t)}_E$ are observable, suggesting that the real-space Chern number can be directly probed.
When we consider the texture of the two central bands, we find $C=1$ for the lower band and $C=-1$ for the upper band.
\\
\section{Band-ensemble induced magnetic field and Landau-level-like wavefunctions.}
\label{sec:bandlandaumapping}
The Fractional Chern insulator (FCI) phase typically occurs in a Chern band when the interaction is larger than the bandwidth, but small compared to the band gaps \cite{zhao_review}. Under this regime, all the wavefunctions in the band are correlated by the interaction, leading to entangled many-body states. 
Looking for additional insights into the relevant band, in this Section we show how real-space band-ensemble textures provide a useful framework for interpreting FCI states observed in recent moir\'e experiments. 
We begin by quantitatively characterizing the polarization of the band to an optimal layer texture (or, more generally, other orbital labels such as sublattice). To this end, we introduce a polarization metric that quantifies the fraction of the band aligned with the texture defined in Sec.~\ref{subsec:texturesofbands}.
We then project the full Hamiltonian onto this texture, obtaining an effective single-component Hamiltonian. We show that this Hamiltonian experiences an emergent magnetic field given by the Berry curvature of the texture and that it reproduces the band dispersion beyond the adiabatic approximation. Finally, we analyze the properties of the wavefunction along the texture direction and identify a natural mapping to (generalized) Landau levels.
In this picture, the
interacting problem within the band reduces to an effective single-component model, paving the way for the construction
of many-body FCI wavefunctions. In the following, we develop the theory for a two-component system relevant to twisted
TMDs, and in the last subsection, we extend the analysis to TBG.

\subsection{Quantifying the band alignment -- are the bands single-component-like?}

We first quantify how well a band is aligned with a texture. We focus on a two-component (or layer) system but the approach is easily generalizable to a larger number of components. 
For the ensemble of a band $b$ in valley $\lambda_b$, its texture is [concretizing Eq.~\eqref{eq:energyresolved}] 
given by the three-dimensional vector
\begin{equation}
\label{eq:bandtext}
\boldsymbol A(\mathbf r)= 
\frac{1}{\nuc} \sum_{\mathbf k \in BZ}u^\dagger_{\mathbf k,b,\lambda_b}(\mathbf r) \boldsymbol{\mu}
u_{\mathbf k,b,\lambda_b}(\mathbf r)  ,
\end{equation}
which allows a definition of the corresponding normalized vector
$\boldsymbol{\hat n}(\mathbf r)=\boldsymbol A(\mathbf r)/|\boldsymbol A(\mathbf r)|$.
The length of the texture vector is upper bounded by the local density 
\begin{equation}
|\mathbf A(\mathbf r)|\leq \rho (\mathbf r) =\frac{1}{\nuc} \sum_{\mathbf k \in BZ}u^\dagger_{\mathbf k,b,\lambda_b}(\mathbf r) 
u_{\mathbf k,b,\lambda_b}(\mathbf r),
\end{equation}
with equality reached when all the Bloch waves in the ensemble are aligned in the same direction. 
To quantify the average alignment across the unit cell, we construct the projection operator to the direction of the texture  
$P_{\boldsymbol{\hat{n}}}(\mathbf r)= \frac{1}{2}(1+\boldsymbol{\mu}  \cdot \nguy)$, which acts in the layer space at each position $\mathbf r$. We then define the polarization index
\begin{eqnarray}
\label{eq:defkappa}
 \kappa&=& \int_{\uc} d\mathbf r
\frac{1}{\nuc}\sum_{\mathbf k \in BZ}u^\dagger_{\mathbf k,b,\lambda_b}(\mathbf r)
P_{\boldsymbol{\hat{n}}}(\mathbf r)
u_{\mathbf k,b,\lambda_b}(\mathbf r)  \\
 &=&\int_{\uc} d\mathbf r
\frac{\rho(\mathbf r)+|\boldsymbol A(\mathbf r)|}{2} \in \left[0,1\right]
\end{eqnarray}
to characterize the band alignment. We show in App.~\ref{app:maximizingtexture} that, for a fixed band, the band-defined texture [defined in Eq.~\eqref{eq:bandtext}] maximizes $\kappa$ over all possible textures, and $\kappa$ then measures how closely the band follows this optimal texture. 

The Bloch wavefunctions can be decomposed as
\begin{equation}
\label{eq:decompose}
u_{\mathbf k,b,\lambda_b}(\mathbf r)=\alpha_{\mathbf k} \chi_{\boldsymbol{\hat{n}}}(\mathbf r)
\psi_{\mathbf k} (\mathbf r) + \beta_{\mathbf k} \chi^\perp_{\boldsymbol{\hat{n}}}(\mathbf r) \varphi_{\mathbf k} (\mathbf r),
\end{equation}
with $|\alpha_{\mathbf k}|^2+|\beta_{\mathbf k}|^2=1$, where $\psi_{\mathbf k}$ and $\varphi_{\mathbf k}$ are normalized, and the spinor $\chi_{\boldsymbol{\hat{n}}}(\mathbf r)$ satisfies $P_{\boldsymbol{\hat{n}}}(\mathbf r)\chi_{\boldsymbol{\hat{n}}}(\mathbf r)=\chi_{\boldsymbol{\hat{n}}}(\mathbf r)$.
In this form, $\kappa =1- \frac{1}{\nuc}\sum_{\mathbf k} |\beta_{\mathbf k}|^2$,
so that large $\kappa$ corresponds to small $\beta_{\mathbf k}$, i.e., wavefunctions dominated by the spinor $\chi_{\boldsymbol{\hat{n}}}$.
This decomposition generalizes Eq.~\eqref{eq:generalwfdecomp}, with a $\mathbf k$-independent spinor, beyond the ideal cases discussed in Sec.~\ref{sec:windinglllikewf}. Accordingly, $\kappa$ quantifies the proximity to Eq.~\eqref{eq:generalwfdecomp}, indicating the fraction of the band following the texture $\hat{n} (\mathbf r)$.

We now provide the calculated values of the alignment strength $\kappa$ for twisted bilayer TMDs,
which turn out to be remarkably large.
For \wse{}, we find that the value of $\kappa$ decreases with twist angle.
However, even at $\theta=2.2^\circ$, it is still very large, having $\kappa = 0.96$.
For the realistic model of \mote{} at $\theta=2.1^\circ$, suggested to host
Landau level analogs, we find good alignment of $\kappa =0.98$ for the topmost band and $\kappa =0.92$ for the second topmost band. The band alignment to the optimal texture is thus excellent for twisted bilayer TMDs.
Below, we present two approaches that exploit this observation to project onto a single-component Hamiltonian or wavefunction and reveal the Landau-level structure of these models.

\subsection{Effective single-component Hamiltonian and energy dispersion}

For $\kappa$ sufficiently close to one, the wavefunction 
$\psi_{\mathbf k} (\mathbf r)$ dominates the decomposition in Eq.~\eqref{eq:decompose} and is approximately a solution of a single-component Hamiltonian
\begin{equation}\label{eq:heff}
    H_{\text{eff}}=\chi^\dagger_{\boldsymbol{\hat{n}}}(\mathbf r) H\chi_{\boldsymbol{\hat{n}}}(\mathbf r),
\end{equation}
which is obtained after projecting the original Hamiltonian onto the optimal texture. 
We derive this result in Appendix~\ref{app:schrieffer-wolff}.
Even though we are not in the adiabatic limit, the derivation follows closely the same logic as in Sec.~\ref{sec:windinglllikewftmd}. 
Specializing to twisted TMDs, we recover for $H_{\text{eff}}$ the Hamiltonian given in Eq.~\eqref{eq:sphammacdoapp}. The main difference is that the effective vector potential 
\begin{equation}\label{eq:vectorAtilde}
    \tilde{A} (\mathbf r) = -i\chi^\dagger_{\boldsymbol{\hat{n}}}(\mathbf r) \nabla_{\br}   \chi_{\boldsymbol{\hat{n}}}(\mathbf r),
\end{equation}
as well as the kinetic potential $D(\bf r)$ are determined by the average band texture $\boldsymbol{\hat n} (\mathbf r)$ and not by the interlayer coupling $\Delta (\mathbf r)$ as in the adiabatic limit of twisted TMDs.
Thus, the effective magnetic field in this Hamiltonian is that of the optimal band texture.

Another consequence of using the optimal texture to construct the effective Hamiltonian is that the validity of this construction does not impose constraints on the rate of variation of the underlying Hamiltonian,
unlike the adiabatic approximation~\cite{yaoYuGiantMagneticField2020,yaoZhaiTheoryTunableFlux2020,macdonaldMorales-DuranMagicAnglesFractional2024}. As a result, it is significantly more broadly applicable, extending the effective Hamiltonian approach to larger twist angles and to higher-harmonic models, where the kinetic energy dominates over interlayer coupling. In particular, the only requirement is that $\kappa$ is close to one for the band of interest.

\begin{figure}[t]
    \centering
    \includegraphics[width=\columnwidth]{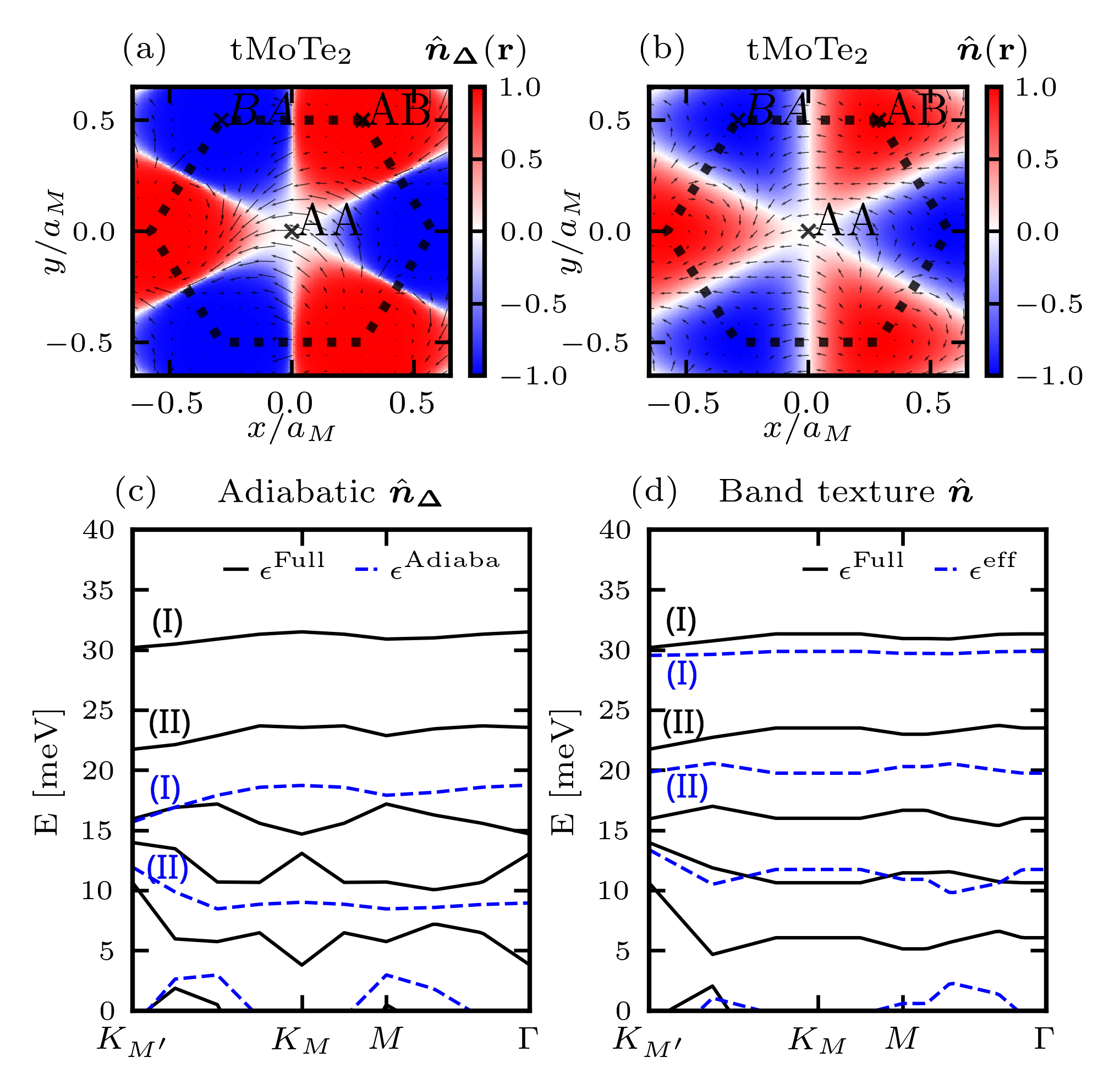}
    \caption{
(a) Texture defined from the Hamiltonian [using Eq.~\eqref{eq:defndelta}] of twisted \mote{} at $\theta=2.1^\circ$. Due to lattice relaxation, the Hamiltonian
contains sharp domains between upper/lower layer polarization, and the adiabatic approximation is inapplicable for it.
(b) Ensemble texture of the upper band [computed using Eq.~\eqref{eq:bandtext}] of twisted \mote{} at $\theta=2.1^\circ$. The variations in the band texture are much
smoother than in (a), and this texture can be used to define a single-component Hamiltonian describing the upper band. 
(c) and (d):
Comparison of the full spectrum (black, top two bands labelled with roman numerals) of \mote{} at $\theta=2.1^\circ$
 with the spectrum of the effective Hamiltonian (blue, top two bands labelled with roman numerals) obtained by projecting onto
(c) the texture defined from the Hamiltonian (adiabatic approximation), and
(d) the texture obtained from the band.
The adiabatic texture is a poor approximation for the energies, while the band-defined texture-defined effective Hamiltonian approximates the upper
part of the spectrum well.} 
    \label{fig:energycompppp}
\end{figure}

To demonstrate the power of this approach, we consider the effective Hamiltonian for a realistic model of \mote{} at $\theta = 2.1^\circ$, which includes higher-harmonic terms. In this regime, the adiabatic approximation breaks down. The texture defined by the Hamiltonian $\boldsymbol{\hat n_{\boldsymbol{\Delta}}}(\mathbf r)$, shown in Fig.~\ref{fig:energycompppp}a, exhibits sharp domain walls arising from lattice relaxation~\cite{choAhnFirstLandauLevel2024,xiaoWangHigherLandauLevelAnalogues2024}. These rapidly varying features make the adiabatic approximation inapplicable. In sharp contrast, the texture defined by the topmost band (Fig.~\ref{fig:energycompppp}b) is much smoother, making it a more suitable starting point for constructing a single-component Hamiltonian.

Consistent with this picture, the numerically computed energies of the adiabatic effective Hamiltonian show strong disagreement with the full spectrum (Fig.~\ref{fig:energycompppp}c).
Remarkably, using instead the average band texture to construct the effective Hamiltonian of Eq.~\eqref{eq:heff}, we reproduce the energies of the topmost band with good accuracy (Fig.~\ref{fig:energycompppp}d).
The agreement is still acceptable even for the second topmost band.

\subsection{Single-component wavefunctions and Landau level content}

In the flat-band regime relevant for FCIs, the physics is largely determined by the wavefunctions and their overlaps. We have just discussed how the energies and eigenstates of a high-$\kappa$ band map onto an effective single-component Hamiltonian with a texture-induced magnetic field. Alternatively, it is possible to directly project the Bloch wavefunction, given in Eq.~\eqref{eq:decompose}, onto the dominant spinor texture $\chi_{\boldsymbol{\hat{n}}}$. 
For the topological real-space textures introduced and discussed in Secs.~\ref{sec:topoofensembles} and~\ref{sec:topoofensembles-appl}, the resulting single-component function can be expanded as $\alpha_{\mathbf k} \scalarpart_{\mathbf k}(\mathbf r) =  \sum_{n=0}^{+\infty} c^{\mathbf k}_{n} \, \llwf^{\Phi=1}_{n,\mathbf k}(\mathbf r)$. Here, 
$\llwf^{\Phi=1}_{n,\mathbf k}$ denotes the $n$-th Landau-level wavefunction with momentum $\mathbf k$ and one flux quantum per unit cell. The coefficients $c^{\mathbf k}_n$ quantify the overlap with each Landau level.
Importantly, their value is bounded by the texture polarization via
\begin{equation}\label{eq:LLdecomposition}
 \frac{1}{\nuc} \, \sum_{\mathbf k}\sum_{n=0}^{+\infty} |c^{\mathbf k}_{n} |^2 = 
\sum_ {n=0}^{+\infty} w_n = \kappa,  
\end{equation}
and can be further separated into the weights $w_n$ of each Landau level. The approach can also be further refined by considering generalized Landau levels which possess ideal but non-uniform quantum geometry 
encoded in a K\"ahler potential~\cite{vishwanathLedwithFractionalChernInsulator2020,yangWangExactLandauLevel2021,parkerLedwithVortexabilityUnifyingCriterion2022,wangLiuTheoryGeneralizedLandau2024,wuLiVariationalMappingChern2024}. In all cases, the projection of the single-component wavefunction onto Landau levels is key for building trial wavefunctions describing different FCI phases.
Crucially, the value of $\kappa$ of our optimal texture gives an upper bound for any such projection.

For large $\kappa$, the interacting physics of the band is essentially
that of a single-component band $\psi_{\mathbf k}$. 
In particular, the spinor structure only affects the interacting terms
to order $d/a_M$ compared to a monolayer problem, where $d$ is the interlayer distance and $a_M$ is the moir\'e wavelength.
Thus, trial states can be constructed just like in a single-component problem. 

In the simplest cases, considering generalized Landau levels is not necessary, and standard Landau levels with a uniform quantum geometry are sufficient to capture the physics in the bands. 
To illustrate the validity of this approach, we consider a realistic model of \mote{} at $\theta = 2.1^\circ$, which has been suggested to host Landau-level analogs~\cite{choAhnFirstLandauLevel2024,xiaoWangHigherLandauLevelAnalogues2024}. We perform a spinor projection followed by the Landau level decomposition of Eq.~\eqref{eq:LLdecomposition}. We find excellent alignment with the average spinor texture for the topmost and second topmost bands, with $\kappa = 0.98$ and $\kappa = 0.92$, respectively. Furthermore, the topmost band has most of its weight in the zeroth Landau level, $w_0 = 0.971$, while the second band predominantly overlaps with the first Landau level, $w_1 = 0.88$. These results indicate that the top two bands of \mote{} at $\theta = 2.1^\circ$ closely correspond to the lowest two Landau levels at an effective flux $\Phi_2 = 1$. This is consistent with previous Hofstadter calculations showing that these bands develop a parabolic dispersion under an external magnetic field $\Phi = -1$~\cite{moraKolarHofstadterSpectrumChern2024}. This correspondence naturally suggests the formation of Laughlin-like states in the topmost band and non-Abelian states in the second band.

\subsection{Application to TBG bands}
We now apply this analysis to sublattice-projected textures in TBG.
In this system, we find a robust fictitious magnetic field, opposite on each sublattice. In particular, all the wavefunctions are essentially aligned, with $\kappa \simeq 0.998$ even for realistic corrugation, see Eq.~\eqref{eq:defkappa}, and follow the texture shown in Figure~\ref{fig:figfour}b.
This suggests a natural decomposition of the TBG flat band eigenfunctions as
\begin{equation}
\label{eq:tbgwavefunctionsdecomposition}
u^{TBG}_{\mathbf k}(\mathbf r) = \chi_{}^{A} (\mathbf r) \tilde\scalarpart^{1}_{\mathbf k} (\mathbf r)+ 
\chi_{}^{B} (\mathbf r) \tilde\scalarpart^{-1}_{\mathbf k} (\mathbf r) + \mathcal O[1-\kappa]
\end{equation}
where the spinors $\chi_{}^{A/B}$ are fully $A/B$-sublattice polarized, 
the unnormalized scalar wavefunctions $\tilde\scalarpart^{\Phi_2}_{\mathbf k} (\mathbf r)$ 
experience a net magnetic field of $\Phi_2=\pm 1$ flux quanta per unit cell,
and the neglected terms are suppressed by a factor of $1-\kappa \ll 1$ for TBG.
The fictitious magnetic field implied by the textures can be made explicit
by projecting the TBG Hamiltonian onto the spinors $\chi^A(\mathbf r)$ and $\chi^B(\mathbf r)$,
turning a four-component problem into a two-component one.

To illustrate this better, Fig.~\ref{fig:figfour}c plots
the value of $|\boldsymbol{A}(\mathbf r)|/\rho(\mathbf r)$ in the unit cell, showing that the tracking is extremely good throughout the unit cell.
However, a crucial difference (as will be discussed below in Section~\ref{sec:heavyfermion}) from
the chiral limit is that the charge density is strongly peaked at the $AA$ point of the unit cell, so
that a large proportion of the unit cell is effectively not sampled.
\\

\section{Heavy fermion model for TBG}
\label{sec:heavyfermion}

Although realistic TBG is relatively far from the analytically accessible  chiral limit~\cite{vishwanathTarnopolskyOriginMagicAngles2019}, our findings show that the real-space topology carries over from the chiral limit, even quantitatively. To better understand this good agreement, we discuss real-space topology in the context of the topological heavy fermion (THF) model~\cite{bernevigSongMagicAngleTwistedBilayer2022,calugaru2023,Shi2022,Yu2023,Calugaru2024,Zhou2024,Rai2024,herzog2025kekul}.
THF is a minimal model involving six fermion species, which captures the band structure of the two central bands of TBG along with their topological properties. Two of these fermion variables, denoted \( f_1 \) and \( f_2 \), describe localized Wannier states centered at the AA points of the moir\'e lattice. These orbitals form a triangular lattice with very weak overlap between neighboring Wannier states, resulting in negligible kinetic energy (dispersion). The overlap between these Wannier states and the two central bands is excellent (\( 96\% \)) throughout most of the Brillouin zone. However, the overlap vanishes at the \( \Gamma \)-point, reflecting a topological obstruction~\cite{bernevigSongTBGIIStable2021,bernevigSongAllMagicAngles2019,vishwanathPoFaithfulTightbindingModels2019,ahn2019,vafekKangSymmetryMaximallyLocalized2018,fuKoshinoMaximallyLocalizedWannier2018,senthilZouBandStructureTwisted2018}, which requires the introduction of four light fermion species, denoted \( c_{j} \) ($j=1,\ldots, 4$). Although their overlap with the central bands is small, these fermions carry the entire topology of the central bands (with Berry curvature concentrated near the \( \Gamma \)-point) and are essentially delocalized in the sense that they are not Wannierizable. Details of the model are reviewed in App.~\ref{app:THF}. The THF model is effective in describing correlation effects in TBG, in particular the coexistence between localized moments and conducting states~\cite{yazdaniWongCascadeElectronicTransitions2020,yazdaniXieSpectroscopicSignaturesManybody2019,youngSaitoIsospinPomeranchukEffect2021,ilaniRozenEntropicEvidencePomeranchuk,khalafLedwithNonlocalMomentsChern2025}. 

\begin{figure}
    \centering
    \includegraphics[width=\columnwidth]{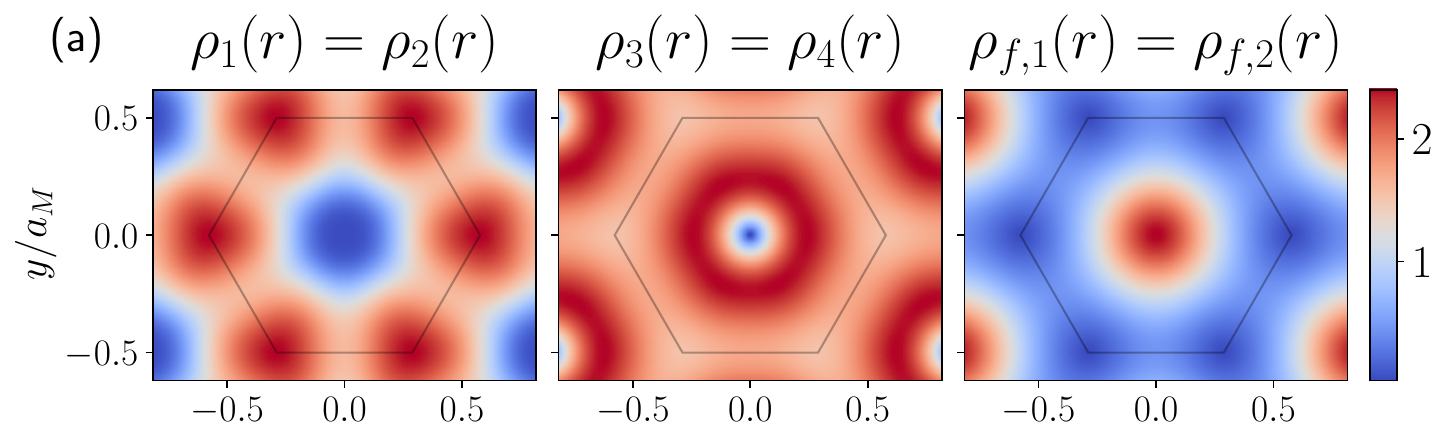} \\[2mm]
    \includegraphics[width=\columnwidth]{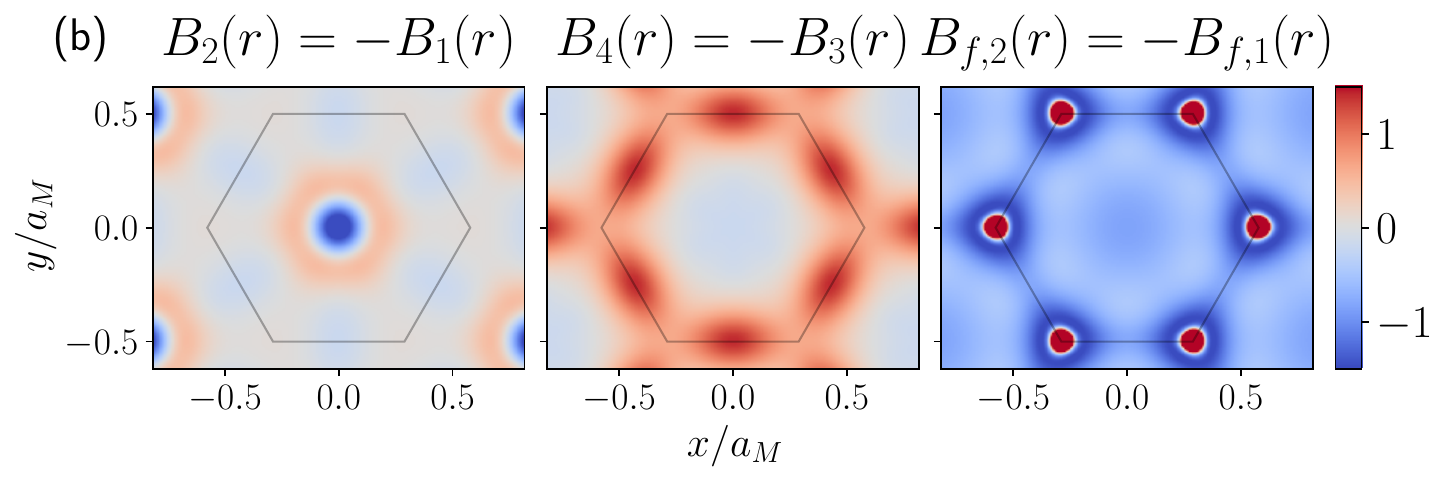}
    \caption{(a) Electronic densities, summed over the sublattice and layer indices, of the six orbital \( (c_1, c_2, c_3, c_4, f_1, f_2) \), two by two equal. The unit cell, centered at the AA site, is depicted as a light hexagon, with its edges corresponding to the BA and AB sites.
    (b) Real-space Berry curvatures for the six orbitals of the topological Heavy Fermion model with the corrugation parameter $w_{AA}/w_{AB} = 0.7$. }
    \label{fig:densitiesGamma}
\end{figure}

As we now elaborate,  regions with strong topological features are concentrated near the AB and BA sites, away from the AA-centered Wannier functions. 
We illustrate the spatial segregation of orbitals 
in Fig.~\ref{fig:densitiesGamma}a. The Wannier orbitals \( f_1 \), \( f_2 \) are well localized around the AA points, exhibiting very small densities at the edges of the unit cell.
In contrast, the orbitals \( c_1 \), \( c_2 \) mainly occupy the regions around the BA and AB points, forming a hexagonal lattice. The remaining conduction electrons, \( c_3 \) and \( c_4 \), exhibit vanishing density at the AA points and display a circular ring structure, which remains significant at the edges of the unit cell, particularly at the BA and AB points.

We contrast these densities with the real-space Berry curvatures of the various orbitals displayed in Fig.~\ref{fig:densitiesGamma}b. The topological features of the Wannier orbitals \( f_1 \), \( f_2 \) emerge in the BA and AB regions, where their densities are negligibly small. Similarly, the light fermions \( c_1 \), \( c_2 \) display topological characteristics in low-density regions and overlap only weakly with the central bands. In contrast, the light fermions \( c_3 \) and \( c_4 \) carry most of the real-space topology, despite having only a small overlap with the central bands overall. \( c_3 \) (resp. \( c_4 \)) carries a real-space Chern number of $-1$ (resp. $+1$). We have verified numerically, by computing the average sublattice polarizations of the different orbitals, that deviations from the chiral limit affect the Wannier states more significantly than the light fermions.  Physically, increasing the interlayer tunneling $w_{AA}$ between A atoms when moving away from the chiral limit is expected to impact AA stacking regions more strongly than AB or BA regions.
Consequently, we find that the real-space topology of TBG is primarily inherited from the \( c_3 \) and \( c_4 \) light conduction electrons and remains largely insensitive to the Wannier orbitals as well as more robust to deviations from the chiral limit.

\begin{figure}[t]
    \centering
    \includegraphics[width=\columnwidth]{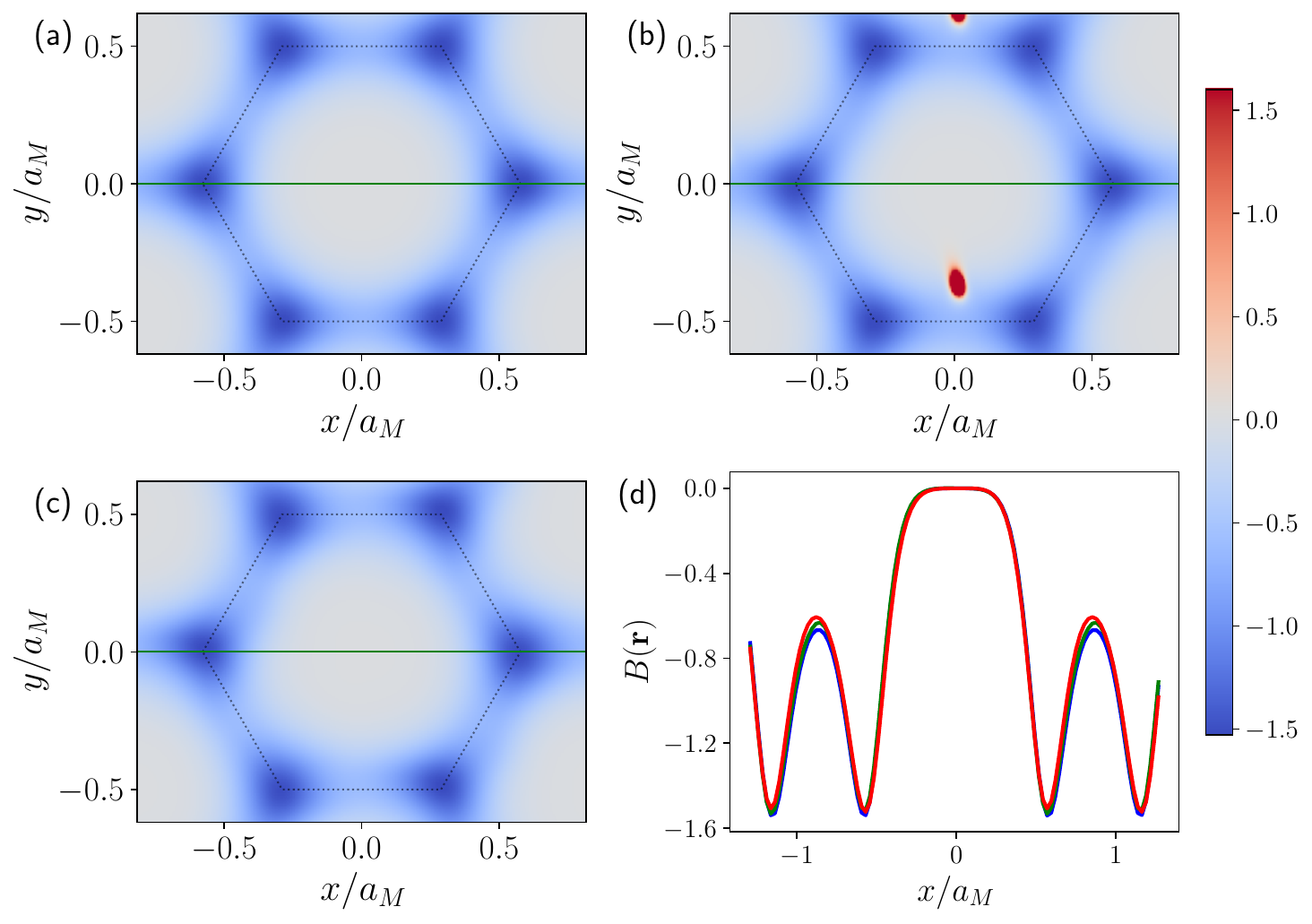}
    \caption{ Real-space topology of twisted bilayer graphene for individual and ensemble of Bloch wavefunctions,
    projected onto the A sublattice. 
    The Berry curvature $B(\mathbf r)$, characterizing the spatial change in orientation of the Bloch spinor wavefunctions, is represented over the unit cell (dotted hexagon) centered at AA sites. (a) In the chiral limit. (b) Individual and (c) ensemble  of wavefunctions for realistic corrugation, $w_{AA}/w_{BA}=0.7$. The real-space Berry curvatures integrate to $-1$ in (a) and (c), and to $0$ in (b) due to the strong positive contribution (dark-red region). (d) Linecuts of the three color maps (a-c) along the green line. (b) has been calculated with ${\bf k} = (\gmoire_3-\gmoire_1)/4$  and (c) with $E = \SI{2.8}{meV}$ but the results are largely insensitive to the value of $E$.
    }
    \label{fig:berrycomparison}
\end{figure}

\section{Discussion}
\label{sec:windingdiscussion}
To conclude, we demonstrated the existence of a robust real-space Chern number characterizing the fictitious magnetic field experienced by electrons in moir\' e systems. This Chern number universally applies to twisted bilayer graphene and twisted bilayer transition metal dichalcogenides without assuming delicate parameter regimes. It extends previous insights obtained for the special limits of chiral twisted bilayer graphene and the adiabatic approximation to transition metal dichalcogenides, which turn out to apply only to fine-tuned cases where the wavefunctions exhibit zeroes in the unit cell.  

We showed that, remarkably, the real-space topology of wavefunction ensembles in twisted bilayer TMDs is always nontrivial. We identified in particular the two spatial symmetries, $\cth$ and $\cty \trs$, as the essential ingredients enforcing a nontrivial real-space Chern number. Extending this framework to twisted bilayer graphene, we showed that the same symmetries protect a nontrivial real-space topology when wavefunctions are projected onto a single sublattice. 
Intriguingly, as shown in Fig.~\ref{fig:berrycomparison} the real-space Berry curvature profiles in the chiral limit and for individual and ensembles of wavefunctions at realistic corrugation are quantitatively very similar. In fact, far from the singularity of Fig.~\ref{fig:berrycomparison}b, the comparison between the three textures is quantitatively excellent as illustrated in Fig.~\ref{fig:berrycomparison}d.

Signatures of nontrivial real-space textures have been recently observed in twisted bilayer TMDs ~\cite{shihZhangExperimentalSignatureLayer2025,yankowitzThompsonMicroscopicSignaturesTopology2025}.
Not only does our theory provide a natural explanation for these experimental observations, 
it also shows that the same $\cth$ and $\cty \trs$-protected real-space topology should in fact also manifest in twisted bilayer graphene,
suggesting the detailed study of real-space textures in twisted bilayer graphene as a promising experimental possibility.
A particularly exciting experimental direction would be the study of layer-projected topology, where all three
components of the vector $\boldsymbol{\hat n}^{(t)}_E$ are observable \cite{yazdaniLiuVisualizingBrokenSymmetry2021}.
Based on our results, we expect this vector to define a topologically nontrivial texture provided the $\ctz\trs$ symmetry is broken.

The existence of a robust fictitious magnetic field in twisted bilayer graphene
raises several interesting theoretical questions. 
For instance, the decomposition in Eq.~\eqref{eq:tbgwavefunctionsdecomposition} suggests that an 
adiabatic approach, akin to the one applied successfully to twisted bilayer TMDs \cite{yaoYuGiantMagneticField2020,yaoZhaiTheoryTunableFlux2020,macdonaldMorales-DuranMagicAnglesFractional2024},
should extend to realistic models of TBG. However, because of the additional sublattice degree of freedom, the resulting adiabatic Hamiltonian will have two components, which experience opposite fictitious magnetic fields. Such an approach could provide additional insights into the analytical structure of the form factors, which is crucial in determining the correlation physics of the system. \cite{khalafLedwithNonlocalMomentsChern2025,bernevigSongMagicAngleTwistedBilayer2022,bernevigHerzog-ArbeitmanTopologicalHeavyFermion2024}.

Our framework 
of band textures developed in Section~\ref{sec:bandlandaumapping}
can also be directly applied to other moir\'e systems 
to study whether a fictitious magnetic field is present, as well as to discover relevant approximation schemes.
Indeed, our method may serve as a simple and general zeroth-order framework for characterizing moir\'e systems. 
Beginning with the single-particle bands, one computes their real-space textures and the associated topology.
These textures can then be used to construct an effective single-component Hamiltonian that features an effective magnetic field, implying ansatz wave functions in the form of a spinor factor multiplied by a Landau-level-like orbital. 
Having pointed out the importance of the $\cth$ and $\cty \trs$ symmetries for the nontrivial quantum anomalous Hall physics in TBG and twisted bilayer TMDs, we expect emerging moir\'e systems possessing these symmetries \cite{bernevigJiang2DTheoreticallyTwistable2024}
to display similar phenomenology.
Other important systems, such as hBN-aligned rhombohedral multilayer graphene, or helical trilayer graphene, lack these symmetries.
The former features an interaction-induced Chern insulator \cite{juLuFractionalQuantumAnomalous2024,Zhengguang2025,youngChoiElectricFieldControl2024,juHanSignaturesChiralSuperconductivity2025,yankowitzWatersChernInsulatorsInteger2025,xie2024fqah}, suggesting the tantalizing possibility of an interaction-induced fictitious magnetic field.

An interesting direction would be to extend our framework to treat intervalley-coherent orders in TBG, and look for the real-space topological properties of the different valley-coherent states~\cite{vafek2019,zaletelBultinckGroundStateHidden2020} as well as the incommensurate Kekul\'e spiral order~\cite{bultinckKwanKekuleSpiralOrder2021}. With inter-valley coherence and preserved spin $S_z$ conservation,
twisted bilayer graphene wavefunctions have $N=8$ components, requiring either a projection onto a layer and sublattice, or the analysis
of the topological properties of an $N=8$-dimensional real-space Hamiltonian, as detailed in Appendix~\ref{app:Ncomponent}.

Finally, the definition of the energy-resolved texture can be extended to many-body states beyond the mean-field approximation by using a layer-resolved local spectral function~\cite{bernevigCalugaruSpectroscopyTwistedBilayer2021}. In this case, our symmetry-based arguments should still hold, 
which raises the possibility that the finite real-space Chern number of twisted bilayer TMDs is relevant even in the fractional Chern insulator phase
\cite{xuCaiSignaturesFractionalQuantum2023,xuParkObservationFractionallyQuantized2023,shanZengThermodynamicEvidenceFractional2023,liXuObservationIntegerFractional2023,feldmanFouttyMappingTwisttunedMultiband2024,makKangEvidenceFractionalQuantum2024}. Moreover, it has recently been predicted~\cite{LinYangLuZhaiYao2025} that an FCI could emerge from an isolated zero-Chern band with interactions that do not couple to other bands. If confirmed, this suggests that nontrivial momentum-space topology may play only a secondary role in the formation of FCIs \cite{PhysRevB.92.195104}, calling for future investigations of the precise roles of real-space topology and fictitious magnetic fields in the formation of FCIs as well as their connection to quantum geometry.

\begin{acknowledgments}
We thank Daniele Guerci, Jie Wang, Ahmed Abouelkomsan, Zhao Liu, Bruno Mera, and Wei Zhu for helpful discussions. We gratefully acknowledge
support by Deutsche Forschungsgemeinschaft through CRC 183 (project C02) in Berlin as well as a joint ANR-DFG project (TWISTGRAPH, ANR-21-CE47-0018) in Berlin and Paris. KK acknowledges  additional support through the Simons Collaboration on New Frontiers in Superconductivity (Simons Foundation grant SFI-MPS-NFS-00006741-12, P.T.). 
\end{acknowledgments}

\appendix

\section{Bloch states in a magnetic field}
\label{sec:BlochMagnField}

We describe the magnetic field using a general linear vector potential 
\begin{eqnarray}
    \mathbf{A} = \beta \mathbf{r} \quad ; \quad \beta = \left(\begin{array}{cc}
        \beta_{xx} & \beta_{xy} \\
      \beta_{yx}   & \beta_{yy}
    \end{array}
    \right).
\end{eqnarray}
Demanding that $\boldsymbol{\nabla}\times \mathbf{A} = B\mathbf{\hat z}$ gives the condition
\begin{eqnarray}
    \beta_{yx} - \beta_{xy} = B.
    \label{eq:betaxybetayxB}
\end{eqnarray}
We define the guiding-center operator
\begin{equation}
    \mathbf{R} = \mathbf{r} - \frac{1}{eB} \mathbf{\hat z}\times \boldsymbol{\pi } 
\end{equation}
with the kinetic momentum $\boldsymbol{\pi} = \mathbf{p} - e\mathbf{A}$. This can be used to express the magnetic translation operators
\begin{equation}
    T_\mathbf{a} = e^{ieB(\mathbf{\hat z}\times \mathbf{R})\cdot \mathbf{a}}.
\end{equation}
In particular, we define the translation operators 
\begin{equation}
    T_j = e^{ieB(\mathbf{\hat z}\times \mathbf{R})\cdot \mathbf{a}_j}
\end{equation}
associated with the two lattice vectors $\mathbf{a}_j$ with $j=1,2$. (In explicit calculations, we choose without loss of generality that $\mathbf{a}_1 \parallel \mathbf{\hat x}$, while the direction of $\mathbf{a}_2$ is arbitrary.) For our choice of gauge, the translation operators take the explicit form
\begin{eqnarray}
    T_1 &=& e^{ieB(\mathbf{a}_1\times\mathbf{\hat z})\cdot \mathbf{r} - ie\mathbf{A} \cdot \mathbf{a}_1+ i\mathbf{p}\cdot \mathbf{a}_1}, 
    \nonumber\\
    T_2 &=& e^{ieB(\mathbf{a}_2\times\mathbf{\hat z})\cdot \mathbf{r} -ie\mathbf{A}\cdot\mathbf{a}_2 + i\mathbf{p}\cdot \mathbf{a}_2}.
\end{eqnarray}
We now choose a gauge such that this simplifies in a manner similar to the Landau gauge.

We first demand that $T_1=e^{i\mathbf{p}\cdot\mathbf{a}_1}$, i.e., that
\begin{align}
    0 &= eB(\mathbf{a}_1\times\mathbf{\hat z})\cdot \mathbf{r} - e \mathbf{a}_1^\top \cdot \beta \cdot \mathbf{r} 
    \nonumber\\
    &= \left[ eB(\mathbf{a}_1\times\mathbf{\hat z})^\top - e \mathbf{a}_1^\top \cdot \beta \right] \cdot \mathbf{r}. 
\end{align}
Thus, the expression in the square brackets must vanish, 
\begin{align}
    0 &= eB(\mathbf{\hat a}_1\times\mathbf{\hat z})^\top - e \mathbf{\hat a}_1^\top \cdot \beta
    \nonumber\\
    &= -eB\mathbf{\hat y}^\top - e \mathbf{\hat x}^\top \cdot \beta, 
\end{align}
which implies (recall we choose our coordinate system to satisfy $\mathbf{a}_1 \parallel \mathbf{\hat x}$)
\begin{eqnarray}
    \beta_{xx} = 0 \quad ; \quad \beta_{xy} = -B.
\end{eqnarray}
Combining with Eq.\ \eqref{eq:betaxybetayxB} gives 
\begin{equation}
    \beta_{yx} = 0.
\end{equation}
We still have $\beta_{yy}$ left at our disposal.

We now consider $T_2$. We demand that $T_2$  separates into a product of two commuting exponential factors involving $\mathbf{p}$ and $\mathbf{r}$, respectively. The term $eB(\mathbf{a}_2\times\mathbf{\hat z})\cdot \mathbf{r}$ in the exponent already contains only the coordinate perpendicular to the $\mathbf{a}_2$ axis, so that it commutes with $\mathbf{p}\cdot\mathbf{a}_2$. We still need to ensure that 
\begin{align}
    e\mathbf{A}\cdot\mathbf{a}_2 &= e \mathbf{a}_2^\top \cdot \beta \cdot \mathbf{r} 
    \nonumber\\
    &= (a_{2x} , a_{2y}) \left(\begin{array}{cc}
      0   & -B \\
      0   &  \beta_{yy}
    \end{array} \right)
    \left(\begin{array}{c}
         x  \\ y
    \end{array}\right)
    \nonumber\\
    &= (- a_{2x} B + a_{2y}\beta_{yy})y
\end{align}
depends only on the coordinate perperdicular to $\mathbf{a}_2$. This is only the case, if we chose the prefactor of $y$ to vanish, which fixes
\begin{equation}
    \beta_{yy} = B\frac{a_{2x}}{a_{2y}}.
\end{equation}
We have now fully determined the Landau-type vector potential, \begin{equation}
    \mathbf{A} =  B\frac{\mathbf{\hat z}\cdot(\mathbf{{ a}_1}\times\mathbf{r})}{\mathbf{\hat z}\cdot (\mathbf{{a}_1}\times\mathbf{ a}_2)} \mathbf{{\hat z}}\times\mathbf{ a}_2,
\end{equation}
as well as the associated  translation operators along the lattice vectors,
\begin{align}
    T_1 &= e^{i\mathbf{p}\cdot\mathbf{a}_1}
    \\
    T_2 &= e^{ieB(\mathbf{a}_2\times\mathbf{\hat z})\cdot \mathbf{r} + i\mathbf{p}\cdot \mathbf{a}_2}.
\end{align}

We can classify eigenstates according to the eigenvalues of $T_1$ and $T_2$ provided that these two operators commute. We find 
\begin{equation}
    T_1 T_2 = e^{ieB( \mathbf{a}_2 \times \mathbf{\hat z})\cdot \mathbf{a}_1} T_2 T_1,
\end{equation}
and read off that $T_1$ and $T_2$ commute if the unit cell is threaded by an integer number $\Phi$ of flux quanta,
\begin{equation}
    B\mathbf{\hat z}\cdot(\mathbf{a}_1\times \mathbf{a}_2) = \Phi \frac{2\pi \hbar}{e}.
\end{equation}
We will now assume that this is the case. 

We define Bloch states as simultaneous eigenstates of $T_1$ and $T_2$,
\begin{equation}
    T_j e^{i\mathbf{k}\cdot \mathbf{r}} u_\mathbf{k}(\mathbf{r}) = e^{i\mathbf{k}\cdot \mathbf{a}_j} e^{i\mathbf{k}\cdot \mathbf{r}} u_\mathbf{k}(\mathbf{r}).
\end{equation}
Using the explicit expressions for the $T_j$, this gives
\begin{eqnarray}
    u_\mathbf{k}(\mathbf{r}+\mathbf{a}_1) &=& u_\mathbf{k}(\mathbf{r}) 
    \nonumber\\
    u_\mathbf{k}(\mathbf{r}+\mathbf{a}_2) &=& e^{-ieB(\mathbf{a}_2\times\mathbf{\hat z})\cdot \mathbf{r}} u_\mathbf{k}(\mathbf{r}).
\end{eqnarray}
Finally expressing the magnetic field through the integer number of flux quanta, we find
\begin{equation}
    \mathbf{A} = \frac{2\pi \hbar \Phi}{e}\frac{\mathbf{\hat z}\cdot(\mathbf{{ a}_1}\times\mathbf{r})}{[\mathbf{\hat z}\cdot (\mathbf{{a}_1}\times\mathbf{ a}_2)]^2} \mathbf{{\hat z}}\times\mathbf{ a}_2
\end{equation}
as well as
\begin{eqnarray}
    u_\mathbf{k}(\mathbf{r}+\mathbf{a}_1) &=& u_\mathbf{k}(\mathbf{r}) 
    \nonumber\\
    u_\mathbf{k}(\mathbf{r}+\mathbf{a}_2) &=& e^{-i2\pi\Phi\frac{\mathbf{\hat z}\cdot(\mathbf{r}\times \mathbf{a}_2)}{\mathbf{\hat z}\cdot(\mathbf{a}_1\times \mathbf{a}_2)}} u_\mathbf{k}(\mathbf{r}).
\end{eqnarray}
We can also define the reciprocal lattice vectors
\begin{equation}
    \mathbf{G}_1 = 2\pi \frac{\mathbf{a}_2\times \mathbf{\hat z}}{\mathbf{\hat z}\cdot(\mathbf{a}_1\times \mathbf{a}_2)} \quad ; \quad \mathbf{G}_2 = 2\pi \frac{ \mathbf{\hat z}\times\mathbf{a}_1}{\mathbf{\hat z}\cdot(\mathbf{a}_1\times \mathbf{a}_2)} 
\end{equation}
to simplify the notation,
\begin{equation}
    \mathbf{A} = -\frac{\Phi}{2\pi e}
    (\mathbf{G}_2\cdot\mathbf{r}) \mathbf{G}_1
\end{equation}
and 
\begin{eqnarray}
    u_\mathbf{k}(\mathbf{r}+\mathbf{a}_1) &=& u_\mathbf{k}(\mathbf{r}) 
    \nonumber\\
    u_\mathbf{k}(\mathbf{r}+\mathbf{a}_2) &=& e^{-i\Phi \mathbf{G}_1\cdot\mathbf{r}}u_\mathbf{k}(\mathbf{r}).
\end{eqnarray}

\section{Details on the TMD model}\label{app:tmdintro}

The Hamiltonian of a twisted bilayer TMD reads
\begin{equation}
\label{eq:sphammanewgaugeapp}
H_{\text{tTMD}}^{K}= -\frac{\hbar^2 (\mathbf k - \frac{\mu_z \mathbf q_1}{2})^2}{2m^*}\mu_0 + \boldsymbol{\Delta}(\mathbf r)\cdot \boldsymbol{\mu} +\Delta_0(\mathbf r)\mu_0,
\end{equation}
where $m^*$ is the effective mass of the holes, and where $\mu_{0,x,y,z}$ are the Pauli matrices in layer space,
and where $\mathbf q_1=(0,4\pi \theta/(3a_0))$ is the momentum offset between the valence band maxima of the two layers,
where $a_0$ is the lattice constant and $\theta$ is the twist angle.
$\boldsymbol{\Delta}(\mathbf r)$  and $\Delta_0(\mathbf r)$ are both periodic functions in the moir\'e unit cell, 
given in terms o layer potentials and interlayer tunneling terms as
\begin{eqnarray}
\Delta_0 =\frac{1}{2}[V_t(\mathbf r)+V_b(\mathbf r)]\\
\Delta_x =\mathrm{Re} \left[ w \left(1 + e^{i\gmoire_5\cdot \mathbf r} + e^{i\gmoire_6\cdot \mathbf r}\right)\right] \\
\Delta_x =-\mathrm{Im} \left[ w \left(1 + e^{i\gmoire_5\cdot \mathbf r} + e^{i\gmoire_6\cdot \mathbf r}\right)\right] \\
\Delta_z =\frac{1}{2}[V_b(\mathbf r)-V_t(\mathbf r)],
\end{eqnarray}
where $w$ is the tunneling strength and
where the layer potentials in the first harmonic approximation are parametrized by an amplitude $V$ and a phase $\psi$:
\begin{eqnarray}
\label{eq:tmdlayerpotbot}
\layerpotreal{b}{\mathbf r} =\left( Ve^{i\psi} \sum_{j=1,3,5} e^{i\mathbf r \cdot \gmoire_j} \right) +\hc \\
\label{eq:tmdlayerpottop}
\layerpotreal{t}{\mathbf r} =\left( Ve^{-i\psi} \sum_{j=1,3,5} e^{i\mathbf r \cdot \gmoire_j}\right)  +\hc,
\end{eqnarray}
where $\hc$ denotes the complex conjugate.
We use the parameters $(a_0,m^*,V,\psi,w) = (\SI{0.332}{nm} ,\SI{0.43 }{m_e},\SI{9}{meV},-128^\circ, \SI{18}{meV}) $,
taken from Ref.~\cite{fuDevakulMagicTwistedTransition2021}.

\section{Non-vanishing spinor wavefunction}\label{app:zero}

The Hamiltonian of graphene moir\'e systems can be expressed, in the chiral limit, as
\begin{equation}
H =
\begin{pmatrix}
0 & \mathcal{D}^\dagger \\
\mathcal{D} & 0
\end{pmatrix},
\end{equation}
when written in the sublattice basis.
The A-polarized ideal bands then correspond to the zero modes of the differential operator
\begin{equation}\label{eq:diff-operator}
\mathcal{D} = -2 i v_F \bar{\partial} + A(\mathbf{r}),
\end{equation}
where $\bar{\partial} = (\partial_x + i \partial_y)/2$, and $A(\mathbf{r})$ is a spatially varying $N \times N$ matrix, with $N$ denoting the number of layers.
Similarly, the B-polarized ideal bands correspond to the zero modes of $\mathcal{D}^\dagger$.

Let us assume that the spinor function $\mathcal{B}(\mathbf{r})$ appearing in the decomposition of Eq.~\eqref{eq:idealwf} vanishes at some position, which we take to be the origin without loss of generality. The Landau-level component of the wavefunction, $\llwf^{\Phi=1}_{0,\mathbf{k}}(\mathbf{r})$, also exhibits a zero that depends on the choice of momentum $\mathbf{k}$. Choosing $\mathbf{k} = \mathbf{k}_0$ such that this zero coincides with the origin, the Bloch part of the ideal wavefunction, $\blochpartidealz$, develops a double zero, and can be expanded as
\begin{eqnarray}\label{eq:low-z-expansion}
\blochpartidealz \simeq C_1 \bar{z} z  + \ldots,
\end{eqnarray}
for $z \to 0$, with $C_1 \ne 0$.
The Landau-level factor contributes a holomorphic term $\sim z$, while the spinor component contributes a $\bar{z}$ dependence corresponding to an antivortex centered at the origin.
Substituting the expansion of Eq.~\eqref{eq:low-z-expansion} into the zero-mode equation $\mathcal{D} \, \blochpartidealz = 0$, we obtain, for small $z$,
\begin{equation}
-2 i v_F C_1 z + \mathcal{O}(z^2, |z|^2, \bar{z}^2) = 0,
\end{equation}
which implies $C_1 = 0$ and thus a contradiction. This completes the proof that the spinor  $\mathcal{B}(\mathbf{r})$ is nowhere vanishing.
In contrast, we note that for a double-Dirac differential operator $\mathcal{D} = \bar{\partial}^2 + A(\mathbf{r})$, the expansion in Eq.~\eqref{eq:low-z-expansion} is compatible~\cite{moraGuerciLayerSkyrmionsIdeal2024} with $C_1 \neq 0$, and the spinor $\mathcal{B}(\mathbf{r})$ may host an antivortex, as predicted in periodically strained graphene~\cite{sunWanTopologicalExactFlat2023,Eugenio_2023}.

\section{Textures defined from a band maximize $\kappa$.}
\label{app:maximizingtexture}
In the text, we claimed that
the texture defined from the band Eq.~\eqref{eq:bandtext} maximizes $\kappa$
[Eq.~\eqref{eq:defkappa}] across all possible projectors on a single component,
thus defining the optimal single-component direction for that band.
To prove this, consider the value of $\kappa_{\mguynor}$ for some arbitrary spinor direction $\mguy$,
defined using the projector onto $\mguy$
$P_{\boldsymbol{\hat{m}}}(\mathbf r)= \frac{1+\boldsymbol{\mu}  \cdot \mguy}{2}$.
We have 
\begin{eqnarray}
 \kappa_{\mguynor}&=& \int_{\uc} d\mathbf r
\sum_{\mathbf k \in BZ}u^\dagger_{\mathbf k,m}(\mathbf r)
P_{\mguynor}(\mathbf r)
u_{\mathbf k,m}(\mathbf r)  \\
 &=& \int_{\uc} d\mathbf r
\sum_{\mathbf k \in BZ}u^\dagger_{\mathbf k,m}(\mathbf r)
\frac{1+\boldsymbol{\mu}  \cdot \mguy}{2}
u_{\mathbf k,m}(\mathbf r)  \\
 &=& \int_{\uc} d\mathbf r
\frac{\rho(\mathbf r)+|\boldsymbol A(\mathbf r)|(\mguy \cdot \nguy)}{2}\\
\label{eq:ourtextureisoptimal}
& \leq & \int_{\uc} d\mathbf r
\frac{\rho(\mathbf r)+|\boldsymbol A(\mathbf r)|}{2} = \kappa.
\end{eqnarray}

\section{Effective Hamiltonian for $\kappa$ near $1$}
\label{app:schrieffer-wolff}

We consider the Bloch Hamiltonian $H_{\bf k} = e^{-i \mathbf k \cdot \mathbf r} H e^{i \mathbf k \cdot \mathbf r}$ for a two-component system. $H$ is the Hamiltonian, given for instance by Eq.~\eqref{eq:sphammanewgauge} for a single valley in the case of TMDs. The periodic Bloch function solves
\begin{equation}\label{eq:blocheigen}
  H_{\bf k}   u_{\mathbf k,b,\lambda_b}(\mathbf r) = \epsilon_{\mathbf k,b,\lambda_b} u_{\mathbf k,b,\lambda_b}(\mathbf r)
\end{equation}
with the energy eigenvalue $\epsilon_{\mathbf k,b,\lambda_b}$. We insert the decomposition Eq.~\eqref{eq:decompose} onto the two spinors $\chi_{\boldsymbol{\hat{n}}}$ and $\chi^\perp_{\boldsymbol{\hat{n}}}$ in Eq.~\eqref{eq:blocheigen}. Using their orthogonality, $\chi_{\boldsymbol{\hat{n}}}^\dagger (\mathbf r) \chi^\perp_{\boldsymbol{\hat{n}}} (\mathbf r) =0$, we multiply on the left Eq.~\eqref{eq:blocheigen} by $\chi_{\boldsymbol{\hat{n}}} (\mathbf r)$ and $\chi^\perp_{\boldsymbol{\hat{n}}} (\mathbf r)$  to arrive at a system of two coupled equations
\begin{equation}
\begin{split}
  & \begin{pmatrix}
      \chi^\dagger_{\boldsymbol{\hat{n}}}(\mathbf r) H_{\bf k} \chi_{\boldsymbol{\hat{n}}}(\mathbf r) &  \chi^\dagger_{\boldsymbol{\hat{n}}}(\mathbf r) H_{\bf k} \chi^\perp_{\boldsymbol{\hat{n}}} (\mathbf r) \\[3mm] [\chi^\perp_{\boldsymbol{\hat{n}}} (\mathbf r)]^\dagger H_{\bf k} \chi_{\boldsymbol{\hat{n}}}(\mathbf r) & [\chi^\perp_{\boldsymbol{\hat{n}}} (\mathbf r) ]^\dagger H_{\bf k} \chi^\perp_{\boldsymbol{\hat{n}}} (\mathbf r) 
  \end{pmatrix} 
  \begin{pmatrix}  \psi_{\mathbf k} (\mathbf r)  \\[3mm]
      \frac{\beta_{\mathbf k}}{\alpha_{\mathbf k}} \varphi_{\mathbf k} (\mathbf r)          
      \end{pmatrix} \\[3mm]
      & =  \epsilon_{\mathbf k,b,\lambda_b} \begin{pmatrix}  \psi_{\mathbf k} (\mathbf r)  \\[3mm]
     \frac{\beta_{\mathbf k}}{\alpha_{\mathbf k}} \varphi_{\mathbf k} (\mathbf r)
      \end{pmatrix}
\end{split}
\end{equation}
The band is nearly aligned for $\kappa$ close to one. In that case, to leading order in
$|\beta_{\mathbf k}|/|\alpha_{\mathbf k}| \ll 1$, the second term is negligible and we obtain an approximate equation for $\psi_{\mathbf k}$,
\begin{equation}
    H_{\text{eff}} \,
 \psi_{\mathbf k} (\mathbf r) \approx \epsilon_{\mathbf k,b,\lambda_b}\psi_{\mathbf k}(\mathbf r),
\end{equation}
corresponding to a {\it single-component} effective Hamiltonian $H_{\text{eff}}=\chi^\dagger_{\boldsymbol{\hat{n}}}(\mathbf r) H\chi_{\boldsymbol{\hat{n}}}(\mathbf r)$,
as given in Eq.~\eqref{eq:heff} of the main text.

These manipulations clarify the emergence of the fictitious magnetic field encoded in the texture of Eq.~\eqref{eq:pontryagin-density}, as the spatial derivative $-i \nabla_{\mathbf r}$ in the Bloch Hamiltonian $H_{\mathbf k}$ generates the Berry connection of the texture magnetic field, see Eq.~\eqref{eq:vectorAtilde}.

\section{Ensemble-averaged topological index}\label{app:Ncomponent}

We detail here the introduction of a robust topological index accounting for the texture of moir\'e bands.
We start with a Bloch spinor wavefunction \( u_{\mathbf k,n,\lambda} (\mathbf{r}) \) with two components (\( N = 2 \)) and define the $2\times 2$ real-space Hamiltonian  
\begin{equation}
\label{eq:energyresolved3}
    H'( \mathbf{r}) = \sum_{\lambda,n,\mathbf k \in BZ} u_{\mathbf k,n,\lambda}(\mathbf{r}) u^\dagger_{\mathbf k,n,\lambda}(\mathbf{r}) \, p_{\mathbf k,n,\lambda}.
\end{equation}
  By construction, \( H'( \mathbf{r}) \) is hermitian and can be expanded in terms of the three Pauli matrices and the identity matrix \( \mu_0 \). This yields:  
\begin{equation}
    H'( \mathbf{r}) = K( \mathbf{r}) \sigma_0 + \frac{1}{2} H( \mathbf{r}),
\end{equation}
with $K( \mathbf{r}) = \frac{1}{2} \sum_{\lambda,n,\mathbf k \in BZ} u^\dagger_{\mathbf k,n,\lambda}(\mathbf{r}) u_{\mathbf k,n,\lambda}(\mathbf{r}) \, p_{\mathbf k,n,\lambda}$  as a global shift and the matrix \( H( \mathbf{r}) \) being equal to the Hamiltonian in  Eq.~\eqref{eq:energyresolved}. The two matrices \( H \) and \( H' \) share the same eigenvectors. We denote the eigenvector corresponding to the highest eigenvalue \( \lambda_{\rm max}(\mathbf{r}) \) as \( \psi_E (\mathbf{r}) \), which we take to be normalized, $\psi_E^\dagger (\mathbf{r}) \psi_E (\mathbf{r})=1$.

In the extreme limit that the sum in Eq.~\eqref{eq:energyresolved3} includes only a single momentum \( \mathbf{k}_0 \), the maximal eigenvalue satisfies \( \lambda_{\rm max}(\mathbf{r}) = 1 \) for all \( \mathbf{r} \), and the corresponding eigenvector is simply \( \psi_E (\mathbf{r}) \propto u_{\mathbf{k}_0} (\mathbf{r}) \). Moreover, the second eigenvalue of \( H' \) is identically zero.
In the general case, \( \lambda_{\rm max}(\mathbf{r}) \) deviates from one and the second eigenvalue from zero. Both eigenvalues form band-like functions of \( \mathbf{r} \), which are periodic over the moir\'e lattice. 
Assuming the gap remains open, recover $C[\boldsymbol{\hat{n}}]$ given in Sec.~\ref{sec:topology-ensemble} as the Chern number of the top band of this Hamiltonian.

The definition of the moir\'e-periodic Hamiltonian Eq.~\eqref{eq:energyresolved3} remains the same with $N$ components, in which case $H'( \mathbf{r})$ is a $N \times N$ matrix. When the sum is restricted to a single momentum $\mathbf{k}_0$, the spectrum of $H'$ is simple: $u_{\mathbf{k}_0} (\mathbf{r})$ is an eigenstate with eigenvalue $1$, the $N-1$ remaining eigenvalues are all vanishing in the subspace orthogonal to $u_{\mathbf{k}_0} (\mathbf{r})$. 
With the ensemble average beyond an individual state, the spectrum is modified and the bands become dispersive. 
Depending on the number of spectral gaps, there are at most $N$ real-space Chern numbers whose sum is constrained to be zero.

Analyzing the properties of this Hamiltonian permits the isolation of a relevant low-energy subspace.
For instance, in TBG, where there are $N=4$ components, such an analysis would reveal the presence of a relevant two-dimensional subspace.
Each dimension would correspond to the chiral limit textures on the A and B sublattices.
Crucially, even without a priori knowledge of the chiral limit, 
such an analysis would allow the presence of fictitous magnetic fields in TBG to be revealed.

\section{Symmetry indicators}
\label{app:symmetryindicators} 
We now use symmetry indicators of band topology \cite{bernevigFangBulkTopologicalInvariants2012}
to show that the properties 
in Eqs.~\eqref{eq:naa},~\eqref{eq:nab}~and~\eqref{eq:nba}~
imply that $C[\boldsymbol{\hat n}_E]$ is necessarily nonzero. 
This approach can be applied to the real-space Hamiltonian in  Eq.~\eqref{eq:energyresolvedh}. The Chern number of its upper band is equal to $C[\boldsymbol{\hat n}_E]$. By symmetry indicators, this Chern number is given in terms of the $\cth$ eigenvalues of the positive-energy eigenstates at the high-symmetry stacking points $AA$, $AB$ and $BA$
as 
\begin{equation}
\label{eq:chernnumberindicator}
   e^{i2 \pi C[\boldsymbol{\hat n}_E]/3} = \theta(AA)\theta(AB) \theta(BA),
\end{equation}
where $\theta(\mathbf r)$ is the $\cth$ eigenvalue at point $\mathbf r$.
For the $AA$ stacking point, we have the trivial eigenvalue $\theta(AA) =1 $.
For the AB and BA points, the eigenvalues depend on the orientation, i.e., the choice of the $\pm$ sign in Eq.~\eqref{eq:nab}.
Using that $e^{i\mathbf g_5 \cdot \mathbf r_{BA}} = e^{i 4\pi/3}$,
we find for $\boldsymbol{\hat n}_E(AB)= (0,0, 1)$ that $\theta (AB) = 1, \theta(BA) = e^{i4\pi  /3}$.
On the other hand, for $\boldsymbol{\hat n}_E(AB)= (0,0, -1)$, 
we use that $e^{i\mathbf g_5 \cdot \mathbf r_{AB}} = e^{i 2 \pi/3}$,
obtaining $\theta (AB) = e^{i 2\pi  /3}, \theta(BA) =1 $.
In both cases, we thus obtain that the real-space Chern number
\begin{equation}
C[\boldsymbol{\hat n}_E] = \pm 1 \mod 3
\end{equation}
\section{Topological Heavy Fermion model}\label{app:THF}

The Bloch hamiltonian reproducing the two central bands is given by the $6\times6$ matrix, in the basis of the six fermion species $(c_1,c_2,c_3,c_4,f_1,f_2)$,
\begin{equation}\label{eq:sixband}
    H_{\bf p} = \begin{pmatrix}
        0 & v_1 (p_x + i \sigma_z p_y) &  f({\bf p}) \\ v_1 (p_x - i \sigma_z p_y) & M \sigma_z & 0 \\
        f({\bf p}) & 0 & 0
    \end{pmatrix}
\end{equation}
where ${\bf p}$ is relative to the $\Gamma$ point. Far from $\Gamma$, the coupling matrix
\begin{equation}
    f({\bf p}) = \Big [ \gamma + v_2 (p_x \sigma_x + p_y \sigma_y) \Big] e^{- \frac{(p \lambda)^2}{2}}
\end{equation}
between the Wannier orbitals \( f_1\), \( f_2 \) and the conduction electrons \( c_1\), \( c_2 \) becomes exponentially small. The central bands thus match the Wannier representation and exhibit very weak energy dispersion. As one approaches the $\Gamma$ point, the orbital content of the two central bands gradually shifts towards \( c_3\), \( c_4 \), with the reduced Hamiltonian at $\Gamma$
\begin{equation}
    \begin{pmatrix}
        0 & M \\ M & 0
    \end{pmatrix}
\end{equation}
having eigenenergies $\pm M$. The overlap between the central bands and the conduction electrons \( c_1\), \( c_2 \) peaks at most at $5\%$.

To better clarify the structure of the six-band hamiltonian~\eqref{eq:sixband}, we focus on the conduction electrons and write the projected Bloch hamiltonian
\begin{equation}
    H^c_{\bf p} = \begin{pmatrix}
        0 & v_1 (p_x + i \sigma_z p_y)  \\ v_1 (p_x - i \sigma_z p_y) & M \sigma_z 
    \end{pmatrix}.
\end{equation}
It is precisely the same as the K-valley effective Hamiltonian of (untwisted) Bernal-stacked bilayer graphene in the absence of trigonal warping. The model exhibits quadratic band touching at the $\Gamma$ point, with a Berry phase winding of $2\pi$ around $\Gamma$. This nontrivial phase winding originates from the K-valley Dirac cones of the two individual monolayers. The two Dirac cones merge to form a quadratic band touching, and they are responsible for the topological obstruction~\cite{bernevigSongMagicAngleTwistedBilayer2022}.

\bibliography{ref}

\begin{thebibliography}{115}%
\makeatletter
\providecommand \@ifxundefined [1]{%
 \@ifx{#1\undefined}
}%
\providecommand \@ifnum [1]{%
 \ifnum #1\expandafter \@firstoftwo
 \else \expandafter \@secondoftwo
 \fi
}%
\providecommand \@ifx [1]{%
 \ifx #1\expandafter \@firstoftwo
 \else \expandafter \@secondoftwo
 \fi
}%
\providecommand \natexlab [1]{#1}%
\providecommand \enquote  [1]{``#1''}%
\providecommand \bibnamefont  [1]{#1}%
\providecommand \bibfnamefont [1]{#1}%
\providecommand \citenamefont [1]{#1}%
\providecommand \href@noop [0]{\@secondoftwo}%
\providecommand \href [0]{\begingroup \@sanitize@url \@href}%
\providecommand \@href[1]{\@@startlink{#1}\@@href}%
\providecommand \@@href[1]{\endgroup#1\@@endlink}%
\providecommand \@sanitize@url [0]{\catcode `\\12\catcode `\$12\catcode `\&12\catcode `\#12\catcode `\^12\catcode `\_12\catcode `\%12\relax}%
\providecommand \@@startlink[1]{}%
\providecommand \@@endlink[0]{}%
\providecommand \url  [0]{\begingroup\@sanitize@url \@url }%
\providecommand \@url [1]{\endgroup\@href {#1}{\urlprefix }}%
\providecommand \urlprefix  [0]{URL }%
\providecommand \Eprint [0]{\href }%
\providecommand \doibase [0]{https://doi.org/}%
\providecommand \selectlanguage [0]{\@gobble}%
\providecommand \bibinfo  [0]{\@secondoftwo}%
\providecommand \bibfield  [0]{\@secondoftwo}%
\providecommand \translation [1]{[#1]}%
\providecommand \BibitemOpen [0]{}%
\providecommand \bibitemStop [0]{}%
\providecommand \bibitemNoStop [0]{.\EOS\space}%
\providecommand \EOS [0]{\spacefactor3000\relax}%
\providecommand \BibitemShut  [1]{\csname bibitem#1\endcsname}%
\let\auto@bib@innerbib\@empty
\bibitem [{\citenamefont {Bistritzer}\ and\ \citenamefont {MacDonald}(2011)}]{macdonaldBistritzerMoireBandsTwisted2011}%
  \BibitemOpen
  \bibfield  {author} {\bibinfo {author} {\bibfnamefont {R.}~\bibnamefont {Bistritzer}}\ and\ \bibinfo {author} {\bibfnamefont {A.~H.}\ \bibnamefont {MacDonald}},\ }\bibfield  {title} {\bibinfo {title} {Moire bands in twisted double-layer graphene},\ }\href {https://doi.org/10.1073/pnas.1108174108} {\bibfield  {journal} {\bibinfo  {journal} {Proceedings of the National Academy of Sciences}\ }\textbf {\bibinfo {volume} {108}},\ \bibinfo {pages} {12233} (\bibinfo {year} {2011})}\BibitemShut {NoStop}%
\bibitem [{\citenamefont {Wu}\ \emph {et~al.}(2019)\citenamefont {Wu}, \citenamefont {Lovorn}, \citenamefont {Tutuc}, \citenamefont {Martin},\ and\ \citenamefont {MacDonald}}]{macdonaldWuTopologicalInsulatorsTwisted2019}%
  \BibitemOpen
  \bibfield  {author} {\bibinfo {author} {\bibfnamefont {F.}~\bibnamefont {Wu}}, \bibinfo {author} {\bibfnamefont {T.}~\bibnamefont {Lovorn}}, \bibinfo {author} {\bibfnamefont {E.}~\bibnamefont {Tutuc}}, \bibinfo {author} {\bibfnamefont {I.}~\bibnamefont {Martin}},\ and\ \bibinfo {author} {\bibfnamefont {A.}~\bibnamefont {MacDonald}},\ }\bibfield  {title} {\bibinfo {title} {Topological insulators in twisted transition metal dichalcogenide homobilayers},\ }\href {https://doi.org/10.1103/PhysRevLett.122.086402} {\bibfield  {journal} {\bibinfo  {journal} {Physical Review Letters}\ }\textbf {\bibinfo {volume} {122}},\ \bibinfo {pages} {086402} (\bibinfo {year} {2019})}\BibitemShut {NoStop}%
\bibitem [{\citenamefont {Cao}\ \emph {et~al.}(2018)\citenamefont {Cao}, \citenamefont {Fatemi}, \citenamefont {Demir}, \citenamefont {Fang}, \citenamefont {Tomarken}, \citenamefont {Luo}, \citenamefont {Sanchez-Yamagishi}, \citenamefont {Watanabe}, \citenamefont {Taniguchi}, \citenamefont {Kaxiras}, \citenamefont {Ashoori},\ and\ \citenamefont {Jarillo-Herrero}}]{jarillo-herreroCaoCorrelatedInsulatorBehaviour2018}%
  \BibitemOpen
  \bibfield  {author} {\bibinfo {author} {\bibfnamefont {Y.}~\bibnamefont {Cao}}, \bibinfo {author} {\bibfnamefont {V.}~\bibnamefont {Fatemi}}, \bibinfo {author} {\bibfnamefont {A.}~\bibnamefont {Demir}}, \bibinfo {author} {\bibfnamefont {S.}~\bibnamefont {Fang}}, \bibinfo {author} {\bibfnamefont {S.~L.}\ \bibnamefont {Tomarken}}, \bibinfo {author} {\bibfnamefont {J.~Y.}\ \bibnamefont {Luo}}, \bibinfo {author} {\bibfnamefont {J.~D.}\ \bibnamefont {Sanchez-Yamagishi}}, \bibinfo {author} {\bibfnamefont {K.}~\bibnamefont {Watanabe}}, \bibinfo {author} {\bibfnamefont {T.}~\bibnamefont {Taniguchi}}, \bibinfo {author} {\bibfnamefont {E.}~\bibnamefont {Kaxiras}}, \bibinfo {author} {\bibfnamefont {R.~C.}\ \bibnamefont {Ashoori}},\ and\ \bibinfo {author} {\bibfnamefont {P.}~\bibnamefont {Jarillo-Herrero}},\ }\bibfield  {title} {\bibinfo {title} {Correlated insulator behaviour at half-filling in magic-angle graphene superlattices},\ }\href {https://doi.org/10.1038/nature26154} {\bibfield  {journal} {\bibinfo  {journal} {Nature}\ }\textbf {\bibinfo {volume} {556}},\ \bibinfo {pages} {80} (\bibinfo {year} {2018})}\BibitemShut {NoStop}%
\bibitem [{\citenamefont {Yankowitz}\ \emph {et~al.}(2019)\citenamefont {Yankowitz}, \citenamefont {Chen}, \citenamefont {Polshyn}, \citenamefont {Zhang}, \citenamefont {Watanabe}, \citenamefont {Taniguchi}, \citenamefont {Graf}, \citenamefont {Young},\ and\ \citenamefont {Dean}}]{deanYankowitzTuningSuperconductivityTwisted2019}%
  \BibitemOpen
  \bibfield  {author} {\bibinfo {author} {\bibfnamefont {M.}~\bibnamefont {Yankowitz}}, \bibinfo {author} {\bibfnamefont {S.}~\bibnamefont {Chen}}, \bibinfo {author} {\bibfnamefont {H.}~\bibnamefont {Polshyn}}, \bibinfo {author} {\bibfnamefont {Y.}~\bibnamefont {Zhang}}, \bibinfo {author} {\bibfnamefont {K.}~\bibnamefont {Watanabe}}, \bibinfo {author} {\bibfnamefont {T.}~\bibnamefont {Taniguchi}}, \bibinfo {author} {\bibfnamefont {D.}~\bibnamefont {Graf}}, \bibinfo {author} {\bibfnamefont {A.~F.}\ \bibnamefont {Young}},\ and\ \bibinfo {author} {\bibfnamefont {C.~R.}\ \bibnamefont {Dean}},\ }\bibfield  {title} {\bibinfo {title} {Tuning superconductivity in twisted bilayer graphene},\ }\href {https://doi.org/10.1126/science.aav1910} {\bibfield  {journal} {\bibinfo  {journal} {Science}\ }\textbf {\bibinfo {volume} {363}},\ \bibinfo {pages} {1059} (\bibinfo {year} {2019})}\BibitemShut {NoStop}%
\bibitem [{\citenamefont {Hao}\ \emph {et~al.}(2021)\citenamefont {Hao}, \citenamefont {Zimmerman}, \citenamefont {Ledwith}, \citenamefont {Khalaf}, \citenamefont {Najafabadi}, \citenamefont {Watanabe}, \citenamefont {Taniguchi}, \citenamefont {Vishwanath},\ and\ \citenamefont {Kim}}]{kimHaoElectricFieldTunable2021}%
  \BibitemOpen
  \bibfield  {author} {\bibinfo {author} {\bibfnamefont {Z.}~\bibnamefont {Hao}}, \bibinfo {author} {\bibfnamefont {A.~M.}\ \bibnamefont {Zimmerman}}, \bibinfo {author} {\bibfnamefont {P.}~\bibnamefont {Ledwith}}, \bibinfo {author} {\bibfnamefont {E.}~\bibnamefont {Khalaf}}, \bibinfo {author} {\bibfnamefont {D.~H.}\ \bibnamefont {Najafabadi}}, \bibinfo {author} {\bibfnamefont {K.}~\bibnamefont {Watanabe}}, \bibinfo {author} {\bibfnamefont {T.}~\bibnamefont {Taniguchi}}, \bibinfo {author} {\bibfnamefont {A.}~\bibnamefont {Vishwanath}},\ and\ \bibinfo {author} {\bibfnamefont {P.}~\bibnamefont {Kim}},\ }\bibfield  {title} {\bibinfo {title} {Electric field–tunable superconductivity in alternating-twist magic-angle trilayer graphene},\ }\href {https://doi.org/10.1126/science.abg0399} {\bibfield  {journal} {\bibinfo  {journal} {Science}\ }\textbf {\bibinfo {volume} {371}},\ \bibinfo {pages} {1133} (\bibinfo {year} {2021})}\BibitemShut {NoStop}%
\bibitem [{\citenamefont {Oh}\ \emph {et~al.}(2021)\citenamefont {Oh}, \citenamefont {Nuckolls}, \citenamefont {Wong}, \citenamefont {Lee}, \citenamefont {Liu}, \citenamefont {Watanabe}, \citenamefont {Taniguchi},\ and\ \citenamefont {Yazdani}}]{yazdaniOhEvidenceUnconventionalSuperconductivity2021}%
  \BibitemOpen
  \bibfield  {author} {\bibinfo {author} {\bibfnamefont {M.}~\bibnamefont {Oh}}, \bibinfo {author} {\bibfnamefont {K.~P.}\ \bibnamefont {Nuckolls}}, \bibinfo {author} {\bibfnamefont {D.}~\bibnamefont {Wong}}, \bibinfo {author} {\bibfnamefont {R.~L.}\ \bibnamefont {Lee}}, \bibinfo {author} {\bibfnamefont {X.}~\bibnamefont {Liu}}, \bibinfo {author} {\bibfnamefont {K.}~\bibnamefont {Watanabe}}, \bibinfo {author} {\bibfnamefont {T.}~\bibnamefont {Taniguchi}},\ and\ \bibinfo {author} {\bibfnamefont {A.}~\bibnamefont {Yazdani}},\ }\bibfield  {title} {\bibinfo {title} {Evidence for unconventional superconductivity in twisted bilayer graphene},\ }\href {https://doi.org/10.1038/s41586-021-04121-x} {\bibfield  {journal} {\bibinfo  {journal} {Nature}\ }\textbf {\bibinfo {volume} {600}},\ \bibinfo {pages} {240} (\bibinfo {year} {2021})}\BibitemShut {NoStop}%
\bibitem [{\citenamefont {Lu}\ \emph {et~al.}(2019)\citenamefont {Lu}, \citenamefont {Stepanov}, \citenamefont {Yang}, \citenamefont {Xie}, \citenamefont {Aamir}, \citenamefont {Das}, \citenamefont {Urgell}, \citenamefont {Watanabe}, \citenamefont {Taniguchi}, \citenamefont {Zhang}, \citenamefont {Bachtold}, \citenamefont {MacDonald},\ and\ \citenamefont {Efetov}}]{efetovLuSuperconductorsOrbitalMagnets2019}%
  \BibitemOpen
  \bibfield  {author} {\bibinfo {author} {\bibfnamefont {X.}~\bibnamefont {Lu}}, \bibinfo {author} {\bibfnamefont {P.}~\bibnamefont {Stepanov}}, \bibinfo {author} {\bibfnamefont {W.}~\bibnamefont {Yang}}, \bibinfo {author} {\bibfnamefont {M.}~\bibnamefont {Xie}}, \bibinfo {author} {\bibfnamefont {M.~A.}\ \bibnamefont {Aamir}}, \bibinfo {author} {\bibfnamefont {I.}~\bibnamefont {Das}}, \bibinfo {author} {\bibfnamefont {C.}~\bibnamefont {Urgell}}, \bibinfo {author} {\bibfnamefont {K.}~\bibnamefont {Watanabe}}, \bibinfo {author} {\bibfnamefont {T.}~\bibnamefont {Taniguchi}}, \bibinfo {author} {\bibfnamefont {G.}~\bibnamefont {Zhang}}, \bibinfo {author} {\bibfnamefont {A.}~\bibnamefont {Bachtold}}, \bibinfo {author} {\bibfnamefont {A.~H.}\ \bibnamefont {MacDonald}},\ and\ \bibinfo {author} {\bibfnamefont {D.~K.}\ \bibnamefont {Efetov}},\ }\bibfield  {title} {\bibinfo {title} {Superconductors, orbital magnets and correlated states in magic-angle bilayer graphene},\ }\href {https://doi.org/10.1038/s41586-019-1695-0} {\bibfield  {journal} {\bibinfo  {journal} {Nature}\ }\textbf {\bibinfo {volume} {574}},\ \bibinfo {pages} {653} (\bibinfo {year} {2019})}\BibitemShut {NoStop}%
\bibitem [{\citenamefont {Cao}\ \emph {et~al.}(2021)\citenamefont {Cao}, \citenamefont {Rodan-Legrain}, \citenamefont {Park}, \citenamefont {Yuan}, \citenamefont {Watanabe}, \citenamefont {Taniguchi}, \citenamefont {Fernandes}, \citenamefont {Fu},\ and\ \citenamefont {Jarillo-Herrero}}]{jarillo-herreroCaoNematicityCompetingOrders2021}%
  \BibitemOpen
  \bibfield  {author} {\bibinfo {author} {\bibfnamefont {Y.}~\bibnamefont {Cao}}, \bibinfo {author} {\bibfnamefont {D.}~\bibnamefont {Rodan-Legrain}}, \bibinfo {author} {\bibfnamefont {J.~M.}\ \bibnamefont {Park}}, \bibinfo {author} {\bibfnamefont {N.~F.~Q.}\ \bibnamefont {Yuan}}, \bibinfo {author} {\bibfnamefont {K.}~\bibnamefont {Watanabe}}, \bibinfo {author} {\bibfnamefont {T.}~\bibnamefont {Taniguchi}}, \bibinfo {author} {\bibfnamefont {R.~M.}\ \bibnamefont {Fernandes}}, \bibinfo {author} {\bibfnamefont {L.}~\bibnamefont {Fu}},\ and\ \bibinfo {author} {\bibfnamefont {P.}~\bibnamefont {Jarillo-Herrero}},\ }\bibfield  {title} {\bibinfo {title} {Nematicity and competing orders in superconducting magic-angle graphene},\ }\href {https://doi.org/10.1126/science.abc2836} {\bibfield  {journal} {\bibinfo  {journal} {Science}\ }\textbf {\bibinfo {volume} {372}},\ \bibinfo {pages} {264} (\bibinfo {year} {2021})}\BibitemShut {NoStop}%
\bibitem [{\citenamefont {Liu}\ \emph {et~al.}(2021{\natexlab{a}})\citenamefont {Liu}, \citenamefont {Wang}, \citenamefont {Watanabe}, \citenamefont {Taniguchi}, \citenamefont {Vafek},\ and\ \citenamefont {Li}}]{liLiuTuningElectronCorrelation2021}%
  \BibitemOpen
  \bibfield  {author} {\bibinfo {author} {\bibfnamefont {X.}~\bibnamefont {Liu}}, \bibinfo {author} {\bibfnamefont {Z.}~\bibnamefont {Wang}}, \bibinfo {author} {\bibfnamefont {K.}~\bibnamefont {Watanabe}}, \bibinfo {author} {\bibfnamefont {T.}~\bibnamefont {Taniguchi}}, \bibinfo {author} {\bibfnamefont {O.}~\bibnamefont {Vafek}},\ and\ \bibinfo {author} {\bibfnamefont {J.~I.~A.}\ \bibnamefont {Li}},\ }\bibfield  {title} {\bibinfo {title} {Tuning electron correlation in magic-angle twisted bilayer graphene using coulomb screening},\ }\href {https://doi.org/10.1126/science.abb8754} {\bibfield  {journal} {\bibinfo  {journal} {Science}\ }\textbf {\bibinfo {volume} {371}},\ \bibinfo {pages} {1261} (\bibinfo {year} {2021}{\natexlab{a}})}\BibitemShut {NoStop}%
\bibitem [{\citenamefont {Arora}\ \emph {et~al.}(2020)\citenamefont {Arora}, \citenamefont {Polski}, \citenamefont {Zhang}, \citenamefont {Thomson}, \citenamefont {Choi}, \citenamefont {Kim}, \citenamefont {Lin}, \citenamefont {Wilson}, \citenamefont {Xu}, \citenamefont {Chu}, \citenamefont {Watanabe}, \citenamefont {Taniguchi}, \citenamefont {Alicea},\ and\ \citenamefont {Nadj-Perge}}]{nadj-pergeAroraSuperconductivityMetallicTwisted2020}%
  \BibitemOpen
  \bibfield  {author} {\bibinfo {author} {\bibfnamefont {H.~S.}\ \bibnamefont {Arora}}, \bibinfo {author} {\bibfnamefont {R.}~\bibnamefont {Polski}}, \bibinfo {author} {\bibfnamefont {Y.}~\bibnamefont {Zhang}}, \bibinfo {author} {\bibfnamefont {A.}~\bibnamefont {Thomson}}, \bibinfo {author} {\bibfnamefont {Y.}~\bibnamefont {Choi}}, \bibinfo {author} {\bibfnamefont {H.}~\bibnamefont {Kim}}, \bibinfo {author} {\bibfnamefont {Z.}~\bibnamefont {Lin}}, \bibinfo {author} {\bibfnamefont {I.~Z.}\ \bibnamefont {Wilson}}, \bibinfo {author} {\bibfnamefont {X.}~\bibnamefont {Xu}}, \bibinfo {author} {\bibfnamefont {J.-H.}\ \bibnamefont {Chu}}, \bibinfo {author} {\bibfnamefont {K.}~\bibnamefont {Watanabe}}, \bibinfo {author} {\bibfnamefont {T.}~\bibnamefont {Taniguchi}}, \bibinfo {author} {\bibfnamefont {J.}~\bibnamefont {Alicea}},\ and\ \bibinfo {author} {\bibfnamefont {S.}~\bibnamefont {Nadj-Perge}},\ }\bibfield  {title} {\bibinfo {title} {Superconductivity in metallic twisted bilayer graphene stabilized by wse2},\ }\href {https://doi.org/10.1038/s41586-020-2473-8} {\bibfield  {journal} {\bibinfo  {journal} {Nature}\ }\textbf {\bibinfo {volume} {583}},\ \bibinfo {pages} {379} (\bibinfo {year} {2020})}\BibitemShut {NoStop}%
\bibitem [{\citenamefont {Stepanov}\ \emph {et~al.}(2020)\citenamefont {Stepanov}, \citenamefont {Das}, \citenamefont {Lu}, \citenamefont {Fahimniya}, \citenamefont {Watanabe}, \citenamefont {Taniguchi}, \citenamefont {Koppens}, \citenamefont {Lischner}, \citenamefont {Levitov},\ and\ \citenamefont {Efetov}}]{efetovStepanovUntyingInsulatingSuperconducting2020}%
  \BibitemOpen
  \bibfield  {author} {\bibinfo {author} {\bibfnamefont {P.}~\bibnamefont {Stepanov}}, \bibinfo {author} {\bibfnamefont {I.}~\bibnamefont {Das}}, \bibinfo {author} {\bibfnamefont {X.}~\bibnamefont {Lu}}, \bibinfo {author} {\bibfnamefont {A.}~\bibnamefont {Fahimniya}}, \bibinfo {author} {\bibfnamefont {K.}~\bibnamefont {Watanabe}}, \bibinfo {author} {\bibfnamefont {T.}~\bibnamefont {Taniguchi}}, \bibinfo {author} {\bibfnamefont {F.~H.~L.}\ \bibnamefont {Koppens}}, \bibinfo {author} {\bibfnamefont {J.}~\bibnamefont {Lischner}}, \bibinfo {author} {\bibfnamefont {L.}~\bibnamefont {Levitov}},\ and\ \bibinfo {author} {\bibfnamefont {D.~K.}\ \bibnamefont {Efetov}},\ }\bibfield  {title} {\bibinfo {title} {Untying the insulating and superconducting orders in magic-angle graphene},\ }\href {https://doi.org/10.1038/s41586-020-2459-6} {\bibfield  {journal} {\bibinfo  {journal} {Nature}\ }\textbf {\bibinfo {volume} {583}},\ \bibinfo {pages} {375} (\bibinfo {year} {2020})}\BibitemShut {NoStop}%
\bibitem [{\citenamefont {Saito}\ \emph {et~al.}(2020)\citenamefont {Saito}, \citenamefont {Ge}, \citenamefont {Watanabe}, \citenamefont {Taniguchi},\ and\ \citenamefont {Young}}]{youngSaitoIndependentSuperconductorsCorrelated2020}%
  \BibitemOpen
  \bibfield  {author} {\bibinfo {author} {\bibfnamefont {Y.}~\bibnamefont {Saito}}, \bibinfo {author} {\bibfnamefont {J.}~\bibnamefont {Ge}}, \bibinfo {author} {\bibfnamefont {K.}~\bibnamefont {Watanabe}}, \bibinfo {author} {\bibfnamefont {T.}~\bibnamefont {Taniguchi}},\ and\ \bibinfo {author} {\bibfnamefont {A.~F.}\ \bibnamefont {Young}},\ }\bibfield  {title} {\bibinfo {title} {Independent superconductors and correlated insulators in twisted bilayer graphene},\ }\href {https://doi.org/10.1038/s41567-020-0928-3} {\bibfield  {journal} {\bibinfo  {journal} {Nature Physics}\ }\textbf {\bibinfo {volume} {16}},\ \bibinfo {pages} {926} (\bibinfo {year} {2020})}\BibitemShut {NoStop}%
\bibitem [{\citenamefont {Zondiner}\ \emph {et~al.}(2020)\citenamefont {Zondiner}, \citenamefont {Rozen}, \citenamefont {Rodan-Legrain}, \citenamefont {Cao}, \citenamefont {Queiroz}, \citenamefont {Taniguchi}, \citenamefont {Watanabe}, \citenamefont {Oreg}, \citenamefont {von Oppen}, \citenamefont {Stern}, \citenamefont {Berg}, \citenamefont {Jarillo-Herrero},\ and\ \citenamefont {Ilani}}]{ilaniZondinerCascadePhaseTransitions2020a}%
  \BibitemOpen
  \bibfield  {author} {\bibinfo {author} {\bibfnamefont {U.}~\bibnamefont {Zondiner}}, \bibinfo {author} {\bibfnamefont {A.}~\bibnamefont {Rozen}}, \bibinfo {author} {\bibfnamefont {D.}~\bibnamefont {Rodan-Legrain}}, \bibinfo {author} {\bibfnamefont {Y.}~\bibnamefont {Cao}}, \bibinfo {author} {\bibfnamefont {R.}~\bibnamefont {Queiroz}}, \bibinfo {author} {\bibfnamefont {T.}~\bibnamefont {Taniguchi}}, \bibinfo {author} {\bibfnamefont {K.}~\bibnamefont {Watanabe}}, \bibinfo {author} {\bibfnamefont {Y.}~\bibnamefont {Oreg}}, \bibinfo {author} {\bibfnamefont {F.}~\bibnamefont {von Oppen}}, \bibinfo {author} {\bibfnamefont {A.}~\bibnamefont {Stern}}, \bibinfo {author} {\bibfnamefont {E.}~\bibnamefont {Berg}}, \bibinfo {author} {\bibfnamefont {P.}~\bibnamefont {Jarillo-Herrero}},\ and\ \bibinfo {author} {\bibfnamefont {S.}~\bibnamefont {Ilani}},\ }\bibfield  {title} {\bibinfo {title} {Cascade of phase transitions and dirac revivals in magic-angle graphene},\ }\href {https://doi.org/10.1038/s41586-020-2373-y} {\bibfield  {journal} {\bibinfo  {journal} {Nature}\ }\textbf {\bibinfo {volume} {582}},\ \bibinfo {pages} {203} (\bibinfo {year} {2020})}\BibitemShut {NoStop}%
\bibitem [{\citenamefont {Wong}\ \emph {et~al.}(2020)\citenamefont {Wong}, \citenamefont {Nuckolls}, \citenamefont {Oh}, \citenamefont {Lian}, \citenamefont {Xie}, \citenamefont {Jeon}, \citenamefont {Watanabe}, \citenamefont {Taniguchi}, \citenamefont {Bernevig},\ and\ \citenamefont {Yazdani}}]{yazdaniWongCascadeElectronicTransitions2020}%
  \BibitemOpen
  \bibfield  {author} {\bibinfo {author} {\bibfnamefont {D.}~\bibnamefont {Wong}}, \bibinfo {author} {\bibfnamefont {K.~P.}\ \bibnamefont {Nuckolls}}, \bibinfo {author} {\bibfnamefont {M.}~\bibnamefont {Oh}}, \bibinfo {author} {\bibfnamefont {B.}~\bibnamefont {Lian}}, \bibinfo {author} {\bibfnamefont {Y.}~\bibnamefont {Xie}}, \bibinfo {author} {\bibfnamefont {S.}~\bibnamefont {Jeon}}, \bibinfo {author} {\bibfnamefont {K.}~\bibnamefont {Watanabe}}, \bibinfo {author} {\bibfnamefont {T.}~\bibnamefont {Taniguchi}}, \bibinfo {author} {\bibfnamefont {B.~A.}\ \bibnamefont {Bernevig}},\ and\ \bibinfo {author} {\bibfnamefont {A.}~\bibnamefont {Yazdani}},\ }\bibfield  {title} {\bibinfo {title} {Cascade of electronic transitions in magic-angle twisted bilayer graphene},\ }\href {https://doi.org/10.1038/s41586-020-2339-0} {\bibfield  {journal} {\bibinfo  {journal} {Nature}\ }\textbf {\bibinfo {volume} {582}},\ \bibinfo {pages} {198} (\bibinfo {year} {2020})}\BibitemShut {NoStop}%
\bibitem [{\citenamefont {Wang}\ \emph {et~al.}(2020)\citenamefont {Wang}, \citenamefont {Shih}, \citenamefont {Ghiotto}, \citenamefont {Xian}, \citenamefont {Rhodes}, \citenamefont {Tan}, \citenamefont {Claassen}, \citenamefont {Kennes}, \citenamefont {Bai}, \citenamefont {Kim}, \citenamefont {Watanabe}, \citenamefont {Taniguchi}, \citenamefont {Zhu}, \citenamefont {Hone}, \citenamefont {Rubio}, \citenamefont {Pasupathy},\ and\ \citenamefont {Dean}}]{deanWangCorrelatedElectronicPhases2020}%
  \BibitemOpen
  \bibfield  {author} {\bibinfo {author} {\bibfnamefont {L.}~\bibnamefont {Wang}}, \bibinfo {author} {\bibfnamefont {E.-M.}\ \bibnamefont {Shih}}, \bibinfo {author} {\bibfnamefont {A.}~\bibnamefont {Ghiotto}}, \bibinfo {author} {\bibfnamefont {L.}~\bibnamefont {Xian}}, \bibinfo {author} {\bibfnamefont {D.~A.}\ \bibnamefont {Rhodes}}, \bibinfo {author} {\bibfnamefont {C.}~\bibnamefont {Tan}}, \bibinfo {author} {\bibfnamefont {M.}~\bibnamefont {Claassen}}, \bibinfo {author} {\bibfnamefont {D.~M.}\ \bibnamefont {Kennes}}, \bibinfo {author} {\bibfnamefont {Y.}~\bibnamefont {Bai}}, \bibinfo {author} {\bibfnamefont {B.}~\bibnamefont {Kim}}, \bibinfo {author} {\bibfnamefont {K.}~\bibnamefont {Watanabe}}, \bibinfo {author} {\bibfnamefont {T.}~\bibnamefont {Taniguchi}}, \bibinfo {author} {\bibfnamefont {X.}~\bibnamefont {Zhu}}, \bibinfo {author} {\bibfnamefont {J.}~\bibnamefont {Hone}}, \bibinfo {author} {\bibfnamefont {A.}~\bibnamefont {Rubio}}, \bibinfo {author} {\bibfnamefont {A.~N.}\ \bibnamefont {Pasupathy}},\ and\ \bibinfo {author} {\bibfnamefont {C.~R.}\ \bibnamefont {Dean}},\ }\bibfield  {title} {\bibinfo {title} {Correlated electronic phases in twisted bilayer transition metal dichalcogenides},\ }\href {https://doi.org/10.1038/s41563-020-0708-6} {\bibfield  {journal} {\bibinfo  {journal} {Nature Materials}\ }\textbf {\bibinfo {volume} {19}},\ \bibinfo {pages} {861} (\bibinfo {year} {2020})}\BibitemShut {NoStop}%
\bibitem [{\citenamefont {Xia}\ \emph {et~al.}(2024)\citenamefont {Xia}, \citenamefont {Han}, \citenamefont {Watanabe}, \citenamefont {Taniguchi}, \citenamefont {Shan},\ and\ \citenamefont {Mak}}]{makXiaUnconventionalSuperconductivityTwisted2024}%
  \BibitemOpen
  \bibfield  {author} {\bibinfo {author} {\bibfnamefont {Y.}~\bibnamefont {Xia}}, \bibinfo {author} {\bibfnamefont {Z.}~\bibnamefont {Han}}, \bibinfo {author} {\bibfnamefont {K.}~\bibnamefont {Watanabe}}, \bibinfo {author} {\bibfnamefont {T.}~\bibnamefont {Taniguchi}}, \bibinfo {author} {\bibfnamefont {J.}~\bibnamefont {Shan}},\ and\ \bibinfo {author} {\bibfnamefont {K.~F.}\ \bibnamefont {Mak}},\ }\href@noop {} {\bibinfo {title} {Unconventional superconductivity in twisted bilayer wse2}} (\bibinfo {year} {2024}),\ \Eprint {https://arxiv.org/abs/2405.14784} {arXiv:2405.14784 [cond-mat]} \BibitemShut {NoStop}%
\bibitem [{\citenamefont {Guo}\ \emph {et~al.}(2024)\citenamefont {Guo}, \citenamefont {Pack}, \citenamefont {Swann}, \citenamefont {Holtzman}, \citenamefont {Cothrine}, \citenamefont {Watanabe}, \citenamefont {Taniguchi}, \citenamefont {Mandrus}, \citenamefont {Barmak}, \citenamefont {Hone}, \citenamefont {Millis}, \citenamefont {Pasupathy},\ and\ \citenamefont {Dean}}]{deanGuoSuperconductivityTwistedBilayer2024}%
  \BibitemOpen
  \bibfield  {author} {\bibinfo {author} {\bibfnamefont {Y.}~\bibnamefont {Guo}}, \bibinfo {author} {\bibfnamefont {J.}~\bibnamefont {Pack}}, \bibinfo {author} {\bibfnamefont {J.}~\bibnamefont {Swann}}, \bibinfo {author} {\bibfnamefont {L.}~\bibnamefont {Holtzman}}, \bibinfo {author} {\bibfnamefont {M.}~\bibnamefont {Cothrine}}, \bibinfo {author} {\bibfnamefont {K.}~\bibnamefont {Watanabe}}, \bibinfo {author} {\bibfnamefont {T.}~\bibnamefont {Taniguchi}}, \bibinfo {author} {\bibfnamefont {D.}~\bibnamefont {Mandrus}}, \bibinfo {author} {\bibfnamefont {K.}~\bibnamefont {Barmak}}, \bibinfo {author} {\bibfnamefont {J.}~\bibnamefont {Hone}}, \bibinfo {author} {\bibfnamefont {A.~J.}\ \bibnamefont {Millis}}, \bibinfo {author} {\bibfnamefont {A.~N.}\ \bibnamefont {Pasupathy}},\ and\ \bibinfo {author} {\bibfnamefont {C.~R.}\ \bibnamefont {Dean}},\ }\href@noop {} {\bibinfo {title} {Superconductivity in twisted bilayer wse$_2$}} (\bibinfo {year} {2024}),\ \Eprint {https://arxiv.org/abs/2406.03418} {arXiv:2406.03418 [cond-mat]} \BibitemShut {NoStop}%
\bibitem [{\citenamefont {Ghiotto}\ \emph {et~al.}(2021)\citenamefont {Ghiotto}, \citenamefont {Shih}, \citenamefont {Pereira}, \citenamefont {Rhodes}, \citenamefont {Kim}, \citenamefont {Zang}, \citenamefont {Millis}, \citenamefont {Watanabe}, \citenamefont {Taniguchi}, \citenamefont {Hone}, \citenamefont {Wang}, \citenamefont {Dean},\ and\ \citenamefont {Pasupathy}}]{pasupathyGhiottoQuantumCriticalityTwisted2021}%
  \BibitemOpen
  \bibfield  {author} {\bibinfo {author} {\bibfnamefont {A.}~\bibnamefont {Ghiotto}}, \bibinfo {author} {\bibfnamefont {E.-M.}\ \bibnamefont {Shih}}, \bibinfo {author} {\bibfnamefont {G.~S. S.~G.}\ \bibnamefont {Pereira}}, \bibinfo {author} {\bibfnamefont {D.~A.}\ \bibnamefont {Rhodes}}, \bibinfo {author} {\bibfnamefont {B.}~\bibnamefont {Kim}}, \bibinfo {author} {\bibfnamefont {J.}~\bibnamefont {Zang}}, \bibinfo {author} {\bibfnamefont {A.~J.}\ \bibnamefont {Millis}}, \bibinfo {author} {\bibfnamefont {K.}~\bibnamefont {Watanabe}}, \bibinfo {author} {\bibfnamefont {T.}~\bibnamefont {Taniguchi}}, \bibinfo {author} {\bibfnamefont {J.~C.}\ \bibnamefont {Hone}}, \bibinfo {author} {\bibfnamefont {L.}~\bibnamefont {Wang}}, \bibinfo {author} {\bibfnamefont {C.~R.}\ \bibnamefont {Dean}},\ and\ \bibinfo {author} {\bibfnamefont {A.~N.}\ \bibnamefont {Pasupathy}},\ }\bibfield  {title} {\bibinfo {title} {Quantum criticality in twisted transition metal dichalcogenides},\ }\href {https://doi.org/10.1038/s41586-021-03815-6} {\bibfield  {journal} {\bibinfo  {journal} {Nature}\ }\textbf {\bibinfo {volume} {597}},\ \bibinfo {pages} {345} (\bibinfo {year} {2021})}\BibitemShut {NoStop}%
\bibitem [{\citenamefont {Lu}\ \emph {et~al.}(2024)\citenamefont {Lu}, \citenamefont {Han}, \citenamefont {Yao}, \citenamefont {Reddy}, \citenamefont {Yang}, \citenamefont {Seo}, \citenamefont {Watanabe}, \citenamefont {Taniguchi}, \citenamefont {Fu},\ and\ \citenamefont {Ju}}]{juLuFractionalQuantumAnomalous2024}%
  \BibitemOpen
  \bibfield  {author} {\bibinfo {author} {\bibfnamefont {Z.}~\bibnamefont {Lu}}, \bibinfo {author} {\bibfnamefont {T.}~\bibnamefont {Han}}, \bibinfo {author} {\bibfnamefont {Y.}~\bibnamefont {Yao}}, \bibinfo {author} {\bibfnamefont {A.~P.}\ \bibnamefont {Reddy}}, \bibinfo {author} {\bibfnamefont {J.}~\bibnamefont {Yang}}, \bibinfo {author} {\bibfnamefont {J.}~\bibnamefont {Seo}}, \bibinfo {author} {\bibfnamefont {K.}~\bibnamefont {Watanabe}}, \bibinfo {author} {\bibfnamefont {T.}~\bibnamefont {Taniguchi}}, \bibinfo {author} {\bibfnamefont {L.}~\bibnamefont {Fu}},\ and\ \bibinfo {author} {\bibfnamefont {L.}~\bibnamefont {Ju}},\ }\bibfield  {title} {\bibinfo {title} {Fractional quantum anomalous hall effect in multilayer graphene},\ }\href {https://doi.org/10.1038/s41586-023-07010-7} {\bibfield  {journal} {\bibinfo  {journal} {Nature}\ }\textbf {\bibinfo {volume} {626}},\ \bibinfo {pages} {759} (\bibinfo {year} {2024})}\BibitemShut {NoStop}%
\bibitem [{\citenamefont {Lu}\ \emph {et~al.}(2025)\citenamefont {Lu}, \citenamefont {Han}, \citenamefont {Yao}, \citenamefont {Hadjri}, \citenamefont {Yang}, \citenamefont {Seo}, \citenamefont {Shi}, \citenamefont {Ye}, \citenamefont {Watanabe}, \citenamefont {Taniguchi},\ and\ \citenamefont {Ju}}]{Zhengguang2025}%
  \BibitemOpen
  \bibfield  {author} {\bibinfo {author} {\bibfnamefont {Z.}~\bibnamefont {Lu}}, \bibinfo {author} {\bibfnamefont {T.}~\bibnamefont {Han}}, \bibinfo {author} {\bibfnamefont {Y.}~\bibnamefont {Yao}}, \bibinfo {author} {\bibfnamefont {Z.}~\bibnamefont {Hadjri}}, \bibinfo {author} {\bibfnamefont {J.}~\bibnamefont {Yang}}, \bibinfo {author} {\bibfnamefont {J.}~\bibnamefont {Seo}}, \bibinfo {author} {\bibfnamefont {L.}~\bibnamefont {Shi}}, \bibinfo {author} {\bibfnamefont {S.}~\bibnamefont {Ye}}, \bibinfo {author} {\bibfnamefont {K.}~\bibnamefont {Watanabe}}, \bibinfo {author} {\bibfnamefont {T.}~\bibnamefont {Taniguchi}},\ and\ \bibinfo {author} {\bibfnamefont {L.}~\bibnamefont {Ju}},\ }\bibfield  {title} {\bibinfo {title} {Extended quantum anomalous hall states in graphene/hbn moir{\'e}superlattices},\ }\href@noop {} {\bibfield  {journal} {\bibinfo  {journal} {Nature}\ }\textbf {\bibinfo {volume} {637}},\ \bibinfo {pages} {1090} (\bibinfo {year} {2025})}\BibitemShut {NoStop}%
\bibitem [{\citenamefont {Xie}\ \emph {et~al.}(2024)\citenamefont {Xie}, \citenamefont {Huo}, \citenamefont {Lu}, \citenamefont {Feng}, \citenamefont {Zhang}, \citenamefont {Wang}, \citenamefont {Yang}, \citenamefont {Watanabe}, \citenamefont {Taniguchi}, \citenamefont {Liu} \emph {et~al.}}]{xie2024fqah}%
  \BibitemOpen
  \bibfield  {author} {\bibinfo {author} {\bibfnamefont {J.}~\bibnamefont {Xie}}, \bibinfo {author} {\bibfnamefont {Z.}~\bibnamefont {Huo}}, \bibinfo {author} {\bibfnamefont {X.}~\bibnamefont {Lu}}, \bibinfo {author} {\bibfnamefont {Z.}~\bibnamefont {Feng}}, \bibinfo {author} {\bibfnamefont {Z.}~\bibnamefont {Zhang}}, \bibinfo {author} {\bibfnamefont {W.}~\bibnamefont {Wang}}, \bibinfo {author} {\bibfnamefont {Q.}~\bibnamefont {Yang}}, \bibinfo {author} {\bibfnamefont {K.}~\bibnamefont {Watanabe}}, \bibinfo {author} {\bibfnamefont {T.}~\bibnamefont {Taniguchi}}, \bibinfo {author} {\bibfnamefont {K.}~\bibnamefont {Liu}}, \emph {et~al.},\ }\bibfield  {title} {\bibinfo {title} {Even-and odd-denominator fractional quantum anomalous hall effect in graphene moire superlattices},\ }\href {https://arxiv.org/abs/2405.16944} {\bibfield  {journal} {\bibinfo  {journal} {arXiv preprint arXiv:2405.16944}\ } (\bibinfo {year} {2024})}\BibitemShut {NoStop}%
\bibitem [{\citenamefont {Cai}\ \emph {et~al.}(2023)\citenamefont {Cai}, \citenamefont {Anderson}, \citenamefont {Wang}, \citenamefont {Zhang}, \citenamefont {Liu}, \citenamefont {Holtzmann}, \citenamefont {Zhang}, \citenamefont {Fan}, \citenamefont {Taniguchi}, \citenamefont {Watanabe}, \citenamefont {Ran}, \citenamefont {Cao}, \citenamefont {Fu}, \citenamefont {Xiao}, \citenamefont {Yao},\ and\ \citenamefont {Xu}}]{xuCaiSignaturesFractionalQuantum2023}%
  \BibitemOpen
  \bibfield  {author} {\bibinfo {author} {\bibfnamefont {J.}~\bibnamefont {Cai}}, \bibinfo {author} {\bibfnamefont {E.}~\bibnamefont {Anderson}}, \bibinfo {author} {\bibfnamefont {C.}~\bibnamefont {Wang}}, \bibinfo {author} {\bibfnamefont {X.}~\bibnamefont {Zhang}}, \bibinfo {author} {\bibfnamefont {X.}~\bibnamefont {Liu}}, \bibinfo {author} {\bibfnamefont {W.}~\bibnamefont {Holtzmann}}, \bibinfo {author} {\bibfnamefont {Y.}~\bibnamefont {Zhang}}, \bibinfo {author} {\bibfnamefont {F.}~\bibnamefont {Fan}}, \bibinfo {author} {\bibfnamefont {T.}~\bibnamefont {Taniguchi}}, \bibinfo {author} {\bibfnamefont {K.}~\bibnamefont {Watanabe}}, \bibinfo {author} {\bibfnamefont {Y.}~\bibnamefont {Ran}}, \bibinfo {author} {\bibfnamefont {T.}~\bibnamefont {Cao}}, \bibinfo {author} {\bibfnamefont {L.}~\bibnamefont {Fu}}, \bibinfo {author} {\bibfnamefont {D.}~\bibnamefont {Xiao}}, \bibinfo {author} {\bibfnamefont {W.}~\bibnamefont {Yao}},\ and\ \bibinfo {author} {\bibfnamefont {X.}~\bibnamefont {Xu}},\ }\bibfield  {title} {\bibinfo {title} {Signatures of fractional quantum anomalous hall states in twisted mote2},\ }\href {https://doi.org/10.1038/s41586-023-06289-w} {\bibfield  {journal} {\bibinfo  {journal} {Nature}\ }\textbf {\bibinfo {volume} {622}},\ \bibinfo {pages} {63} (\bibinfo {year} {2023})}\BibitemShut {NoStop}%
\bibitem [{\citenamefont {Park}\ \emph {et~al.}(2023)\citenamefont {Park}, \citenamefont {Cai}, \citenamefont {Anderson}, \citenamefont {Zhang}, \citenamefont {Zhu}, \citenamefont {Liu}, \citenamefont {Wang}, \citenamefont {Holtzmann}, \citenamefont {Hu}, \citenamefont {Liu}, \citenamefont {Taniguchi}, \citenamefont {Watanabe}, \citenamefont {Chu}, \citenamefont {Cao}, \citenamefont {Fu}, \citenamefont {Yao}, \citenamefont {Chang}, \citenamefont {Cobden}, \citenamefont {Xiao},\ and\ \citenamefont {Xu}}]{xuParkObservationFractionallyQuantized2023}%
  \BibitemOpen
  \bibfield  {author} {\bibinfo {author} {\bibfnamefont {H.}~\bibnamefont {Park}}, \bibinfo {author} {\bibfnamefont {J.}~\bibnamefont {Cai}}, \bibinfo {author} {\bibfnamefont {E.}~\bibnamefont {Anderson}}, \bibinfo {author} {\bibfnamefont {Y.}~\bibnamefont {Zhang}}, \bibinfo {author} {\bibfnamefont {J.}~\bibnamefont {Zhu}}, \bibinfo {author} {\bibfnamefont {X.}~\bibnamefont {Liu}}, \bibinfo {author} {\bibfnamefont {C.}~\bibnamefont {Wang}}, \bibinfo {author} {\bibfnamefont {W.}~\bibnamefont {Holtzmann}}, \bibinfo {author} {\bibfnamefont {C.}~\bibnamefont {Hu}}, \bibinfo {author} {\bibfnamefont {Z.}~\bibnamefont {Liu}}, \bibinfo {author} {\bibfnamefont {T.}~\bibnamefont {Taniguchi}}, \bibinfo {author} {\bibfnamefont {K.}~\bibnamefont {Watanabe}}, \bibinfo {author} {\bibfnamefont {J.-H.}\ \bibnamefont {Chu}}, \bibinfo {author} {\bibfnamefont {T.}~\bibnamefont {Cao}}, \bibinfo {author} {\bibfnamefont {L.}~\bibnamefont {Fu}}, \bibinfo {author} {\bibfnamefont {W.}~\bibnamefont {Yao}}, \bibinfo {author} {\bibfnamefont {C.-Z.}\ \bibnamefont {Chang}}, \bibinfo {author} {\bibfnamefont {D.}~\bibnamefont {Cobden}}, \bibinfo {author} {\bibfnamefont {D.}~\bibnamefont {Xiao}},\ and\ \bibinfo {author} {\bibfnamefont {X.}~\bibnamefont {Xu}},\ }\bibfield  {title} {\bibinfo {title} {Observation of fractionally quantized anomalous hall effect},\ }\href {https://doi.org/10.1038/s41586-023-06536-0} {\bibfield  {journal} {\bibinfo  {journal} {Nature}\ }\textbf {\bibinfo {volume} {622}},\ \bibinfo {pages} {74} (\bibinfo {year} {2023})}\BibitemShut {NoStop}%
\bibitem [{\citenamefont {Zeng}\ \emph {et~al.}(2023)\citenamefont {Zeng}, \citenamefont {Xia}, \citenamefont {Kang}, \citenamefont {Zhu}, \citenamefont {Knüppel}, \citenamefont {Vaswani}, \citenamefont {Watanabe}, \citenamefont {Taniguchi}, \citenamefont {Mak},\ and\ \citenamefont {Shan}}]{shanZengThermodynamicEvidenceFractional2023}%
  \BibitemOpen
  \bibfield  {author} {\bibinfo {author} {\bibfnamefont {Y.}~\bibnamefont {Zeng}}, \bibinfo {author} {\bibfnamefont {Z.}~\bibnamefont {Xia}}, \bibinfo {author} {\bibfnamefont {K.}~\bibnamefont {Kang}}, \bibinfo {author} {\bibfnamefont {J.}~\bibnamefont {Zhu}}, \bibinfo {author} {\bibfnamefont {P.}~\bibnamefont {Knüppel}}, \bibinfo {author} {\bibfnamefont {C.}~\bibnamefont {Vaswani}}, \bibinfo {author} {\bibfnamefont {K.}~\bibnamefont {Watanabe}}, \bibinfo {author} {\bibfnamefont {T.}~\bibnamefont {Taniguchi}}, \bibinfo {author} {\bibfnamefont {K.~F.}\ \bibnamefont {Mak}},\ and\ \bibinfo {author} {\bibfnamefont {J.}~\bibnamefont {Shan}},\ }\bibfield  {title} {\bibinfo {title} {Thermodynamic evidence of fractional chern insulator in moiré mote2},\ }\href {https://doi.org/10.1038/s41586-023-06452-3} {\bibfield  {journal} {\bibinfo  {journal} {Nature}\ }\textbf {\bibinfo {volume} {622}},\ \bibinfo {pages} {69} (\bibinfo {year} {2023})}\BibitemShut {NoStop}%
\bibitem [{\citenamefont {Xu}\ \emph {et~al.}(2023)\citenamefont {Xu}, \citenamefont {Sun}, \citenamefont {Jia}, \citenamefont {Liu}, \citenamefont {Xu}, \citenamefont {Li}, \citenamefont {Gu}, \citenamefont {Watanabe}, \citenamefont {Taniguchi}, \citenamefont {Tong}, \citenamefont {Jia}, \citenamefont {Shi}, \citenamefont {Jiang}, \citenamefont {Zhang}, \citenamefont {Liu},\ and\ \citenamefont {Li}}]{liXuObservationIntegerFractional2023}%
  \BibitemOpen
  \bibfield  {author} {\bibinfo {author} {\bibfnamefont {F.}~\bibnamefont {Xu}}, \bibinfo {author} {\bibfnamefont {Z.}~\bibnamefont {Sun}}, \bibinfo {author} {\bibfnamefont {T.}~\bibnamefont {Jia}}, \bibinfo {author} {\bibfnamefont {C.}~\bibnamefont {Liu}}, \bibinfo {author} {\bibfnamefont {C.}~\bibnamefont {Xu}}, \bibinfo {author} {\bibfnamefont {C.}~\bibnamefont {Li}}, \bibinfo {author} {\bibfnamefont {Y.}~\bibnamefont {Gu}}, \bibinfo {author} {\bibfnamefont {K.}~\bibnamefont {Watanabe}}, \bibinfo {author} {\bibfnamefont {T.}~\bibnamefont {Taniguchi}}, \bibinfo {author} {\bibfnamefont {B.}~\bibnamefont {Tong}}, \bibinfo {author} {\bibfnamefont {J.}~\bibnamefont {Jia}}, \bibinfo {author} {\bibfnamefont {Z.}~\bibnamefont {Shi}}, \bibinfo {author} {\bibfnamefont {S.}~\bibnamefont {Jiang}}, \bibinfo {author} {\bibfnamefont {Y.}~\bibnamefont {Zhang}}, \bibinfo {author} {\bibfnamefont {X.}~\bibnamefont {Liu}},\ and\ \bibinfo {author} {\bibfnamefont {T.}~\bibnamefont {Li}},\ }\bibfield  {title} {\bibinfo {title} {Observation of integer and fractional quantum anomalous hall effects in twisted bilayer ${\mathrm{mote}}_{2}$},\ }\href {https://doi.org/10.1103/PhysRevX.13.031037} {\bibfield  {journal} {\bibinfo  {journal} {Physical Review X}\ }\textbf {\bibinfo {volume} {13}},\ \bibinfo {pages} {031037} (\bibinfo {year} {2023})}\BibitemShut {NoStop}%
\bibitem [{\citenamefont {Foutty}\ \emph {et~al.}(2024)\citenamefont {Foutty}, \citenamefont {Kometter}, \citenamefont {Devakul}, \citenamefont {Reddy}, \citenamefont {Watanabe}, \citenamefont {Taniguchi}, \citenamefont {Fu},\ and\ \citenamefont {Feldman}}]{feldmanFouttyMappingTwisttunedMultiband2024}%
  \BibitemOpen
  \bibfield  {author} {\bibinfo {author} {\bibfnamefont {B.~A.}\ \bibnamefont {Foutty}}, \bibinfo {author} {\bibfnamefont {C.~R.}\ \bibnamefont {Kometter}}, \bibinfo {author} {\bibfnamefont {T.}~\bibnamefont {Devakul}}, \bibinfo {author} {\bibfnamefont {A.~P.}\ \bibnamefont {Reddy}}, \bibinfo {author} {\bibfnamefont {K.}~\bibnamefont {Watanabe}}, \bibinfo {author} {\bibfnamefont {T.}~\bibnamefont {Taniguchi}}, \bibinfo {author} {\bibfnamefont {L.}~\bibnamefont {Fu}},\ and\ \bibinfo {author} {\bibfnamefont {B.~E.}\ \bibnamefont {Feldman}},\ }\bibfield  {title} {\bibinfo {title} {Mapping twist-tuned multiband topology in bilayer wse2},\ }\href {https://doi.org/10.1126/science.adi4728} {\bibfield  {journal} {\bibinfo  {journal} {Science}\ }\textbf {\bibinfo {volume} {384}},\ \bibinfo {pages} {343} (\bibinfo {year} {2024})}\BibitemShut {NoStop}%
\bibitem [{\citenamefont {Kang}\ \emph {et~al.}(2024)\citenamefont {Kang}, \citenamefont {Shen}, \citenamefont {Qiu}, \citenamefont {Zeng}, \citenamefont {Xia}, \citenamefont {Watanabe}, \citenamefont {Taniguchi}, \citenamefont {Shan},\ and\ \citenamefont {Mak}}]{makKangEvidenceFractionalQuantum2024}%
  \BibitemOpen
  \bibfield  {author} {\bibinfo {author} {\bibfnamefont {K.}~\bibnamefont {Kang}}, \bibinfo {author} {\bibfnamefont {B.}~\bibnamefont {Shen}}, \bibinfo {author} {\bibfnamefont {Y.}~\bibnamefont {Qiu}}, \bibinfo {author} {\bibfnamefont {Y.}~\bibnamefont {Zeng}}, \bibinfo {author} {\bibfnamefont {Z.}~\bibnamefont {Xia}}, \bibinfo {author} {\bibfnamefont {K.}~\bibnamefont {Watanabe}}, \bibinfo {author} {\bibfnamefont {T.}~\bibnamefont {Taniguchi}}, \bibinfo {author} {\bibfnamefont {J.}~\bibnamefont {Shan}},\ and\ \bibinfo {author} {\bibfnamefont {K.~F.}\ \bibnamefont {Mak}},\ }\bibfield  {title} {\bibinfo {title} {Evidence of the fractional quantum spin hall effect in moiré mote2},\ }\href {https://doi.org/10.1038/s41586-024-07214-5} {\bibfield  {journal} {\bibinfo  {journal} {Nature}\ }\textbf {\bibinfo {volume} {628}},\ \bibinfo {pages} {522} (\bibinfo {year} {2024})}\BibitemShut {NoStop}%
\bibitem [{\citenamefont {Xie}\ \emph {et~al.}(2021)\citenamefont {Xie}, \citenamefont {Pierce}, \citenamefont {Park}, \citenamefont {Parker}, \citenamefont {Khalaf}, \citenamefont {Ledwith}, \citenamefont {Cao}, \citenamefont {Lee}, \citenamefont {Chen}, \citenamefont {Forrester}, \citenamefont {Watanabe}, \citenamefont {Taniguchi}, \citenamefont {Vishwanath}, \citenamefont {{Jarillo-Herrero}},\ and\ \citenamefont {Yacoby}}]{yacobyXieFractionalChernInsulators2021}%
  \BibitemOpen
  \bibfield  {author} {\bibinfo {author} {\bibfnamefont {Y.}~\bibnamefont {Xie}}, \bibinfo {author} {\bibfnamefont {A.~T.}\ \bibnamefont {Pierce}}, \bibinfo {author} {\bibfnamefont {J.~M.}\ \bibnamefont {Park}}, \bibinfo {author} {\bibfnamefont {D.~E.}\ \bibnamefont {Parker}}, \bibinfo {author} {\bibfnamefont {E.}~\bibnamefont {Khalaf}}, \bibinfo {author} {\bibfnamefont {P.}~\bibnamefont {Ledwith}}, \bibinfo {author} {\bibfnamefont {Y.}~\bibnamefont {Cao}}, \bibinfo {author} {\bibfnamefont {S.~H.}\ \bibnamefont {Lee}}, \bibinfo {author} {\bibfnamefont {S.}~\bibnamefont {Chen}}, \bibinfo {author} {\bibfnamefont {P.~R.}\ \bibnamefont {Forrester}}, \bibinfo {author} {\bibfnamefont {K.}~\bibnamefont {Watanabe}}, \bibinfo {author} {\bibfnamefont {T.}~\bibnamefont {Taniguchi}}, \bibinfo {author} {\bibfnamefont {A.}~\bibnamefont {Vishwanath}}, \bibinfo {author} {\bibfnamefont {P.}~\bibnamefont {{Jarillo-Herrero}}},\ and\ \bibinfo {author} {\bibfnamefont {A.}~\bibnamefont {Yacoby}},\ }\bibfield  {title} {\bibinfo {title} {Fractional {{Chern}} insulators in magic-angle twisted bilayer graphene},\ }\href {https://doi.org/10.1038/s41586-021-04002-3} {\bibfield  {journal} {\bibinfo  {journal} {Nature}\ }\textbf {\bibinfo {volume} {600}},\ \bibinfo {pages} {439} (\bibinfo {year} {2021})}\BibitemShut {NoStop}%
\bibitem [{\citenamefont {Neupert}\ \emph {et~al.}(2011)\citenamefont {Neupert}, \citenamefont {Santos}, \citenamefont {Chamon},\ and\ \citenamefont {Mudry}}]{Neupertprl_2011}%
  \BibitemOpen
  \bibfield  {author} {\bibinfo {author} {\bibfnamefont {T.}~\bibnamefont {Neupert}}, \bibinfo {author} {\bibfnamefont {L.}~\bibnamefont {Santos}}, \bibinfo {author} {\bibfnamefont {C.}~\bibnamefont {Chamon}},\ and\ \bibinfo {author} {\bibfnamefont {C.}~\bibnamefont {Mudry}},\ }\bibfield  {title} {\bibinfo {title} {Fractional quantum hall states at zero magnetic field},\ }\href {https://doi.org/10.1103/PhysRevLett.106.236804} {\bibfield  {journal} {\bibinfo  {journal} {Phys. Rev. Lett.}\ }\textbf {\bibinfo {volume} {106}},\ \bibinfo {pages} {236804} (\bibinfo {year} {2011})}\BibitemShut {NoStop}%
\bibitem [{\citenamefont {Bergholtz}\ and\ \citenamefont {Liu}(2013)}]{zhao_review}%
  \BibitemOpen
  \bibfield  {author} {\bibinfo {author} {\bibfnamefont {E.~J.}\ \bibnamefont {Bergholtz}}\ and\ \bibinfo {author} {\bibfnamefont {Z.}~\bibnamefont {Liu}},\ }\bibfield  {title} {\bibinfo {title} {Topological flat band models and fractional chern insulators},\ }\href@noop {} {\bibfield  {journal} {\bibinfo  {journal} {International Journal of Modern Physics B}\ }\textbf {\bibinfo {volume} {27}},\ \bibinfo {pages} {1330017} (\bibinfo {year} {2013})}\BibitemShut {NoStop}%
\bibitem [{\citenamefont {Regnault}\ and\ \citenamefont {Bernevig}(2011)}]{regnault2011}%
  \BibitemOpen
  \bibfield  {author} {\bibinfo {author} {\bibfnamefont {N.}~\bibnamefont {Regnault}}\ and\ \bibinfo {author} {\bibfnamefont {B.~A.}\ \bibnamefont {Bernevig}},\ }\bibfield  {title} {\bibinfo {title} {Fractional chern insulator},\ }\href {https://doi.org/10.1103/PhysRevX.1.021014} {\bibfield  {journal} {\bibinfo  {journal} {Phys. Rev. X}\ }\textbf {\bibinfo {volume} {1}},\ \bibinfo {pages} {021014} (\bibinfo {year} {2011})}\BibitemShut {NoStop}%
\bibitem [{\citenamefont {Tang}\ \emph {et~al.}(2011)\citenamefont {Tang}, \citenamefont {Mei},\ and\ \citenamefont {Wen}}]{tang2011}%
  \BibitemOpen
  \bibfield  {author} {\bibinfo {author} {\bibfnamefont {E.}~\bibnamefont {Tang}}, \bibinfo {author} {\bibfnamefont {J.-W.}\ \bibnamefont {Mei}},\ and\ \bibinfo {author} {\bibfnamefont {X.-G.}\ \bibnamefont {Wen}},\ }\bibfield  {title} {\bibinfo {title} {High-temperature fractional quantum hall states},\ }\href {https://doi.org/10.1103/PhysRevLett.106.236802} {\bibfield  {journal} {\bibinfo  {journal} {Phys. Rev. Lett.}\ }\textbf {\bibinfo {volume} {106}},\ \bibinfo {pages} {236802} (\bibinfo {year} {2011})}\BibitemShut {NoStop}%
\bibitem [{\citenamefont {Sheng}\ \emph {et~al.}(2011)\citenamefont {Sheng}, \citenamefont {Gu}, \citenamefont {Sun},\ and\ \citenamefont {Sheng}}]{Sheng2011}%
  \BibitemOpen
  \bibfield  {author} {\bibinfo {author} {\bibfnamefont {D.~N.}\ \bibnamefont {Sheng}}, \bibinfo {author} {\bibfnamefont {Z.-C.}\ \bibnamefont {Gu}}, \bibinfo {author} {\bibfnamefont {K.}~\bibnamefont {Sun}},\ and\ \bibinfo {author} {\bibfnamefont {L.}~\bibnamefont {Sheng}},\ }\bibfield  {title} {\bibinfo {title} {Fractional quantum hall effect in the absence of landau levels},\ }\href {https://doi.org/10.1038/ncomms1380} {\bibfield  {journal} {\bibinfo  {journal} {Nature Communications}\ }\textbf {\bibinfo {volume} {2}},\ \bibinfo {pages} {389} (\bibinfo {year} {2011})}\BibitemShut {NoStop}%
\bibitem [{\citenamefont {Qi}(2011)}]{qi2011}%
  \BibitemOpen
  \bibfield  {author} {\bibinfo {author} {\bibfnamefont {X.-L.}\ \bibnamefont {Qi}},\ }\bibfield  {title} {\bibinfo {title} {Generic wave-function description of fractional quantum anomalous hall states and fractional topological insulators},\ }\href {https://doi.org/10.1103/PhysRevLett.107.126803} {\bibfield  {journal} {\bibinfo  {journal} {Phys. Rev. Lett.}\ }\textbf {\bibinfo {volume} {107}},\ \bibinfo {pages} {126803} (\bibinfo {year} {2011})}\BibitemShut {NoStop}%
\bibitem [{\citenamefont {Yu}\ \emph {et~al.}(2020)\citenamefont {Yu}, \citenamefont {Chen},\ and\ \citenamefont {Yao}}]{yaoYuGiantMagneticField2020}%
  \BibitemOpen
  \bibfield  {author} {\bibinfo {author} {\bibfnamefont {H.}~\bibnamefont {Yu}}, \bibinfo {author} {\bibfnamefont {M.}~\bibnamefont {Chen}},\ and\ \bibinfo {author} {\bibfnamefont {W.}~\bibnamefont {Yao}},\ }\bibfield  {title} {\bibinfo {title} {Giant magnetic field from moir{\'e} induced {{Berry}} phase in homobilayer semiconductors},\ }\href {https://doi.org/10.1093/nsr/nwz117} {\bibfield  {journal} {\bibinfo  {journal} {National Science Review}\ }\textbf {\bibinfo {volume} {7}},\ \bibinfo {pages} {12} (\bibinfo {year} {2020})}\BibitemShut {NoStop}%
\bibitem [{\citenamefont {Zhai}\ and\ \citenamefont {Yao}(2020)}]{yaoZhaiTheoryTunableFlux2020}%
  \BibitemOpen
  \bibfield  {author} {\bibinfo {author} {\bibfnamefont {D.}~\bibnamefont {Zhai}}\ and\ \bibinfo {author} {\bibfnamefont {W.}~\bibnamefont {Yao}},\ }\bibfield  {title} {\bibinfo {title} {Theory of tunable flux lattices in the homobilayer moir\'e of twisted and uniformly strained transition metal dichalcogenides},\ }\href {https://doi.org/10.1103/PhysRevMaterials.4.094002} {\bibfield  {journal} {\bibinfo  {journal} {Physical Review Materials}\ }\textbf {\bibinfo {volume} {4}},\ \bibinfo {pages} {094002} (\bibinfo {year} {2020})}\BibitemShut {NoStop}%
\bibitem [{\citenamefont {Morales-Durán}\ \emph {et~al.}(2024)\citenamefont {Morales-Durán}, \citenamefont {Wei}, \citenamefont {Shi},\ and\ \citenamefont {MacDonald}}]{macdonaldMorales-DuranMagicAnglesFractional2024}%
  \BibitemOpen
  \bibfield  {author} {\bibinfo {author} {\bibfnamefont {N.}~\bibnamefont {Morales-Durán}}, \bibinfo {author} {\bibfnamefont {N.}~\bibnamefont {Wei}}, \bibinfo {author} {\bibfnamefont {J.}~\bibnamefont {Shi}},\ and\ \bibinfo {author} {\bibfnamefont {A.~H.}\ \bibnamefont {MacDonald}},\ }\bibfield  {title} {\bibinfo {title} {Magic angles and fractional chern insulators in twisted homobilayer transition metal dichalcogenides},\ }\href {https://doi.org/10.1103/PhysRevLett.132.096602} {\bibfield  {journal} {\bibinfo  {journal} {Physical Review Letters}\ }\textbf {\bibinfo {volume} {132}},\ \bibinfo {pages} {096602} (\bibinfo {year} {2024})}\BibitemShut {NoStop}%
\bibitem [{\citenamefont {Zhang}\ \emph {et~al.}(2025)\citenamefont {Zhang}, \citenamefont {{Morales-Dur{\'a}n}}, \citenamefont {Li}, \citenamefont {Yao}, \citenamefont {Su}, \citenamefont {Lin}, \citenamefont {Dong}, \citenamefont {Liu}, \citenamefont {Chen}, \citenamefont {Kim}, \citenamefont {Watanabe}, \citenamefont {Taniguchi}, \citenamefont {Li}, \citenamefont {Robinson}, \citenamefont {Macdonald},\ and\ \citenamefont {Shih}}]{shihZhangExperimentalSignatureLayer2025}%
  \BibitemOpen
  \bibfield  {author} {\bibinfo {author} {\bibfnamefont {F.}~\bibnamefont {Zhang}}, \bibinfo {author} {\bibfnamefont {N.}~\bibnamefont {{Morales-Dur{\'a}n}}}, \bibinfo {author} {\bibfnamefont {Y.}~\bibnamefont {Li}}, \bibinfo {author} {\bibfnamefont {W.}~\bibnamefont {Yao}}, \bibinfo {author} {\bibfnamefont {J.-J.}\ \bibnamefont {Su}}, \bibinfo {author} {\bibfnamefont {Y.-C.}\ \bibnamefont {Lin}}, \bibinfo {author} {\bibfnamefont {C.}~\bibnamefont {Dong}}, \bibinfo {author} {\bibfnamefont {X.}~\bibnamefont {Liu}}, \bibinfo {author} {\bibfnamefont {F.-X.~R.}\ \bibnamefont {Chen}}, \bibinfo {author} {\bibfnamefont {H.}~\bibnamefont {Kim}}, \bibinfo {author} {\bibfnamefont {K.}~\bibnamefont {Watanabe}}, \bibinfo {author} {\bibfnamefont {T.}~\bibnamefont {Taniguchi}}, \bibinfo {author} {\bibfnamefont {X.}~\bibnamefont {Li}}, \bibinfo {author} {\bibfnamefont {J.~A.}\ \bibnamefont {Robinson}}, \bibinfo {author} {\bibfnamefont {A.~H.}\ \bibnamefont {Macdonald}},\ and\ \bibinfo {author} {\bibfnamefont {C.-K.}\ \bibnamefont {Shih}},\ }\bibfield  {title} {\bibinfo {title} {Experimental signature of layer skyrmions and implications for band topology in twisted {{WSe2}} bilayers},\ }\href {https://doi.org/10.1038/s41567-025-02876-y} {\bibfield  {journal} {\bibinfo  {journal} {Nature Physics}\ ,\ \bibinfo {pages} {1}} (\bibinfo {year} {2025})}\BibitemShut {NoStop}%
\bibitem [{\citenamefont {Thompson}\ \emph {et~al.}(2025)\citenamefont {Thompson}, \citenamefont {Chu}, \citenamefont {Mesple}, \citenamefont {Zhang}, \citenamefont {Hu}, \citenamefont {Zhao}, \citenamefont {Park}, \citenamefont {Cai}, \citenamefont {Anderson}, \citenamefont {Watanabe}, \citenamefont {Taniguchi}, \citenamefont {Yang}, \citenamefont {Chu}, \citenamefont {Xu}, \citenamefont {Cao}, \citenamefont {Xiao},\ and\ \citenamefont {Yankowitz}}]{yankowitzThompsonMicroscopicSignaturesTopology2025}%
  \BibitemOpen
  \bibfield  {author} {\bibinfo {author} {\bibfnamefont {E.}~\bibnamefont {Thompson}}, \bibinfo {author} {\bibfnamefont {K.~T.}\ \bibnamefont {Chu}}, \bibinfo {author} {\bibfnamefont {F.}~\bibnamefont {Mesple}}, \bibinfo {author} {\bibfnamefont {X.-W.}\ \bibnamefont {Zhang}}, \bibinfo {author} {\bibfnamefont {C.}~\bibnamefont {Hu}}, \bibinfo {author} {\bibfnamefont {Y.}~\bibnamefont {Zhao}}, \bibinfo {author} {\bibfnamefont {H.}~\bibnamefont {Park}}, \bibinfo {author} {\bibfnamefont {J.}~\bibnamefont {Cai}}, \bibinfo {author} {\bibfnamefont {E.}~\bibnamefont {Anderson}}, \bibinfo {author} {\bibfnamefont {K.}~\bibnamefont {Watanabe}}, \bibinfo {author} {\bibfnamefont {T.}~\bibnamefont {Taniguchi}}, \bibinfo {author} {\bibfnamefont {J.}~\bibnamefont {Yang}}, \bibinfo {author} {\bibfnamefont {J.-H.}\ \bibnamefont {Chu}}, \bibinfo {author} {\bibfnamefont {X.}~\bibnamefont {Xu}}, \bibinfo {author} {\bibfnamefont {T.}~\bibnamefont {Cao}}, \bibinfo {author} {\bibfnamefont {D.}~\bibnamefont {Xiao}},\ and\ \bibinfo {author} {\bibfnamefont {M.}~\bibnamefont {Yankowitz}},\ }\bibfield  {title} {\bibinfo {title} {Microscopic signatures of topology in twisted {{MoTe2}}},\ }\href {https://doi.org/10.1038/s41567-025-02877-x} {\bibfield  {journal} {\bibinfo  {journal} {Nature Physics}\ ,\ \bibinfo {pages} {1}} (\bibinfo {year} {2025})}\BibitemShut {NoStop}%
\bibitem [{\citenamefont {Guerci}\ \emph {et~al.}(2024{\natexlab{a}})\citenamefont {Guerci}, \citenamefont {Wang},\ and\ \citenamefont {Mora}}]{moraGuerciLayerSkyrmionsIdeal2024}%
  \BibitemOpen
  \bibfield  {author} {\bibinfo {author} {\bibfnamefont {D.}~\bibnamefont {Guerci}}, \bibinfo {author} {\bibfnamefont {J.}~\bibnamefont {Wang}},\ and\ \bibinfo {author} {\bibfnamefont {C.}~\bibnamefont {Mora}},\ }\href {https://doi.org/10.48550/arXiv.2408.12652} {\bibinfo {title} {Layer skyrmions for ideal {{Chern}} bands and twisted bilayer graphene}} (\bibinfo {year} {2024}{\natexlab{a}}),\ \Eprint {https://arxiv.org/abs/2408.12652} {arXiv:2408.12652} \BibitemShut {NoStop}%
\bibitem [{\citenamefont {Tarnopolsky}\ \emph {et~al.}(2019)\citenamefont {Tarnopolsky}, \citenamefont {Kruchkov},\ and\ \citenamefont {Vishwanath}}]{vishwanathTarnopolskyOriginMagicAngles2019}%
  \BibitemOpen
  \bibfield  {author} {\bibinfo {author} {\bibfnamefont {G.}~\bibnamefont {Tarnopolsky}}, \bibinfo {author} {\bibfnamefont {A.~J.}\ \bibnamefont {Kruchkov}},\ and\ \bibinfo {author} {\bibfnamefont {A.}~\bibnamefont {Vishwanath}},\ }\bibfield  {title} {\bibinfo {title} {Origin of {{Magic Angles}} in {{Twisted Bilayer Graphene}}},\ }\href {https://doi.org/10.1103/PhysRevLett.122.106405} {\bibfield  {journal} {\bibinfo  {journal} {Physical Review Letters}\ }\textbf {\bibinfo {volume} {122}},\ \bibinfo {pages} {106405} (\bibinfo {year} {2019})},\ \Eprint {https://arxiv.org/abs/1808.05250} {arXiv:1808.05250} \BibitemShut {NoStop}%
\bibitem [{\citenamefont {Wang}\ \emph {et~al.}(2021{\natexlab{a}})\citenamefont {Wang}, \citenamefont {Cano}, \citenamefont {Millis}, \citenamefont {Liu},\ and\ \citenamefont {Yang}}]{yangWangExactLandauLevel2021}%
  \BibitemOpen
  \bibfield  {author} {\bibinfo {author} {\bibfnamefont {J.}~\bibnamefont {Wang}}, \bibinfo {author} {\bibfnamefont {J.}~\bibnamefont {Cano}}, \bibinfo {author} {\bibfnamefont {A.~J.}\ \bibnamefont {Millis}}, \bibinfo {author} {\bibfnamefont {Z.}~\bibnamefont {Liu}},\ and\ \bibinfo {author} {\bibfnamefont {B.}~\bibnamefont {Yang}},\ }\bibfield  {title} {\bibinfo {title} {Exact landau level description of geometry and interaction in a flatband},\ }\href {https://doi.org/10.1103/PhysRevLett.127.246403} {\bibfield  {journal} {\bibinfo  {journal} {Physical Review Letters}\ }\textbf {\bibinfo {volume} {127}},\ \bibinfo {pages} {246403} (\bibinfo {year} {2021}{\natexlab{a}})},\ \Eprint {https://arxiv.org/abs/2105.07491} {arXiv:2105.07491 [cond-mat]} \BibitemShut {NoStop}%
\bibitem [{\citenamefont {Wang}\ \emph {et~al.}(2021{\natexlab{b}})\citenamefont {Wang}, \citenamefont {Zheng}, \citenamefont {Millis},\ and\ \citenamefont {Cano}}]{canoWangChiralApproximationTwisted2021}%
  \BibitemOpen
  \bibfield  {author} {\bibinfo {author} {\bibfnamefont {J.}~\bibnamefont {Wang}}, \bibinfo {author} {\bibfnamefont {Y.}~\bibnamefont {Zheng}}, \bibinfo {author} {\bibfnamefont {A.~J.}\ \bibnamefont {Millis}},\ and\ \bibinfo {author} {\bibfnamefont {J.}~\bibnamefont {Cano}},\ }\bibfield  {title} {\bibinfo {title} {Chiral approximation to twisted bilayer graphene: {{Exact}} intravalley inversion symmetry, nodal structure, and implications for higher magic angles},\ }\href {https://doi.org/10.1103/PhysRevResearch.3.023155} {\bibfield  {journal} {\bibinfo  {journal} {Physical Review Research}\ }\textbf {\bibinfo {volume} {3}},\ \bibinfo {pages} {023155} (\bibinfo {year} {2021}{\natexlab{b}})}\BibitemShut {NoStop}%
\bibitem [{\citenamefont {Ledwith}\ \emph {et~al.}(2020)\citenamefont {Ledwith}, \citenamefont {Tarnopolsky}, \citenamefont {Khalaf},\ and\ \citenamefont {Vishwanath}}]{vishwanathLedwithFractionalChernInsulator2020}%
  \BibitemOpen
  \bibfield  {author} {\bibinfo {author} {\bibfnamefont {P.~J.}\ \bibnamefont {Ledwith}}, \bibinfo {author} {\bibfnamefont {G.}~\bibnamefont {Tarnopolsky}}, \bibinfo {author} {\bibfnamefont {E.}~\bibnamefont {Khalaf}},\ and\ \bibinfo {author} {\bibfnamefont {A.}~\bibnamefont {Vishwanath}},\ }\bibfield  {title} {\bibinfo {title} {Fractional chern insulator states in twisted bilayer graphene: An analytical approach},\ }\href {https://doi.org/10.1103/PhysRevResearch.2.023237} {\bibfield  {journal} {\bibinfo  {journal} {Physical Review Research}\ }\textbf {\bibinfo {volume} {2}},\ \bibinfo {pages} {023237} (\bibinfo {year} {2020})},\ \Eprint {https://arxiv.org/abs/1912.09634} {arXiv:1912.09634} \BibitemShut {NoStop}%
\bibitem [{\citenamefont {Wang}\ \emph {et~al.}(2023)\citenamefont {Wang}, \citenamefont {Klevtsov},\ and\ \citenamefont {Liu}}]{liuWangOriginModelFractional2023}%
  \BibitemOpen
  \bibfield  {author} {\bibinfo {author} {\bibfnamefont {J.}~\bibnamefont {Wang}}, \bibinfo {author} {\bibfnamefont {S.}~\bibnamefont {Klevtsov}},\ and\ \bibinfo {author} {\bibfnamefont {Z.}~\bibnamefont {Liu}},\ }\bibfield  {title} {\bibinfo {title} {Origin of model fractional chern insulators in all topological ideal flatbands: Explicit color-entangled wave function and exact density algebra},\ }\href {https://doi.org/10.1103/PhysRevResearch.5.023167} {\bibfield  {journal} {\bibinfo  {journal} {Physical Review Research}\ }\textbf {\bibinfo {volume} {5}},\ \bibinfo {pages} {023167} (\bibinfo {year} {2023})}\BibitemShut {NoStop}%
\bibitem [{\citenamefont {Dong}\ \emph {et~al.}(2023{\natexlab{a}})\citenamefont {Dong}, \citenamefont {Wang}, \citenamefont {Ledwith}, \citenamefont {Vishwanath},\ and\ \citenamefont {Parker}}]{parkerDongCompositeFermiLiquid2023}%
  \BibitemOpen
  \bibfield  {author} {\bibinfo {author} {\bibfnamefont {J.}~\bibnamefont {Dong}}, \bibinfo {author} {\bibfnamefont {J.}~\bibnamefont {Wang}}, \bibinfo {author} {\bibfnamefont {P.~J.}\ \bibnamefont {Ledwith}}, \bibinfo {author} {\bibfnamefont {A.}~\bibnamefont {Vishwanath}},\ and\ \bibinfo {author} {\bibfnamefont {D.~E.}\ \bibnamefont {Parker}},\ }\bibfield  {title} {\bibinfo {title} {Composite fermi liquid at zero magnetic field in twisted ${\mathrm{mote}}_{2}$},\ }\href {https://doi.org/10.1103/PhysRevLett.131.136502} {\bibfield  {journal} {\bibinfo  {journal} {Physical Review Letters}\ }\textbf {\bibinfo {volume} {131}},\ \bibinfo {pages} {136502} (\bibinfo {year} {2023}{\natexlab{a}})}\BibitemShut {NoStop}%
\bibitem [{\citenamefont {Ahn}\ \emph {et~al.}(2024)\citenamefont {Ahn}, \citenamefont {Lee}, \citenamefont {Yananose}, \citenamefont {Kim},\ and\ \citenamefont {Cho}}]{choAhnFirstLandauLevel2024}%
  \BibitemOpen
  \bibfield  {author} {\bibinfo {author} {\bibfnamefont {C.-E.}\ \bibnamefont {Ahn}}, \bibinfo {author} {\bibfnamefont {W.}~\bibnamefont {Lee}}, \bibinfo {author} {\bibfnamefont {K.}~\bibnamefont {Yananose}}, \bibinfo {author} {\bibfnamefont {Y.}~\bibnamefont {Kim}},\ and\ \bibinfo {author} {\bibfnamefont {G.~Y.}\ \bibnamefont {Cho}},\ }\href {https://doi.org/10.48550/arXiv.2403.19155} {\bibinfo {title} {First landau level physics in second moir\'e band of $2.1^\circ$ twisted bilayer mote${}_2$}} (\bibinfo {year} {2024}),\ \Eprint {https://arxiv.org/abs/2403.19155} {arXiv:2403.19155 [cond-mat]} \BibitemShut {NoStop}%
\bibitem [{\citenamefont {Wang}\ \emph {et~al.}(2024)\citenamefont {Wang}, \citenamefont {Zhang}, \citenamefont {Liu}, \citenamefont {Wang}, \citenamefont {Cao},\ and\ \citenamefont {Xiao}}]{xiaoWangHigherLandauLevelAnalogues2024}%
  \BibitemOpen
  \bibfield  {author} {\bibinfo {author} {\bibfnamefont {C.}~\bibnamefont {Wang}}, \bibinfo {author} {\bibfnamefont {X.-W.}\ \bibnamefont {Zhang}}, \bibinfo {author} {\bibfnamefont {X.}~\bibnamefont {Liu}}, \bibinfo {author} {\bibfnamefont {J.}~\bibnamefont {Wang}}, \bibinfo {author} {\bibfnamefont {T.}~\bibnamefont {Cao}},\ and\ \bibinfo {author} {\bibfnamefont {D.}~\bibnamefont {Xiao}},\ }\href@noop {} {\bibinfo {title} {Higher landau-level analogues and signatures of non-abelian states in twisted bilayer mote$_2$}} (\bibinfo {year} {2024}),\ \Eprint {https://arxiv.org/abs/2404.05697} {arXiv:2404.05697 [cond-mat]} \BibitemShut {NoStop}%
\bibitem [{\citenamefont {Song}\ and\ \citenamefont {Bernevig}(2022)}]{bernevigSongMagicAngleTwistedBilayer2022}%
  \BibitemOpen
  \bibfield  {author} {\bibinfo {author} {\bibfnamefont {Z.-D.}\ \bibnamefont {Song}}\ and\ \bibinfo {author} {\bibfnamefont {B.~A.}\ \bibnamefont {Bernevig}},\ }\bibfield  {title} {\bibinfo {title} {Magic-angle twisted bilayer graphene as a topological heavy fermion problem},\ }\href {https://doi.org/10.1103/PhysRevLett.129.047601} {\bibfield  {journal} {\bibinfo  {journal} {Physical Review Letters}\ }\textbf {\bibinfo {volume} {129}},\ \bibinfo {pages} {047601} (\bibinfo {year} {2022})}\BibitemShut {NoStop}%
\bibitem [{\citenamefont {Călugăru}\ \emph {et~al.}(2023)\citenamefont {Călugăru}, \citenamefont {Borovkov}, \citenamefont {Lau}, \citenamefont {Coleman}, \citenamefont {Song},\ and\ \citenamefont {Bernevig}}]{calugaru2023}%
  \BibitemOpen
  \bibfield  {author} {\bibinfo {author} {\bibfnamefont {D.}~\bibnamefont {Călugăru}}, \bibinfo {author} {\bibfnamefont {M.}~\bibnamefont {Borovkov}}, \bibinfo {author} {\bibfnamefont {L.~L.~H.}\ \bibnamefont {Lau}}, \bibinfo {author} {\bibfnamefont {P.}~\bibnamefont {Coleman}}, \bibinfo {author} {\bibfnamefont {Z.-D.}\ \bibnamefont {Song}},\ and\ \bibinfo {author} {\bibfnamefont {B.~A.}\ \bibnamefont {Bernevig}},\ }\bibfield  {title} {\bibinfo {title} {Twisted bilayer graphene as topological heavy fermion: Ii. analytical approximations of the model parameters},\ }\href {https://doi.org/10.1063/10.0019421} {\bibfield  {journal} {\bibinfo  {journal} {Low Temperature Physics}\ }\textbf {\bibinfo {volume} {49}},\ \bibinfo {pages} {640} (\bibinfo {year} {2023})}\BibitemShut {NoStop}%
\bibitem [{\citenamefont {Shi}\ and\ \citenamefont {Dai}(2022)}]{Shi2022}%
  \BibitemOpen
  \bibfield  {author} {\bibinfo {author} {\bibfnamefont {H.}~\bibnamefont {Shi}}\ and\ \bibinfo {author} {\bibfnamefont {X.}~\bibnamefont {Dai}},\ }\bibfield  {title} {\bibinfo {title} {Heavy-fermion representation for twisted bilayer graphene systems},\ }\href {https://doi.org/10.1103/PhysRevB.106.245129} {\bibfield  {journal} {\bibinfo  {journal} {Physical Review B}\ }\textbf {\bibinfo {volume} {106}},\ \bibinfo {pages} {245129} (\bibinfo {year} {2022})}\BibitemShut {NoStop}%
\bibitem [{Note1()}]{Note1}%
  \BibitemOpen
  \bibinfo {note} {More precisely, the wavefunction $u_{\protect \mathbf k}(\protect \mathbf r)$ defines a nonperiodic map from the unit cell to complex numbers $\protect \mathbb C$. Outside vortex cores where $u_{\protect \mathbf k}(\protect \mathbf r)$ vanishes, it defines a map from part of the unit cell to nonzero complex numbers $\protect \mathbb C^\times =\protect \mathbb C-\{0\}$. The fundamental group of $\protect \mathbb C^\times $, given $\pi _1(\protect \mathbb C^\times )=\protect \mathbb Z$, captures the phase winding along a closed loop. Since we can deform the loop along the unit cell border to a set of loops encircling each vortex core, it follows that the total vorticity $\Phi $ is given by the sum of the vorticities associated with each zero of $u_{\protect \mathbf k}(\protect \mathbf r)$.}\BibitemShut {Stop}%
\bibitem [{\citenamefont {Liu}\ \emph {et~al.}(2024)\citenamefont {Liu}, \citenamefont {Mera}, \citenamefont {Fujimoto}, \citenamefont {Ozawa},\ and\ \citenamefont {Wang}}]{wangLiuTheoryGeneralizedLandau2024}%
  \BibitemOpen
  \bibfield  {author} {\bibinfo {author} {\bibfnamefont {Z.}~\bibnamefont {Liu}}, \bibinfo {author} {\bibfnamefont {B.}~\bibnamefont {Mera}}, \bibinfo {author} {\bibfnamefont {M.}~\bibnamefont {Fujimoto}}, \bibinfo {author} {\bibfnamefont {T.}~\bibnamefont {Ozawa}},\ and\ \bibinfo {author} {\bibfnamefont {J.}~\bibnamefont {Wang}},\ }\href@noop {} {\bibinfo {title} {Theory of generalized landau levels and implication for non-abelian states}} (\bibinfo {year} {2024}),\ \Eprint {https://arxiv.org/abs/2405.14479} {arXiv:2405.14479 [cond-mat, physics:hep-th, physics:math-ph]} \BibitemShut {NoStop}%
\bibitem [{Note2()}]{Note2}%
  \BibitemOpen
  \bibinfo {note} {There is some arbitrariness (gauge-freedom) in the decomposition of Eq.~\protect \eqref {eq:generalwfdecomp} as a phase term $e^{i \alpha _{\protect \mathbf k} (\protect \mathbf r)}$ can always be transferred between the spinor and the scalar parts. Although this shift of phase can change the Berry connection, it does not affect the gauge-invariant Berry curvature, yielding a uniquely defined real-space Chern number $C_{\protect \mathbf k}$, fixed by the winding $\Phi _1$ of $\protect \boldsymbol {\protect \hat n_{\protect \mathbf k}}$.}\BibitemShut {Stop}%
\bibitem [{Note3()}]{Note3}%
  \BibitemOpen
  \bibinfo {note} {In contrast to $\protect \frac {u_{\protect \mathbf k}(\protect \mathbf r)}{\left |u_{\protect \mathbf k}(\protect \mathbf r)\right |}$ and its phase vortices, the ratio $\protect \frac {u_{\protect \mathbf k}^\dagger (\protect \mathbf r) \protect \boldsymbol {\mu } u_{\protect \mathbf k}(\protect \mathbf r)}{|u_{\protect \mathbf k}^\dagger (\protect \mathbf r) u_{\protect \mathbf k}(\protect \mathbf r)|}$ can be extended continuously at the zeroes of $u_{\protect \mathbf k}(\protect \mathbf r)$. We also note that the scalar part $\psi ^{\Phi _2}_{\protect \mathbf k}$ drops out from this ratio.}\BibitemShut {Stop}%
\bibitem [{\citenamefont {Graf}\ and\ \citenamefont {Pi{\'e}chon}(2021)}]{piechonGrafBerryCurvatureQuantum2021}%
  \BibitemOpen
  \bibfield  {author} {\bibinfo {author} {\bibfnamefont {A.}~\bibnamefont {Graf}}\ and\ \bibinfo {author} {\bibfnamefont {F.}~\bibnamefont {Pi{\'e}chon}},\ }\bibfield  {title} {\bibinfo {title} {Berry curvature and quantum metric in \${{N}}\$-band systems: {{An}} eigenprojector approach},\ }\href {https://doi.org/10.1103/PhysRevB.104.085114} {\bibfield  {journal} {\bibinfo  {journal} {Physical Review B}\ }\textbf {\bibinfo {volume} {104}},\ \bibinfo {pages} {085114} (\bibinfo {year} {2021})}\BibitemShut {NoStop}%
\bibitem [{\citenamefont {Kemp}\ \emph {et~al.}(2022)\citenamefont {Kemp}, \citenamefont {Cooper},\ and\ \citenamefont {{\"U}nal}}]{unalKempNestedsphereDescription$N$level2022}%
  \BibitemOpen
  \bibfield  {author} {\bibinfo {author} {\bibfnamefont {C.~J.~D.}\ \bibnamefont {Kemp}}, \bibinfo {author} {\bibfnamefont {N.~R.}\ \bibnamefont {Cooper}},\ and\ \bibinfo {author} {\bibfnamefont {F.~N.}\ \bibnamefont {{\"U}nal}},\ }\bibfield  {title} {\bibinfo {title} {Nested-sphere description of the \${{N}}\$-level {{Chern}} number and the generalized {{Bloch}} hypersphere},\ }\href {https://doi.org/10.1103/PhysRevResearch.4.023120} {\bibfield  {journal} {\bibinfo  {journal} {Physical Review Research}\ }\textbf {\bibinfo {volume} {4}},\ \bibinfo {pages} {023120} (\bibinfo {year} {2022})}\BibitemShut {NoStop}%
\bibitem [{\citenamefont {Devakul}\ \emph {et~al.}(2021)\citenamefont {Devakul}, \citenamefont {Crépel}, \citenamefont {Zhang},\ and\ \citenamefont {Fu}}]{fuDevakulMagicTwistedTransition2021}%
  \BibitemOpen
  \bibfield  {author} {\bibinfo {author} {\bibfnamefont {T.}~\bibnamefont {Devakul}}, \bibinfo {author} {\bibfnamefont {V.}~\bibnamefont {Crépel}}, \bibinfo {author} {\bibfnamefont {Y.}~\bibnamefont {Zhang}},\ and\ \bibinfo {author} {\bibfnamefont {L.}~\bibnamefont {Fu}},\ }\bibfield  {title} {\bibinfo {title} {Magic in twisted transition metal dichalcogenide bilayers},\ }\href {https://doi.org/10.1038/s41467-021-27042-9} {\bibfield  {journal} {\bibinfo  {journal} {Nature Communications}\ }\textbf {\bibinfo {volume} {12}},\ \bibinfo {pages} {6730} (\bibinfo {year} {2021})},\ \Eprint {https://arxiv.org/abs/2106.11954} {arXiv:2106.11954} \BibitemShut {NoStop}%
\bibitem [{\citenamefont {Li}\ and\ \citenamefont {Wu}(2025)}]{wuLiVariationalMappingChern2024}%
  \BibitemOpen
  \bibfield  {author} {\bibinfo {author} {\bibfnamefont {B.}~\bibnamefont {Li}}\ and\ \bibinfo {author} {\bibfnamefont {F.}~\bibnamefont {Wu}},\ }\bibfield  {title} {\bibinfo {title} {Variational mapping of chern bands to landau levels: Application to fractional chern insulators in twisted ${\mathrm{mote}}_{2}$},\ }\href {https://doi.org/10.1103/PhysRevB.111.125122} {\bibfield  {journal} {\bibinfo  {journal} {Phys. Rev. B}\ }\textbf {\bibinfo {volume} {111}},\ \bibinfo {pages} {125122} (\bibinfo {year} {2025})}\BibitemShut {NoStop}%
\bibitem [{\citenamefont {Shi}\ \emph {et~al.}(2024)\citenamefont {Shi}, \citenamefont {Morales-Durán}, \citenamefont {Khalaf},\ and\ \citenamefont {MacDonald}}]{macdonaldShiAdiabaticApproximationAharonovCasher2024}%
  \BibitemOpen
  \bibfield  {author} {\bibinfo {author} {\bibfnamefont {J.}~\bibnamefont {Shi}}, \bibinfo {author} {\bibfnamefont {N.}~\bibnamefont {Morales-Durán}}, \bibinfo {author} {\bibfnamefont {E.}~\bibnamefont {Khalaf}},\ and\ \bibinfo {author} {\bibfnamefont {A.~H.}\ \bibnamefont {MacDonald}},\ }\bibfield  {title} {\bibinfo {title} {Adiabatic approximation and aharonov-casher bands in twisted homobilayer transition metal dichalcogenides},\ }\href {https://doi.org/10.1103/PhysRevB.110.035130} {\bibfield  {journal} {\bibinfo  {journal} {Physical Review B}\ }\textbf {\bibinfo {volume} {110}},\ \bibinfo {pages} {035130} (\bibinfo {year} {2024})}\BibitemShut {NoStop}%
\bibitem [{\citenamefont {Bruno}\ \emph {et~al.}(2004)\citenamefont {Bruno}, \citenamefont {Dugaev},\ and\ \citenamefont {Taillefumier}}]{taillefumierBrunoTopologicalHallEffect2004}%
  \BibitemOpen
  \bibfield  {author} {\bibinfo {author} {\bibfnamefont {P.}~\bibnamefont {Bruno}}, \bibinfo {author} {\bibfnamefont {V.~K.}\ \bibnamefont {Dugaev}},\ and\ \bibinfo {author} {\bibfnamefont {M.}~\bibnamefont {Taillefumier}},\ }\bibfield  {title} {\bibinfo {title} {Topological hall effect and berry phase in magnetic nanostructures},\ }\href {https://doi.org/10.1103/PhysRevLett.93.096806} {\bibfield  {journal} {\bibinfo  {journal} {Physical Review Letters}\ }\textbf {\bibinfo {volume} {93}},\ \bibinfo {pages} {096806} (\bibinfo {year} {2004})}\BibitemShut {NoStop}%
\bibitem [{\citenamefont {Nakazawa}\ \emph {et~al.}(2019)\citenamefont {Nakazawa}, \citenamefont {Bibes},\ and\ \citenamefont {Kohno}}]{kohnoNakazawaTopologicalHallEffect2019}%
  \BibitemOpen
  \bibfield  {author} {\bibinfo {author} {\bibfnamefont {K.}~\bibnamefont {Nakazawa}}, \bibinfo {author} {\bibfnamefont {M.}~\bibnamefont {Bibes}},\ and\ \bibinfo {author} {\bibfnamefont {H.}~\bibnamefont {Kohno}},\ }\bibfield  {title} {\bibinfo {title} {Topological hall effect from strong to weak coupling},\ }\href {https://doi.org/10.1103/PhysRevB.99.174425} {\bibfield  {journal} {\bibinfo  {journal} {Physical Review B}\ }\textbf {\bibinfo {volume} {99}},\ \bibinfo {pages} {174425} (\bibinfo {year} {2019})},\ \Eprint {https://arxiv.org/abs/1711.08366} {arXiv:1711.08366 [cond-mat]} \BibitemShut {NoStop}%
\bibitem [{\citenamefont {Matsui}\ \emph {et~al.}(2021)\citenamefont {Matsui}, \citenamefont {Nomoto},\ and\ \citenamefont {Arita}}]{aritaMatsuiSkyrmionsizeDependenceTopological2021}%
  \BibitemOpen
  \bibfield  {author} {\bibinfo {author} {\bibfnamefont {A.}~\bibnamefont {Matsui}}, \bibinfo {author} {\bibfnamefont {T.}~\bibnamefont {Nomoto}},\ and\ \bibinfo {author} {\bibfnamefont {R.}~\bibnamefont {Arita}},\ }\bibfield  {title} {\bibinfo {title} {Skyrmion-size dependence of the topological hall effect: A real-space calculation},\ }\href {https://doi.org/10.1103/PhysRevB.104.174432} {\bibfield  {journal} {\bibinfo  {journal} {Physical Review B}\ }\textbf {\bibinfo {volume} {104}},\ \bibinfo {pages} {174432} (\bibinfo {year} {2021})}\BibitemShut {NoStop}%
\bibitem [{\citenamefont {Kol{\'a}{\v{r}}}\ \emph {et~al.}(2024)\citenamefont {Kol{\'a}{\v{r}}}, \citenamefont {Yang}, \citenamefont {von Oppen},\ and\ \citenamefont {Mora}}]{moraKolarHofstadterSpectrumChern2024}%
  \BibitemOpen
  \bibfield  {author} {\bibinfo {author} {\bibfnamefont {K.}~\bibnamefont {Kol{\'a}{\v{r}}}}, \bibinfo {author} {\bibfnamefont {K.}~\bibnamefont {Yang}}, \bibinfo {author} {\bibfnamefont {F.}~\bibnamefont {von Oppen}},\ and\ \bibinfo {author} {\bibfnamefont {C.}~\bibnamefont {Mora}},\ }\bibfield  {title} {\bibinfo {title} {Hofstadter spectrum of chern bands in twisted transition metal dichalcogenides},\ }\href {https://doi.org/10.1103/PhysRevB.110.115114} {\bibfield  {journal} {\bibinfo  {journal} {Phys. Rev. B}\ }\textbf {\bibinfo {volume} {110}},\ \bibinfo {pages} {115114} (\bibinfo {year} {2024})}\BibitemShut {NoStop}%
\bibitem [{\citenamefont {Roy}(2014)}]{royRoyBandGeometryFractional2014}%
  \BibitemOpen
  \bibfield  {author} {\bibinfo {author} {\bibfnamefont {R.}~\bibnamefont {Roy}},\ }\bibfield  {title} {\bibinfo {title} {Band geometry of fractional topological insulators},\ }\href {https://doi.org/10.1103/PhysRevB.90.165139} {\bibfield  {journal} {\bibinfo  {journal} {Physical Review B}\ }\textbf {\bibinfo {volume} {90}},\ \bibinfo {pages} {165139} (\bibinfo {year} {2014})}\BibitemShut {NoStop}%
\bibitem [{\citenamefont {Abouelkomsan}\ \emph {et~al.}(2023)\citenamefont {Abouelkomsan}, \citenamefont {Yang},\ and\ \citenamefont {Bergholtz}}]{bergholtzAbouelkomsanQuantumMetricInduced2023}%
  \BibitemOpen
  \bibfield  {author} {\bibinfo {author} {\bibfnamefont {A.}~\bibnamefont {Abouelkomsan}}, \bibinfo {author} {\bibfnamefont {K.}~\bibnamefont {Yang}},\ and\ \bibinfo {author} {\bibfnamefont {E.~J.}\ \bibnamefont {Bergholtz}},\ }\bibfield  {title} {\bibinfo {title} {Quantum metric induced phases in moir\'e materials},\ }\href {https://doi.org/10.1103/PhysRevResearch.5.L012015} {\bibfield  {journal} {\bibinfo  {journal} {Physical Review Research}\ }\textbf {\bibinfo {volume} {5}},\ \bibinfo {pages} {L012015} (\bibinfo {year} {2023})}\BibitemShut {NoStop}%
\bibitem [{\citenamefont {Fujimoto}\ \emph {et~al.}(2024)\citenamefont {Fujimoto}, \citenamefont {Parker}, \citenamefont {Dong}, \citenamefont {Khalaf}, \citenamefont {Vishwanath},\ and\ \citenamefont {Ledwith}}]{ledwithFujimotoHigherVortexabilityZero2024}%
  \BibitemOpen
  \bibfield  {author} {\bibinfo {author} {\bibfnamefont {M.}~\bibnamefont {Fujimoto}}, \bibinfo {author} {\bibfnamefont {D.~E.}\ \bibnamefont {Parker}}, \bibinfo {author} {\bibfnamefont {J.}~\bibnamefont {Dong}}, \bibinfo {author} {\bibfnamefont {E.}~\bibnamefont {Khalaf}}, \bibinfo {author} {\bibfnamefont {A.}~\bibnamefont {Vishwanath}},\ and\ \bibinfo {author} {\bibfnamefont {P.}~\bibnamefont {Ledwith}},\ }\href {https://doi.org/10.48550/arXiv.2403.00856} {\bibinfo {title} {Higher vortexability: zero field realization of higher landau levels}} (\bibinfo {year} {2024}),\ \Eprint {https://arxiv.org/abs/2403.00856} {arXiv:2403.00856 [cond-mat]} \BibitemShut {NoStop}%
\bibitem [{\citenamefont {Estienne}\ \emph {et~al.}(2023)\citenamefont {Estienne}, \citenamefont {Regnault},\ and\ \citenamefont {Crépel}}]{crepelEstienneIdealChernBands2023}%
  \BibitemOpen
  \bibfield  {author} {\bibinfo {author} {\bibfnamefont {B.}~\bibnamefont {Estienne}}, \bibinfo {author} {\bibfnamefont {N.}~\bibnamefont {Regnault}},\ and\ \bibinfo {author} {\bibfnamefont {V.}~\bibnamefont {Crépel}},\ }\bibfield  {title} {\bibinfo {title} {Ideal chern bands as landau levels in curved space},\ }\href {https://doi.org/10.1103/PhysRevResearch.5.L032048} {\bibfield  {journal} {\bibinfo  {journal} {Physical Review Research}\ }\textbf {\bibinfo {volume} {5}},\ \bibinfo {pages} {L032048} (\bibinfo {year} {2023})}\BibitemShut {NoStop}%
\bibitem [{\citenamefont {Liu}\ \emph {et~al.}(2025)\citenamefont {Liu}, \citenamefont {Yang}, \citenamefont {Abouelkomsan}, \citenamefont {Liu},\ and\ \citenamefont {Bergholtz}}]{PhysRevB.111.L201105}%
  \BibitemOpen
  \bibfield  {author} {\bibinfo {author} {\bibfnamefont {H.}~\bibnamefont {Liu}}, \bibinfo {author} {\bibfnamefont {K.}~\bibnamefont {Yang}}, \bibinfo {author} {\bibfnamefont {A.}~\bibnamefont {Abouelkomsan}}, \bibinfo {author} {\bibfnamefont {Z.}~\bibnamefont {Liu}},\ and\ \bibinfo {author} {\bibfnamefont {E.~J.}\ \bibnamefont {Bergholtz}},\ }\bibfield  {title} {\bibinfo {title} {Broken symmetry in ideal chern bands},\ }\href {https://doi.org/10.1103/PhysRevB.111.L201105} {\bibfield  {journal} {\bibinfo  {journal} {Phys. Rev. B}\ }\textbf {\bibinfo {volume} {111}},\ \bibinfo {pages} {L201105} (\bibinfo {year} {2025})}\BibitemShut {NoStop}%
\bibitem [{\citenamefont {Wan}\ \emph {et~al.}(2023)\citenamefont {Wan}, \citenamefont {Sarkar}, \citenamefont {Lin},\ and\ \citenamefont {Sun}}]{sunWanTopologicalExactFlat2023}%
  \BibitemOpen
  \bibfield  {author} {\bibinfo {author} {\bibfnamefont {X.}~\bibnamefont {Wan}}, \bibinfo {author} {\bibfnamefont {S.}~\bibnamefont {Sarkar}}, \bibinfo {author} {\bibfnamefont {S.-Z.}\ \bibnamefont {Lin}},\ and\ \bibinfo {author} {\bibfnamefont {K.}~\bibnamefont {Sun}},\ }\bibfield  {title} {\bibinfo {title} {Topological exact flat bands in two-dimensional materials under periodic strain},\ }\href {https://doi.org/10.1103/PhysRevLett.130.216401} {\bibfield  {journal} {\bibinfo  {journal} {Physical Review Letters}\ }\textbf {\bibinfo {volume} {130}},\ \bibinfo {pages} {216401} (\bibinfo {year} {2023})}\BibitemShut {NoStop}%
\bibitem [{\citenamefont {Eugenio}\ and\ \citenamefont {Vafek}(2023)}]{Eugenio_2023}%
  \BibitemOpen
  \bibfield  {author} {\bibinfo {author} {\bibfnamefont {P.~M.}\ \bibnamefont {Eugenio}}\ and\ \bibinfo {author} {\bibfnamefont {O.}~\bibnamefont {Vafek}},\ }\bibfield  {title} {\bibinfo {title} {{Twisted-bilayer FeSe and the Fe-based superlattices}},\ }\href {https://doi.org/10.21468/SciPostPhys.15.3.081} {\bibfield  {journal} {\bibinfo  {journal} {SciPost Phys.}\ }\textbf {\bibinfo {volume} {15}},\ \bibinfo {pages} {081} (\bibinfo {year} {2023})}\BibitemShut {NoStop}%
\bibitem [{\citenamefont {Sheffer}\ and\ \citenamefont {Stern}(2021)}]{sternShefferChiralMagicangleTwisted2021}%
  \BibitemOpen
  \bibfield  {author} {\bibinfo {author} {\bibfnamefont {Y.}~\bibnamefont {Sheffer}}\ and\ \bibinfo {author} {\bibfnamefont {A.}~\bibnamefont {Stern}},\ }\bibfield  {title} {\bibinfo {title} {Chiral magic-angle twisted bilayer graphene in a magnetic field: Landau level correspondence, exact wave functions, and fractional chern insulators},\ }\href {https://doi.org/10.1103/PhysRevB.104.L121405} {\bibfield  {journal} {\bibinfo  {journal} {Physical Review B}\ }\textbf {\bibinfo {volume} {104}},\ \bibinfo {pages} {L121405} (\bibinfo {year} {2021})}\BibitemShut {NoStop}%
\bibitem [{\citenamefont {Dong}\ \emph {et~al.}(2023{\natexlab{b}})\citenamefont {Dong}, \citenamefont {Ledwith}, \citenamefont {Khalaf}, \citenamefont {Lee},\ and\ \citenamefont {Vishwanath}}]{vishwanathDongExactManyBodyGround2022}%
  \BibitemOpen
  \bibfield  {author} {\bibinfo {author} {\bibfnamefont {J.}~\bibnamefont {Dong}}, \bibinfo {author} {\bibfnamefont {P.~J.}\ \bibnamefont {Ledwith}}, \bibinfo {author} {\bibfnamefont {E.}~\bibnamefont {Khalaf}}, \bibinfo {author} {\bibfnamefont {J.~Y.}\ \bibnamefont {Lee}},\ and\ \bibinfo {author} {\bibfnamefont {A.}~\bibnamefont {Vishwanath}},\ }\bibfield  {title} {\bibinfo {title} {Many-body ground states from decomposition of ideal higher chern bands: Applications to chirally twisted graphene multilayers},\ }\href {https://doi.org/10.1103/PhysRevResearch.5.023166} {\bibfield  {journal} {\bibinfo  {journal} {Phys. Rev. Res.}\ }\textbf {\bibinfo {volume} {5}},\ \bibinfo {pages} {023166} (\bibinfo {year} {2023}{\natexlab{b}})}\BibitemShut {NoStop}%
\bibitem [{\citenamefont {Ledwith}\ \emph {et~al.}(2022{\natexlab{a}})\citenamefont {Ledwith}, \citenamefont {Vishwanath},\ and\ \citenamefont {Khalaf}}]{khalafLedwithFamilyIdealChern2022}%
  \BibitemOpen
  \bibfield  {author} {\bibinfo {author} {\bibfnamefont {P.~J.}\ \bibnamefont {Ledwith}}, \bibinfo {author} {\bibfnamefont {A.}~\bibnamefont {Vishwanath}},\ and\ \bibinfo {author} {\bibfnamefont {E.}~\bibnamefont {Khalaf}},\ }\bibfield  {title} {\bibinfo {title} {Family of {{Ideal Chern Flatbands}} with {{Arbitrary Chern Number}} in {{Chiral Twisted Graphene Multilayers}}},\ }\href {https://doi.org/10.1103/PhysRevLett.128.176404} {\bibfield  {journal} {\bibinfo  {journal} {Physical Review Letters}\ }\textbf {\bibinfo {volume} {128}},\ \bibinfo {pages} {176404} (\bibinfo {year} {2022}{\natexlab{a}})}\BibitemShut {NoStop}%
\bibitem [{\citenamefont {Guerci}\ \emph {et~al.}(2024{\natexlab{b}})\citenamefont {Guerci}, \citenamefont {Mao},\ and\ \citenamefont {Mora}}]{moraGuerciChernMosaicIdeal2024}%
  \BibitemOpen
  \bibfield  {author} {\bibinfo {author} {\bibfnamefont {D.}~\bibnamefont {Guerci}}, \bibinfo {author} {\bibfnamefont {Y.}~\bibnamefont {Mao}},\ and\ \bibinfo {author} {\bibfnamefont {C.}~\bibnamefont {Mora}},\ }\bibfield  {title} {\bibinfo {title} {Chern mosaic and ideal flat bands in equal-twist trilayer graphene},\ }\href {https://doi.org/10.1103/PhysRevResearch.6.L022025} {\bibfield  {journal} {\bibinfo  {journal} {Physical Review Research}\ }\textbf {\bibinfo {volume} {6}},\ \bibinfo {pages} {L022025} (\bibinfo {year} {2024}{\natexlab{b}})}\BibitemShut {NoStop}%
\bibitem [{\citenamefont {Guerci}\ \emph {et~al.}(2024{\natexlab{c}})\citenamefont {Guerci}, \citenamefont {Mao},\ and\ \citenamefont {Mora}}]{guerci2023nature}%
  \BibitemOpen
  \bibfield  {author} {\bibinfo {author} {\bibfnamefont {D.}~\bibnamefont {Guerci}}, \bibinfo {author} {\bibfnamefont {Y.}~\bibnamefont {Mao}},\ and\ \bibinfo {author} {\bibfnamefont {C.}~\bibnamefont {Mora}},\ }\bibfield  {title} {\bibinfo {title} {Nature of even and odd magic angles in helical twisted trilayer graphene},\ }\href {https://doi.org/10.1103/PhysRevB.109.205411} {\bibfield  {journal} {\bibinfo  {journal} {Phys. Rev. B}\ }\textbf {\bibinfo {volume} {109}},\ \bibinfo {pages} {205411} (\bibinfo {year} {2024}{\natexlab{c}})}\BibitemShut {NoStop}%
\bibitem [{\citenamefont {Choi}\ \emph {et~al.}(2019)\citenamefont {Choi}, \citenamefont {Kemmer}, \citenamefont {Peng}, \citenamefont {Thomson}, \citenamefont {Arora}, \citenamefont {Polski}, \citenamefont {Zhang}, \citenamefont {Ren}, \citenamefont {Alicea}, \citenamefont {Refael}, \citenamefont {von Oppen}, \citenamefont {Watanabe}, \citenamefont {Taniguchi},\ and\ \citenamefont {Nadj-Perge}}]{nadj-pergeChoiElectronicCorrelationsTwisted2019}%
  \BibitemOpen
  \bibfield  {author} {\bibinfo {author} {\bibfnamefont {Y.}~\bibnamefont {Choi}}, \bibinfo {author} {\bibfnamefont {J.}~\bibnamefont {Kemmer}}, \bibinfo {author} {\bibfnamefont {Y.}~\bibnamefont {Peng}}, \bibinfo {author} {\bibfnamefont {A.}~\bibnamefont {Thomson}}, \bibinfo {author} {\bibfnamefont {H.}~\bibnamefont {Arora}}, \bibinfo {author} {\bibfnamefont {R.}~\bibnamefont {Polski}}, \bibinfo {author} {\bibfnamefont {Y.}~\bibnamefont {Zhang}}, \bibinfo {author} {\bibfnamefont {H.}~\bibnamefont {Ren}}, \bibinfo {author} {\bibfnamefont {J.}~\bibnamefont {Alicea}}, \bibinfo {author} {\bibfnamefont {G.}~\bibnamefont {Refael}}, \bibinfo {author} {\bibfnamefont {F.}~\bibnamefont {von Oppen}}, \bibinfo {author} {\bibfnamefont {K.}~\bibnamefont {Watanabe}}, \bibinfo {author} {\bibfnamefont {T.}~\bibnamefont {Taniguchi}},\ and\ \bibinfo {author} {\bibfnamefont {S.}~\bibnamefont {Nadj-Perge}},\ }\bibfield  {title} {\bibinfo {title} {Electronic correlations in twisted bilayer graphene near the magic angle},\ }\href {https://doi.org/10.1038/s41567-019-0606-5} {\bibfield  {journal} {\bibinfo  {journal} {Nature Physics}\ }\textbf {\bibinfo {volume} {15}},\ \bibinfo {pages} {1174} (\bibinfo {year} {2019})}\BibitemShut {NoStop}%
\bibitem [{\citenamefont {Kerelsky}\ \emph {et~al.}(2019)\citenamefont {Kerelsky}, \citenamefont {McGilly}, \citenamefont {Kennes}, \citenamefont {Xian}, \citenamefont {Yankowitz}, \citenamefont {Chen}, \citenamefont {Watanabe}, \citenamefont {Taniguchi}, \citenamefont {Hone}, \citenamefont {Dean}, \citenamefont {Rubio},\ and\ \citenamefont {Pasupathy}}]{pasupathyKerelskyMaximizedElectronInteractions2019}%
  \BibitemOpen
  \bibfield  {author} {\bibinfo {author} {\bibfnamefont {A.}~\bibnamefont {Kerelsky}}, \bibinfo {author} {\bibfnamefont {L.~J.}\ \bibnamefont {McGilly}}, \bibinfo {author} {\bibfnamefont {D.~M.}\ \bibnamefont {Kennes}}, \bibinfo {author} {\bibfnamefont {L.}~\bibnamefont {Xian}}, \bibinfo {author} {\bibfnamefont {M.}~\bibnamefont {Yankowitz}}, \bibinfo {author} {\bibfnamefont {S.}~\bibnamefont {Chen}}, \bibinfo {author} {\bibfnamefont {K.}~\bibnamefont {Watanabe}}, \bibinfo {author} {\bibfnamefont {T.}~\bibnamefont {Taniguchi}}, \bibinfo {author} {\bibfnamefont {J.}~\bibnamefont {Hone}}, \bibinfo {author} {\bibfnamefont {C.}~\bibnamefont {Dean}}, \bibinfo {author} {\bibfnamefont {A.}~\bibnamefont {Rubio}},\ and\ \bibinfo {author} {\bibfnamefont {A.~N.}\ \bibnamefont {Pasupathy}},\ }\bibfield  {title} {\bibinfo {title} {Maximized electron interactions at the magic angle in twisted bilayer graphene},\ }\href {https://doi.org/10.1038/s41586-019-1431-9} {\bibfield  {journal} {\bibinfo  {journal} {Nature}\ }\textbf {\bibinfo {volume} {572}},\ \bibinfo {pages} {95} (\bibinfo {year} {2019})}\BibitemShut {NoStop}%
\bibitem [{\citenamefont {C\ifmmode \u{a}\else \u{a}\fi{}lug\ifmmode~\u{a}\else \u{a}\fi{}ru}\ \emph {et~al.}(2022)\citenamefont {C\ifmmode \u{a}\else \u{a}\fi{}lug\ifmmode~\u{a}\else \u{a}\fi{}ru}, \citenamefont {Regnault}, \citenamefont {Oh}, \citenamefont {Nuckolls}, \citenamefont {Wong}, \citenamefont {Lee}, \citenamefont {Yazdani}, \citenamefont {Vafek},\ and\ \citenamefont {Bernevig}}]{bernevigCalugaruSpectroscopyTwistedBilayer2021}%
  \BibitemOpen
  \bibfield  {author} {\bibinfo {author} {\bibfnamefont {D.}~\bibnamefont {C\ifmmode \u{a}\else \u{a}\fi{}lug\ifmmode~\u{a}\else \u{a}\fi{}ru}}, \bibinfo {author} {\bibfnamefont {N.}~\bibnamefont {Regnault}}, \bibinfo {author} {\bibfnamefont {M.}~\bibnamefont {Oh}}, \bibinfo {author} {\bibfnamefont {K.~P.}\ \bibnamefont {Nuckolls}}, \bibinfo {author} {\bibfnamefont {D.}~\bibnamefont {Wong}}, \bibinfo {author} {\bibfnamefont {R.~L.}\ \bibnamefont {Lee}}, \bibinfo {author} {\bibfnamefont {A.}~\bibnamefont {Yazdani}}, \bibinfo {author} {\bibfnamefont {O.}~\bibnamefont {Vafek}},\ and\ \bibinfo {author} {\bibfnamefont {B.~A.}\ \bibnamefont {Bernevig}},\ }\bibfield  {title} {\bibinfo {title} {Spectroscopy of twisted bilayer graphene correlated insulators},\ }\href {https://doi.org/10.1103/PhysRevLett.129.117602} {\bibfield  {journal} {\bibinfo  {journal} {Phys. Rev. Lett.}\ }\textbf {\bibinfo {volume} {129}},\ \bibinfo {pages} {117602} (\bibinfo {year} {2022})}\BibitemShut {NoStop}%
\bibitem [{\citenamefont {Hong}\ \emph {et~al.}(2021)\citenamefont {Hong}, \citenamefont {Soejima},\ and\ \citenamefont {Zaletel}}]{zaletelHongDetectingSymmetryBreaking2021}%
  \BibitemOpen
  \bibfield  {author} {\bibinfo {author} {\bibfnamefont {J.~P.}\ \bibnamefont {Hong}}, \bibinfo {author} {\bibfnamefont {T.}~\bibnamefont {Soejima}},\ and\ \bibinfo {author} {\bibfnamefont {M.~P.}\ \bibnamefont {Zaletel}},\ }\bibfield  {title} {\bibinfo {title} {Detecting symmetry breaking in magic angle graphene using scanning tunneling microscopy},\ }\href@noop {} {\bibfield  {journal} {\bibinfo  {journal} {arXiv:2110.14674 [cond-mat]}\ } (\bibinfo {year} {2021})},\ \Eprint {https://arxiv.org/abs/2110.14674} {arXiv:2110.14674 [cond-mat]} \BibitemShut {NoStop}%
\bibitem [{\citenamefont {Nuckolls}\ \emph {et~al.}(2023)\citenamefont {Nuckolls}, \citenamefont {Lee}, \citenamefont {Oh}, \citenamefont {Wong}, \citenamefont {Soejima}, \citenamefont {Hong}, \citenamefont {Călugăru}, \citenamefont {Herzog-Arbeitman}, \citenamefont {Bernevig}, \citenamefont {Watanabe}, \citenamefont {Taniguchi}, \citenamefont {Regnault}, \citenamefont {Zaletel},\ and\ \citenamefont {Yazdani}}]{yazdaniNuckollsQuantumTexturesManybody2023}%
  \BibitemOpen
  \bibfield  {author} {\bibinfo {author} {\bibfnamefont {K.~P.}\ \bibnamefont {Nuckolls}}, \bibinfo {author} {\bibfnamefont {R.~L.}\ \bibnamefont {Lee}}, \bibinfo {author} {\bibfnamefont {M.}~\bibnamefont {Oh}}, \bibinfo {author} {\bibfnamefont {D.}~\bibnamefont {Wong}}, \bibinfo {author} {\bibfnamefont {T.}~\bibnamefont {Soejima}}, \bibinfo {author} {\bibfnamefont {J.~P.}\ \bibnamefont {Hong}}, \bibinfo {author} {\bibfnamefont {D.}~\bibnamefont {Călugăru}}, \bibinfo {author} {\bibfnamefont {J.}~\bibnamefont {Herzog-Arbeitman}}, \bibinfo {author} {\bibfnamefont {B.~A.}\ \bibnamefont {Bernevig}}, \bibinfo {author} {\bibfnamefont {K.}~\bibnamefont {Watanabe}}, \bibinfo {author} {\bibfnamefont {T.}~\bibnamefont {Taniguchi}}, \bibinfo {author} {\bibfnamefont {N.}~\bibnamefont {Regnault}}, \bibinfo {author} {\bibfnamefont {M.~P.}\ \bibnamefont {Zaletel}},\ and\ \bibinfo {author} {\bibfnamefont {A.}~\bibnamefont {Yazdani}},\ }\bibfield  {title} {\bibinfo {title} {Quantum textures of the many-body wavefunctions in magic-angle graphene},\ }\href {https://doi.org/10.1038/s41586-023-06226-x} {\bibfield  {journal} {\bibinfo  {journal} {Nature}\ }\textbf {\bibinfo {volume} {620}},\ \bibinfo {pages} {525} (\bibinfo {year} {2023})}\BibitemShut {NoStop}%
\bibitem [{\citenamefont {Kim}\ \emph {et~al.}(2023)\citenamefont {Kim}, \citenamefont {Choi}, \citenamefont {Lantagne-Hurtubise}, \citenamefont {Lewandowski}, \citenamefont {Thomson}, \citenamefont {Kong}, \citenamefont {Zhou}, \citenamefont {Baum}, \citenamefont {Zhang}, \citenamefont {Holleis}, \citenamefont {Watanabe}, \citenamefont {Taniguchi}, \citenamefont {Young}, \citenamefont {Alicea},\ and\ \citenamefont {Nadj-Perge}}]{nadj-pergeKimImagingIntervalleyCoherent2023}%
  \BibitemOpen
  \bibfield  {author} {\bibinfo {author} {\bibfnamefont {H.}~\bibnamefont {Kim}}, \bibinfo {author} {\bibfnamefont {Y.}~\bibnamefont {Choi}}, \bibinfo {author} {\bibfnamefont {Ã.}~\bibnamefont {Lantagne-Hurtubise}}, \bibinfo {author} {\bibfnamefont {C.}~\bibnamefont {Lewandowski}}, \bibinfo {author} {\bibfnamefont {A.}~\bibnamefont {Thomson}}, \bibinfo {author} {\bibfnamefont {L.}~\bibnamefont {Kong}}, \bibinfo {author} {\bibfnamefont {H.}~\bibnamefont {Zhou}}, \bibinfo {author} {\bibfnamefont {E.}~\bibnamefont {Baum}}, \bibinfo {author} {\bibfnamefont {Y.}~\bibnamefont {Zhang}}, \bibinfo {author} {\bibfnamefont {L.}~\bibnamefont {Holleis}}, \bibinfo {author} {\bibfnamefont {K.}~\bibnamefont {Watanabe}}, \bibinfo {author} {\bibfnamefont {T.}~\bibnamefont {Taniguchi}}, \bibinfo {author} {\bibfnamefont {A.~F.}\ \bibnamefont {Young}}, \bibinfo {author} {\bibfnamefont {J.}~\bibnamefont {Alicea}},\ and\ \bibinfo {author} {\bibfnamefont {S.}~\bibnamefont {Nadj-Perge}},\ }\bibfield  {title} {\bibinfo {title} {Imaging inter-valley coherent order in magic-angle twisted trilayer graphene},\ }\href {https://doi.org/10.1038/s41586-023-06663-8} {\bibfield  {journal} {\bibinfo  {journal} {Nature}\ }\textbf {\bibinfo {volume} {623}},\ \bibinfo {pages} {1} (\bibinfo {year} {2023})}\BibitemShut {NoStop}%
\bibitem [{\citenamefont {Liu}\ \emph {et~al.}(2021{\natexlab{b}})\citenamefont {Liu}, \citenamefont {Farahi}, \citenamefont {Chiu}, \citenamefont {Papic}, \citenamefont {Watanabe}, \citenamefont {Taniguchi}, \citenamefont {Zaletel},\ and\ \citenamefont {Yazdani}}]{yazdaniLiuVisualizingBrokenSymmetry2021}%
  \BibitemOpen
  \bibfield  {author} {\bibinfo {author} {\bibfnamefont {X.}~\bibnamefont {Liu}}, \bibinfo {author} {\bibfnamefont {G.}~\bibnamefont {Farahi}}, \bibinfo {author} {\bibfnamefont {C.-L.}\ \bibnamefont {Chiu}}, \bibinfo {author} {\bibfnamefont {Z.}~\bibnamefont {Papic}}, \bibinfo {author} {\bibfnamefont {K.}~\bibnamefont {Watanabe}}, \bibinfo {author} {\bibfnamefont {T.}~\bibnamefont {Taniguchi}}, \bibinfo {author} {\bibfnamefont {M.~P.}\ \bibnamefont {Zaletel}},\ and\ \bibinfo {author} {\bibfnamefont {A.}~\bibnamefont {Yazdani}},\ }\bibfield  {title} {\bibinfo {title} {Visualizing {{Broken Symmetry}} and {{Topological Defects}} in a {{Quantum Hall Ferromagnet}}},\ }\href@noop {} {\bibfield  {journal} {\bibinfo  {journal} {arXiv:2109.11555 [cond-mat]}\ } (\bibinfo {year} {2021}{\natexlab{b}})},\ \Eprint {https://arxiv.org/abs/2109.11555} {arXiv:2109.11555 [cond-mat]} \BibitemShut {NoStop}%
\bibitem [{\citenamefont {Fang}\ \emph {et~al.}(2012)\citenamefont {Fang}, \citenamefont {Gilbert},\ and\ \citenamefont {Bernevig}}]{bernevigFangBulkTopologicalInvariants2012}%
  \BibitemOpen
  \bibfield  {author} {\bibinfo {author} {\bibfnamefont {C.}~\bibnamefont {Fang}}, \bibinfo {author} {\bibfnamefont {M.~J.}\ \bibnamefont {Gilbert}},\ and\ \bibinfo {author} {\bibfnamefont {B.~A.}\ \bibnamefont {Bernevig}},\ }\bibfield  {title} {\bibinfo {title} {Bulk topological invariants in noninteracting point group symmetric insulators},\ }\href {https://doi.org/10.1103/PhysRevB.86.115112} {\bibfield  {journal} {\bibinfo  {journal} {Physical Review B}\ }\textbf {\bibinfo {volume} {86}},\ \bibinfo {pages} {115112} (\bibinfo {year} {2012})}\BibitemShut {NoStop}%
\bibitem [{\citenamefont {Reddy}\ \emph {et~al.}(2024)\citenamefont {Reddy}, \citenamefont {Paul}, \citenamefont {Abouelkomsan},\ and\ \citenamefont {Fu}}]{fuReddyNonAbelianFractionalizationTopological2024}%
  \BibitemOpen
  \bibfield  {author} {\bibinfo {author} {\bibfnamefont {A.~P.}\ \bibnamefont {Reddy}}, \bibinfo {author} {\bibfnamefont {N.}~\bibnamefont {Paul}}, \bibinfo {author} {\bibfnamefont {A.}~\bibnamefont {Abouelkomsan}},\ and\ \bibinfo {author} {\bibfnamefont {L.}~\bibnamefont {Fu}},\ }\bibfield  {title} {\bibinfo {title} {Non-abelian fractionalization in topological minibands},\ }\href {https://doi.org/10.1103/PhysRevLett.133.166503} {\bibfield  {journal} {\bibinfo  {journal} {Phys. Rev. Lett.}\ }\textbf {\bibinfo {volume} {133}},\ \bibinfo {pages} {166503} (\bibinfo {year} {2024})}\BibitemShut {NoStop}%
\bibitem [{\citenamefont {Xu}\ \emph {et~al.}(2024)\citenamefont {Xu}, \citenamefont {Mao}, \citenamefont {Zeng},\ and\ \citenamefont {Zhang}}]{zhangXuMultipleChernBands2024}%
  \BibitemOpen
  \bibfield  {author} {\bibinfo {author} {\bibfnamefont {C.}~\bibnamefont {Xu}}, \bibinfo {author} {\bibfnamefont {N.}~\bibnamefont {Mao}}, \bibinfo {author} {\bibfnamefont {T.}~\bibnamefont {Zeng}},\ and\ \bibinfo {author} {\bibfnamefont {Y.}~\bibnamefont {Zhang}},\ }\href {https://doi.org/10.48550/arXiv.2403.17003} {\bibinfo {title} {Multiple chern bands in twisted mote$_2$ and possible non-abelian states}} (\bibinfo {year} {2024}),\ \Eprint {https://arxiv.org/abs/2403.17003} {arXiv:2403.17003 [cond-mat]} \BibitemShut {NoStop}%
\bibitem [{\citenamefont {Zhang}\ \emph {et~al.}(2024)\citenamefont {Zhang}, \citenamefont {Wang}, \citenamefont {Liu}, \citenamefont {Fan}, \citenamefont {Cao},\ and\ \citenamefont {Xiao}}]{xiaoZhangPolarizationdrivenBandTopology2024}%
  \BibitemOpen
  \bibfield  {author} {\bibinfo {author} {\bibfnamefont {X.-W.}\ \bibnamefont {Zhang}}, \bibinfo {author} {\bibfnamefont {C.}~\bibnamefont {Wang}}, \bibinfo {author} {\bibfnamefont {X.}~\bibnamefont {Liu}}, \bibinfo {author} {\bibfnamefont {Y.}~\bibnamefont {Fan}}, \bibinfo {author} {\bibfnamefont {T.}~\bibnamefont {Cao}},\ and\ \bibinfo {author} {\bibfnamefont {D.}~\bibnamefont {Xiao}},\ }\href {https://doi.org/10.48550/arXiv.2311.12776} {\bibinfo {title} {Polarization-driven band topology evolution in twisted mote$_2$ and wse$_2$}} (\bibinfo {year} {2024}),\ \Eprint {https://arxiv.org/abs/2311.12776} {arXiv:2311.12776 [cond-mat]} \BibitemShut {NoStop}%
\bibitem [{\citenamefont {Chen}\ \emph {et~al.}(2025)\citenamefont {Chen}, \citenamefont {Luo}, \citenamefont {Zhu},\ and\ \citenamefont {Sheng}}]{chen2025robust}%
  \BibitemOpen
  \bibfield  {author} {\bibinfo {author} {\bibfnamefont {F.}~\bibnamefont {Chen}}, \bibinfo {author} {\bibfnamefont {W.-W.}\ \bibnamefont {Luo}}, \bibinfo {author} {\bibfnamefont {W.}~\bibnamefont {Zhu}},\ and\ \bibinfo {author} {\bibfnamefont {D.}~\bibnamefont {Sheng}},\ }\bibfield  {title} {\bibinfo {title} {Robust non-abelian even-denominator fractional chern insulator in twisted bilayer mote2},\ }\href {https://www.nature.com/articles/s41467-025-57326-3} {\bibfield  {journal} {\bibinfo  {journal} {Nature Communications}\ }\textbf {\bibinfo {volume} {16}},\ \bibinfo {pages} {2115} (\bibinfo {year} {2025})}\BibitemShut {NoStop}%
\bibitem [{\citenamefont {Ledwith}\ \emph {et~al.}(2022{\natexlab{b}})\citenamefont {Ledwith}, \citenamefont {Vishwanath},\ and\ \citenamefont {Parker}}]{parkerLedwithVortexabilityUnifyingCriterion2022}%
  \BibitemOpen
  \bibfield  {author} {\bibinfo {author} {\bibfnamefont {P.~J.}\ \bibnamefont {Ledwith}}, \bibinfo {author} {\bibfnamefont {A.}~\bibnamefont {Vishwanath}},\ and\ \bibinfo {author} {\bibfnamefont {D.~E.}\ \bibnamefont {Parker}},\ }\href {https://doi.org/10.48550/arXiv.2209.15023} {\bibinfo {title} {Vortexability: {{A Unifying Criterion}} for {{Ideal Fractional Chern Insulators}}}} (\bibinfo {year} {2022}{\natexlab{b}}),\ \Eprint {https://arxiv.org/abs/2209.15023} {arXiv:2209.15023 [cond-mat]} \BibitemShut {NoStop}%
\bibitem [{\citenamefont {Yu}\ \emph {et~al.}(2023)\citenamefont {Yu}, \citenamefont {Xie}, \citenamefont {Bernevig},\ and\ \citenamefont {Sarma}}]{Yu2023}%
  \BibitemOpen
  \bibfield  {author} {\bibinfo {author} {\bibfnamefont {J.}~\bibnamefont {Yu}}, \bibinfo {author} {\bibfnamefont {M.}~\bibnamefont {Xie}}, \bibinfo {author} {\bibfnamefont {B.~A.}\ \bibnamefont {Bernevig}},\ and\ \bibinfo {author} {\bibfnamefont {S.~D.}\ \bibnamefont {Sarma}},\ }\bibfield  {title} {\bibinfo {title} {Magic-angle twisted symmetric trilayer graphene as topological heavy fermion problem},\ }\href {https://doi.org/10.1103/PhysRevB.108.035129} {\bibfield  {journal} {\bibinfo  {journal} {Physical Review B}\ }\textbf {\bibinfo {volume} {108}},\ \bibinfo {pages} {035129} (\bibinfo {year} {2023})}\BibitemShut {NoStop}%
\bibitem [{\citenamefont {Călugăru}\ \emph {et~al.}(2024)\citenamefont {Călugăru}, \citenamefont {Hu}, \citenamefont {Merino}, \citenamefont {Regnault}, \citenamefont {Efetov},\ and\ \citenamefont {Bernevig}}]{Calugaru2024}%
  \BibitemOpen
  \bibfield  {author} {\bibinfo {author} {\bibfnamefont {D.}~\bibnamefont {Călugăru}}, \bibinfo {author} {\bibfnamefont {H.}~\bibnamefont {Hu}}, \bibinfo {author} {\bibfnamefont {R.~L.}\ \bibnamefont {Merino}}, \bibinfo {author} {\bibfnamefont {N.}~\bibnamefont {Regnault}}, \bibinfo {author} {\bibfnamefont {D.~K.}\ \bibnamefont {Efetov}},\ and\ \bibinfo {author} {\bibfnamefont {B.~A.}\ \bibnamefont {Bernevig}},\ }\href@noop {} {\bibinfo {title} {The thermoelectric effect and its natural heavy fermion explanation in twisted bilayer and trilayer graphene}},\ \bibinfo {howpublished} {arXiv e-prints} (\bibinfo {year} {2024}),\ \Eprint {https://arxiv.org/abs/2402.14057} {arXiv:2402.14057 [cond-mat.str-el]} \BibitemShut {NoStop}%
\bibitem [{\citenamefont {Zhou}\ \emph {et~al.}(2024)\citenamefont {Zhou}, \citenamefont {Wang}, \citenamefont {Tong},\ and\ \citenamefont {Song}}]{Zhou2024}%
  \BibitemOpen
  \bibfield  {author} {\bibinfo {author} {\bibfnamefont {G.-D.}\ \bibnamefont {Zhou}}, \bibinfo {author} {\bibfnamefont {Y.-J.}\ \bibnamefont {Wang}}, \bibinfo {author} {\bibfnamefont {N.}~\bibnamefont {Tong}},\ and\ \bibinfo {author} {\bibfnamefont {Z.-D.}\ \bibnamefont {Song}},\ }\bibfield  {title} {\bibinfo {title} {Kondo phase in twisted bilayer graphene},\ }\href {https://doi.org/10.1103/PhysRevB.109.045419} {\bibfield  {journal} {\bibinfo  {journal} {Physical Review B}\ }\textbf {\bibinfo {volume} {109}},\ \bibinfo {pages} {045419} (\bibinfo {year} {2024})}\BibitemShut {NoStop}%
\bibitem [{\citenamefont {Rai}\ \emph {et~al.}(2024)\citenamefont {Rai}, \citenamefont {Crippa}, \citenamefont {Călugăru}, \citenamefont {Hu}, \citenamefont {Paoletti}, \citenamefont {de' Medici}, \citenamefont {Georges}, \citenamefont {Bernevig}, \citenamefont {Valentí}, \citenamefont {Sangiovanni},\ and\ \citenamefont {Wehling}}]{Rai2024}%
  \BibitemOpen
  \bibfield  {author} {\bibinfo {author} {\bibfnamefont {G.}~\bibnamefont {Rai}}, \bibinfo {author} {\bibfnamefont {L.}~\bibnamefont {Crippa}}, \bibinfo {author} {\bibfnamefont {D.}~\bibnamefont {Călugăru}}, \bibinfo {author} {\bibfnamefont {H.}~\bibnamefont {Hu}}, \bibinfo {author} {\bibfnamefont {F.}~\bibnamefont {Paoletti}}, \bibinfo {author} {\bibfnamefont {L.}~\bibnamefont {de' Medici}}, \bibinfo {author} {\bibfnamefont {A.}~\bibnamefont {Georges}}, \bibinfo {author} {\bibfnamefont {B.~A.}\ \bibnamefont {Bernevig}}, \bibinfo {author} {\bibfnamefont {R.}~\bibnamefont {Valentí}}, \bibinfo {author} {\bibfnamefont {G.}~\bibnamefont {Sangiovanni}},\ and\ \bibinfo {author} {\bibfnamefont {T.}~\bibnamefont {Wehling}},\ }\bibfield  {title} {\bibinfo {title} {Dynamical correlations and order in magic-angle twisted bilayer graphene},\ }\href {https://doi.org/10.1103/PhysRevX.14.031045} {\bibfield  {journal} {\bibinfo  {journal} {Physical Review X}\ }\textbf {\bibinfo {volume} {14}},\ \bibinfo {pages} {031045} (\bibinfo {year} {2024})}\BibitemShut {NoStop}%
\bibitem [{\citenamefont {Herzog-Arbeitman}\ \emph {et~al.}(2025)\citenamefont {Herzog-Arbeitman}, \citenamefont {C{\u{a}}lug{\u{a}}ru}, \citenamefont {Hu}, \citenamefont {Yu}, \citenamefont {Regnault}, \citenamefont {Kang}, \citenamefont {Bernevig},\ and\ \citenamefont {Vafek}}]{herzog2025kekul}%
  \BibitemOpen
  \bibfield  {author} {\bibinfo {author} {\bibfnamefont {J.}~\bibnamefont {Herzog-Arbeitman}}, \bibinfo {author} {\bibfnamefont {D.}~\bibnamefont {C{\u{a}}lug{\u{a}}ru}}, \bibinfo {author} {\bibfnamefont {H.}~\bibnamefont {Hu}}, \bibinfo {author} {\bibfnamefont {J.}~\bibnamefont {Yu}}, \bibinfo {author} {\bibfnamefont {N.}~\bibnamefont {Regnault}}, \bibinfo {author} {\bibfnamefont {J.}~\bibnamefont {Kang}}, \bibinfo {author} {\bibfnamefont {B.~A.}\ \bibnamefont {Bernevig}},\ and\ \bibinfo {author} {\bibfnamefont {O.}~\bibnamefont {Vafek}},\ }\bibfield  {title} {\bibinfo {title} {Kekul$\backslash$'e spiral order from strained topological heavy fermions},\ }\href@noop {} {\bibfield  {journal} {\bibinfo  {journal} {arXiv preprint arXiv:2502.08700}\ } (\bibinfo {year} {2025})}\BibitemShut {NoStop}%
\bibitem [{\citenamefont {Song}\ \emph {et~al.}(2021)\citenamefont {Song}, \citenamefont {Lian}, \citenamefont {Regnault},\ and\ \citenamefont {Bernevig}}]{bernevigSongTBGIIStable2021}%
  \BibitemOpen
  \bibfield  {author} {\bibinfo {author} {\bibfnamefont {Z.-D.}\ \bibnamefont {Song}}, \bibinfo {author} {\bibfnamefont {B.}~\bibnamefont {Lian}}, \bibinfo {author} {\bibfnamefont {N.}~\bibnamefont {Regnault}},\ and\ \bibinfo {author} {\bibfnamefont {B.~A.}\ \bibnamefont {Bernevig}},\ }\bibfield  {title} {\bibinfo {title} {{{TBG II}}: {{Stable Symmetry Anomaly}} in {{Twisted Bilayer Graphene}}},\ }\href {https://doi.org/10.1103/PhysRevB.103.205412} {\bibfield  {journal} {\bibinfo  {journal} {Physical Review B}\ }\textbf {\bibinfo {volume} {103}},\ \bibinfo {pages} {205412} (\bibinfo {year} {2021})},\ \Eprint {https://arxiv.org/abs/2009.11872} {arXiv:2009.11872} \BibitemShut {NoStop}%
\bibitem [{\citenamefont {Song}\ \emph {et~al.}(2019)\citenamefont {Song}, \citenamefont {Wang}, \citenamefont {Shi}, \citenamefont {Li}, \citenamefont {Fang},\ and\ \citenamefont {Bernevig}}]{bernevigSongAllMagicAngles2019}%
  \BibitemOpen
  \bibfield  {author} {\bibinfo {author} {\bibfnamefont {Z.}~\bibnamefont {Song}}, \bibinfo {author} {\bibfnamefont {Z.}~\bibnamefont {Wang}}, \bibinfo {author} {\bibfnamefont {W.}~\bibnamefont {Shi}}, \bibinfo {author} {\bibfnamefont {G.}~\bibnamefont {Li}}, \bibinfo {author} {\bibfnamefont {C.}~\bibnamefont {Fang}},\ and\ \bibinfo {author} {\bibfnamefont {B.~A.}\ \bibnamefont {Bernevig}},\ }\bibfield  {title} {\bibinfo {title} {All "{{Magic Angles}}" {{Are}} "{{Stable}}" {{Topological}}},\ }\href {https://doi.org/10.1103/PhysRevLett.123.036401} {\bibfield  {journal} {\bibinfo  {journal} {Physical Review Letters}\ }\textbf {\bibinfo {volume} {123}},\ \bibinfo {pages} {036401} (\bibinfo {year} {2019})},\ \Eprint {https://arxiv.org/abs/1807.10676} {arXiv:1807.10676} \BibitemShut {NoStop}%
\bibitem [{\citenamefont {Po}\ \emph {et~al.}(2019)\citenamefont {Po}, \citenamefont {Zou}, \citenamefont {Senthil},\ and\ \citenamefont {Vishwanath}}]{vishwanathPoFaithfulTightbindingModels2019}%
  \BibitemOpen
  \bibfield  {author} {\bibinfo {author} {\bibfnamefont {H.~C.}\ \bibnamefont {Po}}, \bibinfo {author} {\bibfnamefont {L.}~\bibnamefont {Zou}}, \bibinfo {author} {\bibfnamefont {T.}~\bibnamefont {Senthil}},\ and\ \bibinfo {author} {\bibfnamefont {A.}~\bibnamefont {Vishwanath}},\ }\bibfield  {title} {\bibinfo {title} {Faithful {{Tight-binding Models}} and {{Fragile Topology}} of {{Magic-angle Bilayer Graphene}}},\ }\href {https://doi.org/10.1103/PhysRevB.99.195455} {\bibfield  {journal} {\bibinfo  {journal} {Physical Review B}\ }\textbf {\bibinfo {volume} {99}},\ \bibinfo {pages} {195455} (\bibinfo {year} {2019})},\ \Eprint {https://arxiv.org/abs/1808.02482} {arXiv:1808.02482} \BibitemShut {NoStop}%
\bibitem [{\citenamefont {Ahn}\ \emph {et~al.}(2019)\citenamefont {Ahn}, \citenamefont {Park},\ and\ \citenamefont {Yang}}]{ahn2019}%
  \BibitemOpen
  \bibfield  {author} {\bibinfo {author} {\bibfnamefont {J.}~\bibnamefont {Ahn}}, \bibinfo {author} {\bibfnamefont {S.}~\bibnamefont {Park}},\ and\ \bibinfo {author} {\bibfnamefont {B.-J.}\ \bibnamefont {Yang}},\ }\bibfield  {title} {\bibinfo {title} {Failure of nielsen-ninomiya theorem and fragile topology in two-dimensional systems with space-time inversion symmetry: Application to twisted bilayer graphene at magic angle},\ }\href {https://doi.org/10.1103/PhysRevX.9.021013} {\bibfield  {journal} {\bibinfo  {journal} {Phys. Rev. X}\ }\textbf {\bibinfo {volume} {9}},\ \bibinfo {pages} {021013} (\bibinfo {year} {2019})}\BibitemShut {NoStop}%
\bibitem [{\citenamefont {Kang}\ and\ \citenamefont {Vafek}(2018)}]{vafekKangSymmetryMaximallyLocalized2018}%
  \BibitemOpen
  \bibfield  {author} {\bibinfo {author} {\bibfnamefont {J.}~\bibnamefont {Kang}}\ and\ \bibinfo {author} {\bibfnamefont {O.}~\bibnamefont {Vafek}},\ }\bibfield  {title} {\bibinfo {title} {Symmetry, maximally localized {{Wannier}} states, and low energy model for the twisted bilayer graphene narrow bands},\ }\href {https://doi.org/10.1103/PhysRevX.8.031088} {\bibfield  {journal} {\bibinfo  {journal} {Physical Review X}\ }\textbf {\bibinfo {volume} {8}},\ \bibinfo {pages} {031088} (\bibinfo {year} {2018})},\ \Eprint {https://arxiv.org/abs/1805.04918} {arXiv:1805.04918} \BibitemShut {NoStop}%
\bibitem [{\citenamefont {Koshino}\ \emph {et~al.}(2018)\citenamefont {Koshino}, \citenamefont {Yuan}, \citenamefont {Koretsune}, \citenamefont {Ochi}, \citenamefont {Kuroki},\ and\ \citenamefont {Fu}}]{fuKoshinoMaximallyLocalizedWannier2018}%
  \BibitemOpen
  \bibfield  {author} {\bibinfo {author} {\bibfnamefont {M.}~\bibnamefont {Koshino}}, \bibinfo {author} {\bibfnamefont {N.~F.~Q.}\ \bibnamefont {Yuan}}, \bibinfo {author} {\bibfnamefont {T.}~\bibnamefont {Koretsune}}, \bibinfo {author} {\bibfnamefont {M.}~\bibnamefont {Ochi}}, \bibinfo {author} {\bibfnamefont {K.}~\bibnamefont {Kuroki}},\ and\ \bibinfo {author} {\bibfnamefont {L.}~\bibnamefont {Fu}},\ }\bibfield  {title} {\bibinfo {title} {Maximally {{Localized Wannier Orbitals}} and the {{Extended Hubbard Model}} for {{Twisted Bilayer Graphene}}},\ }\href {https://doi.org/10.1103/PhysRevX.8.031087} {\bibfield  {journal} {\bibinfo  {journal} {Physical Review X}\ }\textbf {\bibinfo {volume} {8}},\ \bibinfo {pages} {031087} (\bibinfo {year} {2018})}\BibitemShut {NoStop}%
\bibitem [{\citenamefont {Zou}\ \emph {et~al.}(2018)\citenamefont {Zou}, \citenamefont {Po}, \citenamefont {Vishwanath},\ and\ \citenamefont {Senthil}}]{senthilZouBandStructureTwisted2018}%
  \BibitemOpen
  \bibfield  {author} {\bibinfo {author} {\bibfnamefont {L.}~\bibnamefont {Zou}}, \bibinfo {author} {\bibfnamefont {H.~C.}\ \bibnamefont {Po}}, \bibinfo {author} {\bibfnamefont {A.}~\bibnamefont {Vishwanath}},\ and\ \bibinfo {author} {\bibfnamefont {T.}~\bibnamefont {Senthil}},\ }\bibfield  {title} {\bibinfo {title} {Band structure of twisted bilayer graphene: {{Emergent}} symmetries, commensurate approximants, and {{Wannier}} obstructions},\ }\href {https://doi.org/10.1103/PhysRevB.98.085435} {\bibfield  {journal} {\bibinfo  {journal} {Physical Review B}\ }\textbf {\bibinfo {volume} {98}},\ \bibinfo {pages} {085435} (\bibinfo {year} {2018})}\BibitemShut {NoStop}%
\bibitem [{\citenamefont {Xie}\ \emph {et~al.}(2019)\citenamefont {Xie}, \citenamefont {Lian}, \citenamefont {Jäck}, \citenamefont {Liu}, \citenamefont {Chiu}, \citenamefont {Watanabe}, \citenamefont {Taniguchi}, \citenamefont {Bernevig},\ and\ \citenamefont {Yazdani}}]{yazdaniXieSpectroscopicSignaturesManybody2019}%
  \BibitemOpen
  \bibfield  {author} {\bibinfo {author} {\bibfnamefont {Y.}~\bibnamefont {Xie}}, \bibinfo {author} {\bibfnamefont {B.}~\bibnamefont {Lian}}, \bibinfo {author} {\bibfnamefont {B.}~\bibnamefont {Jäck}}, \bibinfo {author} {\bibfnamefont {X.}~\bibnamefont {Liu}}, \bibinfo {author} {\bibfnamefont {C.-L.}\ \bibnamefont {Chiu}}, \bibinfo {author} {\bibfnamefont {K.}~\bibnamefont {Watanabe}}, \bibinfo {author} {\bibfnamefont {T.}~\bibnamefont {Taniguchi}}, \bibinfo {author} {\bibfnamefont {B.~A.}\ \bibnamefont {Bernevig}},\ and\ \bibinfo {author} {\bibfnamefont {A.}~\bibnamefont {Yazdani}},\ }\bibfield  {title} {\bibinfo {title} {Spectroscopic signatures of many-body correlations in magic-angle twisted bilayer graphene},\ }\href {https://doi.org/10.1038/s41586-019-1422-x} {\bibfield  {journal} {\bibinfo  {journal} {Nature}\ }\textbf {\bibinfo {volume} {572}},\ \bibinfo {pages} {101} (\bibinfo {year} {2019})}\BibitemShut {NoStop}%
\bibitem [{\citenamefont {Saito}\ \emph {et~al.}(2021)\citenamefont {Saito}, \citenamefont {Yang}, \citenamefont {Ge}, \citenamefont {Liu}, \citenamefont {Watanabe}, \citenamefont {Taniguchi}, \citenamefont {Li}, \citenamefont {Berg},\ and\ \citenamefont {Young}}]{youngSaitoIsospinPomeranchukEffect2021}%
  \BibitemOpen
  \bibfield  {author} {\bibinfo {author} {\bibfnamefont {Y.}~\bibnamefont {Saito}}, \bibinfo {author} {\bibfnamefont {F.}~\bibnamefont {Yang}}, \bibinfo {author} {\bibfnamefont {J.}~\bibnamefont {Ge}}, \bibinfo {author} {\bibfnamefont {X.}~\bibnamefont {Liu}}, \bibinfo {author} {\bibfnamefont {K.}~\bibnamefont {Watanabe}}, \bibinfo {author} {\bibfnamefont {T.}~\bibnamefont {Taniguchi}}, \bibinfo {author} {\bibfnamefont {J.~I.~A.}\ \bibnamefont {Li}}, \bibinfo {author} {\bibfnamefont {E.}~\bibnamefont {Berg}},\ and\ \bibinfo {author} {\bibfnamefont {A.~F.}\ \bibnamefont {Young}},\ }\bibfield  {title} {\bibinfo {title} {Isospin {{Pomeranchuk}} effect and the entropy of collective excitations in twisted bilayer graphene},\ }\href {https://doi.org/10.1038/s41586-021-03409-2} {\bibfield  {journal} {\bibinfo  {journal} {Nature}\ }\textbf {\bibinfo {volume} {592}},\ \bibinfo {pages} {220} (\bibinfo {year} {2021})},\ \Eprint {https://arxiv.org/abs/2008.10830} {arXiv:2008.10830} \BibitemShut {NoStop}%
\bibitem [{\citenamefont {Rozen}\ \emph {et~al.}(2021)\citenamefont {Rozen}, \citenamefont {Park}, \citenamefont {Zondiner}, \citenamefont {Cao}, \citenamefont {Rodan-Legrain}, \citenamefont {Taniguchi}, \citenamefont {Watanabe}, \citenamefont {Oreg}, \citenamefont {Stern}, \citenamefont {Berg}, \citenamefont {Jarillo-Herrero},\ and\ \citenamefont {Ilani}}]{ilaniRozenEntropicEvidencePomeranchuk}%
  \BibitemOpen
  \bibfield  {author} {\bibinfo {author} {\bibfnamefont {A.}~\bibnamefont {Rozen}}, \bibinfo {author} {\bibfnamefont {J.~M.}\ \bibnamefont {Park}}, \bibinfo {author} {\bibfnamefont {U.}~\bibnamefont {Zondiner}}, \bibinfo {author} {\bibfnamefont {Y.}~\bibnamefont {Cao}}, \bibinfo {author} {\bibfnamefont {D.}~\bibnamefont {Rodan-Legrain}}, \bibinfo {author} {\bibfnamefont {T.}~\bibnamefont {Taniguchi}}, \bibinfo {author} {\bibfnamefont {K.}~\bibnamefont {Watanabe}}, \bibinfo {author} {\bibfnamefont {Y.}~\bibnamefont {Oreg}}, \bibinfo {author} {\bibfnamefont {A.}~\bibnamefont {Stern}}, \bibinfo {author} {\bibfnamefont {E.}~\bibnamefont {Berg}}, \bibinfo {author} {\bibfnamefont {P.}~\bibnamefont {Jarillo-Herrero}},\ and\ \bibinfo {author} {\bibfnamefont {S.}~\bibnamefont {Ilani}},\ }\bibfield  {title} {\bibinfo {title} {Entropic evidence for a pomeranchuk effect in magic-angle graphene},\ }\href {https://doi.org/10.1038/s41586-021-03319-3} {\bibfield  {journal} {\bibinfo  {journal} {Nature}\ }\textbf {\bibinfo {volume} {592}},\ \bibinfo {pages} {214} (\bibinfo {year} {2021})}\BibitemShut {NoStop}%
\bibitem [{\citenamefont {Ledwith}\ \emph {et~al.}(2025)\citenamefont {Ledwith}, \citenamefont {Dong}, \citenamefont {Vishwanath},\ and\ \citenamefont {Khalaf}}]{khalafLedwithNonlocalMomentsChern2025}%
  \BibitemOpen
  \bibfield  {author} {\bibinfo {author} {\bibfnamefont {P.~J.}\ \bibnamefont {Ledwith}}, \bibinfo {author} {\bibfnamefont {J.}~\bibnamefont {Dong}}, \bibinfo {author} {\bibfnamefont {A.}~\bibnamefont {Vishwanath}},\ and\ \bibinfo {author} {\bibfnamefont {E.}~\bibnamefont {Khalaf}},\ }\href {https://doi.org/10.48550/arXiv.2408.16761} {\bibinfo {title} {Nonlocal {{Moments}} in the {{Chern Bands}} of {{Twisted Bilayer Graphene}}}} (\bibinfo {year} {2025}),\ \Eprint {https://arxiv.org/abs/2408.16761} {arXiv:2408.16761 [cond-mat]} \BibitemShut {NoStop}%
\bibitem [{\citenamefont {{Herzog-Arbeitman}}\ \emph {et~al.}(2024)\citenamefont {{Herzog-Arbeitman}}, \citenamefont {Yu}, \citenamefont {C{\u a}lug{\u a}ru}, \citenamefont {Hu}, \citenamefont {Regnault}, \citenamefont {Liu}, \citenamefont {Vafek}, \citenamefont {Coleman}, \citenamefont {Tsvelik}, \citenamefont {Song},\ and\ \citenamefont {Bernevig}}]{bernevigHerzog-ArbeitmanTopologicalHeavyFermion2024}%
  \BibitemOpen
  \bibfield  {author} {\bibinfo {author} {\bibfnamefont {J.}~\bibnamefont {{Herzog-Arbeitman}}}, \bibinfo {author} {\bibfnamefont {J.}~\bibnamefont {Yu}}, \bibinfo {author} {\bibfnamefont {D.}~\bibnamefont {C{\u a}lug{\u a}ru}}, \bibinfo {author} {\bibfnamefont {H.}~\bibnamefont {Hu}}, \bibinfo {author} {\bibfnamefont {N.}~\bibnamefont {Regnault}}, \bibinfo {author} {\bibfnamefont {C.}~\bibnamefont {Liu}}, \bibinfo {author} {\bibfnamefont {O.}~\bibnamefont {Vafek}}, \bibinfo {author} {\bibfnamefont {P.}~\bibnamefont {Coleman}}, \bibinfo {author} {\bibfnamefont {A.}~\bibnamefont {Tsvelik}}, \bibinfo {author} {\bibfnamefont {Z.-d.}\ \bibnamefont {Song}},\ and\ \bibinfo {author} {\bibfnamefont {B.~A.}\ \bibnamefont {Bernevig}},\ }\href {https://doi.org/10.48550/arXiv.2404.07253} {\bibinfo {title} {Topological {{Heavy Fermion Principle For Flat}} ({{Narrow}}) {{Bands With Concentrated Quantum Geometry}}}} (\bibinfo {year} {2024}),\ \Eprint {https://arxiv.org/abs/2404.07253} {arXiv:2404.07253 [cond-mat]} \BibitemShut {NoStop}%
\bibitem [{\citenamefont {Jiang}\ \emph {et~al.}(2024)\citenamefont {Jiang}, \citenamefont {Petralanda}, \citenamefont {Skorupskii}, \citenamefont {Xu}, \citenamefont {Pi}, \citenamefont {C{\u a}lug{\u a}ru}, \citenamefont {Hu}, \citenamefont {Xie}, \citenamefont {Mustaf}, \citenamefont {H{\"o}hn}, \citenamefont {Haase}, \citenamefont {Vergniory}, \citenamefont {Claassen}, \citenamefont {Elcoro}, \citenamefont {Regnault}, \citenamefont {Shan}, \citenamefont {Mak}, \citenamefont {Efetov}, \citenamefont {Morosan}, \citenamefont {Kennes}, \citenamefont {Rubio}, \citenamefont {Xian}, \citenamefont {Felser}, \citenamefont {Schoop},\ and\ \citenamefont {Bernevig}}]{bernevigJiang2DTheoreticallyTwistable2024}%
  \BibitemOpen
  \bibfield  {author} {\bibinfo {author} {\bibfnamefont {Y.}~\bibnamefont {Jiang}}, \bibinfo {author} {\bibfnamefont {U.}~\bibnamefont {Petralanda}}, \bibinfo {author} {\bibfnamefont {G.}~\bibnamefont {Skorupskii}}, \bibinfo {author} {\bibfnamefont {Q.}~\bibnamefont {Xu}}, \bibinfo {author} {\bibfnamefont {H.}~\bibnamefont {Pi}}, \bibinfo {author} {\bibfnamefont {D.}~\bibnamefont {C{\u a}lug{\u a}ru}}, \bibinfo {author} {\bibfnamefont {H.}~\bibnamefont {Hu}}, \bibinfo {author} {\bibfnamefont {J.}~\bibnamefont {Xie}}, \bibinfo {author} {\bibfnamefont {R.~A.}\ \bibnamefont {Mustaf}}, \bibinfo {author} {\bibfnamefont {P.}~\bibnamefont {H{\"o}hn}}, \bibinfo {author} {\bibfnamefont {V.}~\bibnamefont {Haase}}, \bibinfo {author} {\bibfnamefont {M.~G.}\ \bibnamefont {Vergniory}}, \bibinfo {author} {\bibfnamefont {M.}~\bibnamefont {Claassen}}, \bibinfo {author} {\bibfnamefont {L.}~\bibnamefont {Elcoro}}, \bibinfo {author} {\bibfnamefont {N.}~\bibnamefont {Regnault}}, \bibinfo {author} {\bibfnamefont {J.}~\bibnamefont {Shan}}, \bibinfo {author} {\bibfnamefont {K.~F.}\ \bibnamefont {Mak}}, \bibinfo {author} {\bibfnamefont {D.~K.}\ \bibnamefont {Efetov}}, \bibinfo {author} {\bibfnamefont {E.}~\bibnamefont {Morosan}}, \bibinfo {author} {\bibfnamefont {D.~M.}\ \bibnamefont {Kennes}}, \bibinfo {author} {\bibfnamefont {A.}~\bibnamefont {Rubio}}, \bibinfo {author} {\bibfnamefont {L.}~\bibnamefont {Xian}}, \bibinfo {author} {\bibfnamefont {C.}~\bibnamefont {Felser}}, \bibinfo {author} {\bibfnamefont {L.~M.}\ \bibnamefont {Schoop}},\ and\ \bibinfo {author} {\bibfnamefont {B.~A.}\ \bibnamefont {Bernevig}},\ }\href {https://doi.org/10.48550/arXiv.2411.09741} {\bibinfo {title} {{{2D Theoretically Twistable Material Database}}}} (\bibinfo {year} {2024}),\ \Eprint {https://arxiv.org/abs/2411.09741} {arXiv:2411.09741 [cond-mat]} \BibitemShut {NoStop}%
\bibitem [{\citenamefont {Choi}\ \emph {et~al.}(2024)\citenamefont {Choi}, \citenamefont {Choi}, \citenamefont {Valentini}, \citenamefont {Patterson}, \citenamefont {Holleis}, \citenamefont {Sheekey}, \citenamefont {Stoyanov}, \citenamefont {Cheng}, \citenamefont {Taniguchi}, \citenamefont {Watanabe},\ and\ \citenamefont {Young}}]{youngChoiElectricFieldControl2024}%
  \BibitemOpen
  \bibfield  {author} {\bibinfo {author} {\bibfnamefont {Y.}~\bibnamefont {Choi}}, \bibinfo {author} {\bibfnamefont {Y.}~\bibnamefont {Choi}}, \bibinfo {author} {\bibfnamefont {M.}~\bibnamefont {Valentini}}, \bibinfo {author} {\bibfnamefont {C.~L.}\ \bibnamefont {Patterson}}, \bibinfo {author} {\bibfnamefont {L.~F.~W.}\ \bibnamefont {Holleis}}, \bibinfo {author} {\bibfnamefont {O.~I.}\ \bibnamefont {Sheekey}}, \bibinfo {author} {\bibfnamefont {H.}~\bibnamefont {Stoyanov}}, \bibinfo {author} {\bibfnamefont {X.}~\bibnamefont {Cheng}}, \bibinfo {author} {\bibfnamefont {T.}~\bibnamefont {Taniguchi}}, \bibinfo {author} {\bibfnamefont {K.}~\bibnamefont {Watanabe}},\ and\ \bibinfo {author} {\bibfnamefont {A.~F.}\ \bibnamefont {Young}},\ }\href {https://doi.org/10.48550/arXiv.2408.12584} {\bibinfo {title} {Electric field control of superconductivity and quantized anomalous {{Hall}} effects in rhombohedral tetralayer graphene}} (\bibinfo {year} {2024}),\ \Eprint {https://arxiv.org/abs/2408.12584} {arXiv:2408.12584} \BibitemShut {NoStop}%
\bibitem [{\citenamefont {Han}\ \emph {et~al.}(2025)\citenamefont {Han}, \citenamefont {Lu}, \citenamefont {Hadjri}, \citenamefont {Shi}, \citenamefont {Wu}, \citenamefont {Xu}, \citenamefont {Yao}, \citenamefont {Cotten}, \citenamefont {Sedeh}, \citenamefont {Weldeyesus}, \citenamefont {Yang}, \citenamefont {Seo}, \citenamefont {Ye}, \citenamefont {Zhou}, \citenamefont {Liu}, \citenamefont {Shi}, \citenamefont {Hua}, \citenamefont {Watanabe}, \citenamefont {Taniguchi}, \citenamefont {Xiong}, \citenamefont {Zumb{\"u}hl}, \citenamefont {Fu},\ and\ \citenamefont {Ju}}]{juHanSignaturesChiralSuperconductivity2025}%
  \BibitemOpen
  \bibfield  {author} {\bibinfo {author} {\bibfnamefont {T.}~\bibnamefont {Han}}, \bibinfo {author} {\bibfnamefont {Z.}~\bibnamefont {Lu}}, \bibinfo {author} {\bibfnamefont {Z.}~\bibnamefont {Hadjri}}, \bibinfo {author} {\bibfnamefont {L.}~\bibnamefont {Shi}}, \bibinfo {author} {\bibfnamefont {Z.}~\bibnamefont {Wu}}, \bibinfo {author} {\bibfnamefont {W.}~\bibnamefont {Xu}}, \bibinfo {author} {\bibfnamefont {Y.}~\bibnamefont {Yao}}, \bibinfo {author} {\bibfnamefont {A.~A.}\ \bibnamefont {Cotten}}, \bibinfo {author} {\bibfnamefont {O.~S.}\ \bibnamefont {Sedeh}}, \bibinfo {author} {\bibfnamefont {H.}~\bibnamefont {Weldeyesus}}, \bibinfo {author} {\bibfnamefont {J.}~\bibnamefont {Yang}}, \bibinfo {author} {\bibfnamefont {J.}~\bibnamefont {Seo}}, \bibinfo {author} {\bibfnamefont {S.}~\bibnamefont {Ye}}, \bibinfo {author} {\bibfnamefont {M.}~\bibnamefont {Zhou}}, \bibinfo {author} {\bibfnamefont {H.}~\bibnamefont {Liu}}, \bibinfo {author} {\bibfnamefont {G.}~\bibnamefont {Shi}}, \bibinfo {author} {\bibfnamefont {Z.}~\bibnamefont {Hua}}, \bibinfo {author} {\bibfnamefont {K.}~\bibnamefont {Watanabe}}, \bibinfo {author} {\bibfnamefont {T.}~\bibnamefont {Taniguchi}}, \bibinfo {author} {\bibfnamefont {P.}~\bibnamefont {Xiong}}, \bibinfo {author} {\bibfnamefont {D.~M.}\ \bibnamefont {Zumb{\"u}hl}}, \bibinfo {author} {\bibfnamefont {L.}~\bibnamefont {Fu}},\ and\ \bibinfo {author} {\bibfnamefont {L.}~\bibnamefont {Ju}},\ }\bibfield  {title} {\bibinfo {title} {Signatures of chiral superconductivity in rhombohedral graphene},\ }\href {https://doi.org/10.1038/s41586-025-09169-7} {\bibfield  {journal} {\bibinfo  {journal} {Nature}\ ,\ \bibinfo {pages} {1}} (\bibinfo {year} {2025})}\BibitemShut {NoStop}%
\bibitem [{\citenamefont {Waters}\ \emph {et~al.}(2025)\citenamefont {Waters}, \citenamefont {Okounkova}, \citenamefont {Su}, \citenamefont {Zhou}, \citenamefont {Yao}, \citenamefont {Watanabe}, \citenamefont {Taniguchi}, \citenamefont {Xu}, \citenamefont {Zhang}, \citenamefont {Folk},\ and\ \citenamefont {Yankowitz}}]{yankowitzWatersChernInsulatorsInteger2025}%
  \BibitemOpen
  \bibfield  {author} {\bibinfo {author} {\bibfnamefont {D.}~\bibnamefont {Waters}}, \bibinfo {author} {\bibfnamefont {A.}~\bibnamefont {Okounkova}}, \bibinfo {author} {\bibfnamefont {R.}~\bibnamefont {Su}}, \bibinfo {author} {\bibfnamefont {B.}~\bibnamefont {Zhou}}, \bibinfo {author} {\bibfnamefont {J.}~\bibnamefont {Yao}}, \bibinfo {author} {\bibfnamefont {K.}~\bibnamefont {Watanabe}}, \bibinfo {author} {\bibfnamefont {T.}~\bibnamefont {Taniguchi}}, \bibinfo {author} {\bibfnamefont {X.}~\bibnamefont {Xu}}, \bibinfo {author} {\bibfnamefont {Y.-H.}\ \bibnamefont {Zhang}}, \bibinfo {author} {\bibfnamefont {J.}~\bibnamefont {Folk}},\ and\ \bibinfo {author} {\bibfnamefont {M.}~\bibnamefont {Yankowitz}},\ }\bibfield  {title} {\bibinfo {title} {Chern {{Insulators}} at {{Integer}} and {{Fractional Filling}} in {{Moir}}{\textbackslash}'e {{Pentalayer Graphene}}},\ }\href {https://doi.org/10.1103/PhysRevX.15.011045} {\bibfield  {journal} {\bibinfo  {journal} {Physical Review X}\ }\textbf {\bibinfo {volume} {15}},\ \bibinfo {pages} {011045} (\bibinfo {year} {2025})}\BibitemShut {NoStop}%
\bibitem [{\citenamefont {Kang}\ and\ \citenamefont {Vafek}(2019)}]{vafek2019}%
  \BibitemOpen
  \bibfield  {author} {\bibinfo {author} {\bibfnamefont {J.}~\bibnamefont {Kang}}\ and\ \bibinfo {author} {\bibfnamefont {O.}~\bibnamefont {Vafek}},\ }\bibfield  {title} {\bibinfo {title} {Strong coupling phases of partially filled twisted bilayer graphene narrow bands},\ }\href {https://doi.org/10.1103/PhysRevLett.122.246401} {\bibfield  {journal} {\bibinfo  {journal} {Phys. Rev. Lett.}\ }\textbf {\bibinfo {volume} {122}},\ \bibinfo {pages} {246401} (\bibinfo {year} {2019})}\BibitemShut {NoStop}%
\bibitem [{\citenamefont {Bultinck}\ \emph {et~al.}(2020)\citenamefont {Bultinck}, \citenamefont {Khalaf}, \citenamefont {Liu}, \citenamefont {Chatterjee}, \citenamefont {Vishwanath},\ and\ \citenamefont {Zaletel}}]{zaletelBultinckGroundStateHidden2020}%
  \BibitemOpen
  \bibfield  {author} {\bibinfo {author} {\bibfnamefont {N.}~\bibnamefont {Bultinck}}, \bibinfo {author} {\bibfnamefont {E.}~\bibnamefont {Khalaf}}, \bibinfo {author} {\bibfnamefont {S.}~\bibnamefont {Liu}}, \bibinfo {author} {\bibfnamefont {S.}~\bibnamefont {Chatterjee}}, \bibinfo {author} {\bibfnamefont {A.}~\bibnamefont {Vishwanath}},\ and\ \bibinfo {author} {\bibfnamefont {M.~P.}\ \bibnamefont {Zaletel}},\ }\bibfield  {title} {\bibinfo {title} {Ground {{State}} and {{Hidden Symmetry}} of {{Magic Angle Graphene}} at {{Even Integer Filling}}},\ }\href {https://doi.org/10.1103/PhysRevX.10.031034} {\bibfield  {journal} {\bibinfo  {journal} {Physical Review X}\ }\textbf {\bibinfo {volume} {10}},\ \bibinfo {pages} {031034} (\bibinfo {year} {2020})},\ \Eprint {https://arxiv.org/abs/1911.02045} {arXiv:1911.02045} \BibitemShut {NoStop}%
\bibitem [{\citenamefont {Kwan}\ \emph {et~al.}(2021)\citenamefont {Kwan}, \citenamefont {Wagner}, \citenamefont {Soejima}, \citenamefont {Zaletel}, \citenamefont {Simon}, \citenamefont {Parameswaran},\ and\ \citenamefont {Bultinck}}]{bultinckKwanKekuleSpiralOrder2021}%
  \BibitemOpen
  \bibfield  {author} {\bibinfo {author} {\bibfnamefont {Y.~H.}\ \bibnamefont {Kwan}}, \bibinfo {author} {\bibfnamefont {G.}~\bibnamefont {Wagner}}, \bibinfo {author} {\bibfnamefont {T.}~\bibnamefont {Soejima}}, \bibinfo {author} {\bibfnamefont {M.~P.}\ \bibnamefont {Zaletel}}, \bibinfo {author} {\bibfnamefont {S.~H.}\ \bibnamefont {Simon}}, \bibinfo {author} {\bibfnamefont {S.~A.}\ \bibnamefont {Parameswaran}},\ and\ \bibinfo {author} {\bibfnamefont {N.}~\bibnamefont {Bultinck}},\ }\bibfield  {title} {\bibinfo {title} {Kekul\'e spiral order at all nonzero integer fillings in twisted bilayer graphene},\ }\href {https://doi.org/10.1103/PhysRevX.11.041063} {\bibfield  {journal} {\bibinfo  {journal} {Phys. Rev. X}\ }\textbf {\bibinfo {volume} {11}},\ \bibinfo {pages} {041063} (\bibinfo {year} {2021})}\BibitemShut {NoStop}%
\bibitem [{\citenamefont {Lin}\ \emph {et~al.}(2025)\citenamefont {Lin}, \citenamefont {Yang}, \citenamefont {Lu}, \citenamefont {Zhai},\ and\ \citenamefont {Yao}}]{LinYangLuZhaiYao2025}%
  \BibitemOpen
  \bibfield  {author} {\bibinfo {author} {\bibfnamefont {Z.}~\bibnamefont {Lin}}, \bibinfo {author} {\bibfnamefont {W.}~\bibnamefont {Yang}}, \bibinfo {author} {\bibfnamefont {H.}~\bibnamefont {Lu}}, \bibinfo {author} {\bibfnamefont {D.}~\bibnamefont {Zhai}},\ and\ \bibinfo {author} {\bibfnamefont {W.}~\bibnamefont {Yao}},\ }\bibfield  {title} {\bibinfo {title} {Fractional chern insulator states in an isolated flat band of zero chern number},\ }\href {https://arxiv.org/abs/2505.09009} {\bibfield  {journal} {\bibinfo  {journal} {arXiv:2505.09009}\ } (\bibinfo {year} {2025})}\BibitemShut {NoStop}%
\bibitem [{\citenamefont {Simon}\ \emph {et~al.}(2015)\citenamefont {Simon}, \citenamefont {Harper},\ and\ \citenamefont {Read}}]{PhysRevB.92.195104}%
  \BibitemOpen
  \bibfield  {author} {\bibinfo {author} {\bibfnamefont {S.~H.}\ \bibnamefont {Simon}}, \bibinfo {author} {\bibfnamefont {F.}~\bibnamefont {Harper}},\ and\ \bibinfo {author} {\bibfnamefont {N.}~\bibnamefont {Read}},\ }\bibfield  {title} {\bibinfo {title} {Fractional chern insulators in bands with zero berry curvature},\ }\href {https://doi.org/10.1103/PhysRevB.92.195104} {\bibfield  {journal} {\bibinfo  {journal} {Phys. Rev. B}\ }\textbf {\bibinfo {volume} {92}},\ \bibinfo {pages} {195104} (\bibinfo {year} {2015})}\BibitemShut {NoStop}%
\end{thebibliography}%

\end{document}